
\def\epsfpreprint{Y}   

\def\draftversion{N}   
\def\preprint{Y}       

\input jnl \refstylenp \def\cit#1{[#1]}

\def\physica{\journal Physica,} \def\zfp{\journal Z. Phys.}

\if \draftversion N \input reforder \input eqnorder \citeall\cit \fi

\if \preprint N \def\epsfpreprint{N} \fi
\if \epsfpreprint Y \input epsf \fi

\def\fp{f_\pi } \def\zpar{\cal Z }  \def\fhi{\vec \phi}
\def\Fhi{\vec \Phi} \def\ratio{{{m_R}\over{f_\pi}}} \def\half{{1 \over
2}} \def\fs{~~.} \def\cm{~~,}
\def\d{$^\dagger$}
\def\b{$^\#$}

\def\figure#1#2#3{\if \epsfpreprint Y \midinsert \epsfxsize=#3truein
\centerline{\epsffile{fig_#1_eps}} \halign{##\hfill\quad
&\vtop{\parindent=0pt \hsize=5.5in \strut## \strut}\cr {\bf Figure
#1}&#2 \cr} \endinsert \fi}

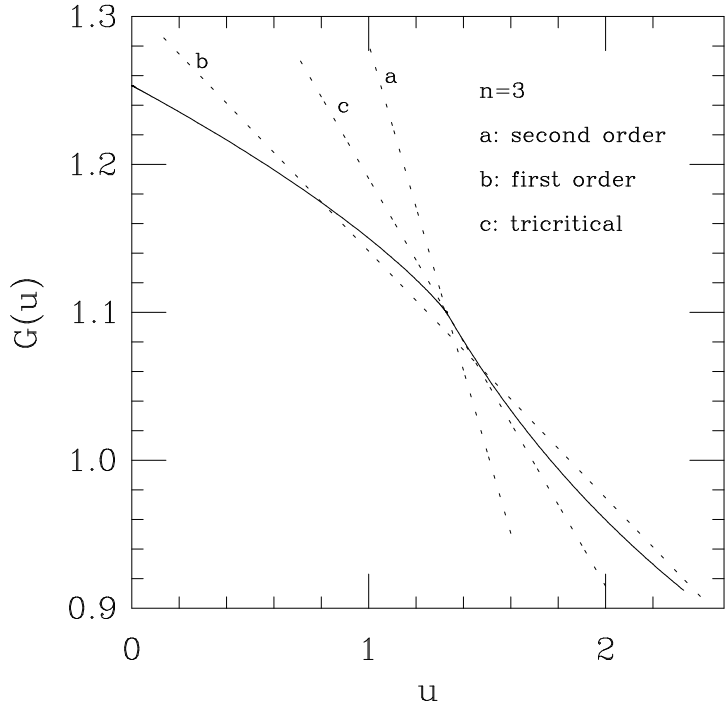
\def\captionthreeone{The solid line is the function $G(u)$ for $n=3$ and
the straight  dotted lines represent  $-g^*(u-u^*)$ for three typical
transition points.  All candidate saddles  are found at intersections
between the straight line and the function $G$;p the coupling constants in
the action only affect $g^*$ and $u^*$.}

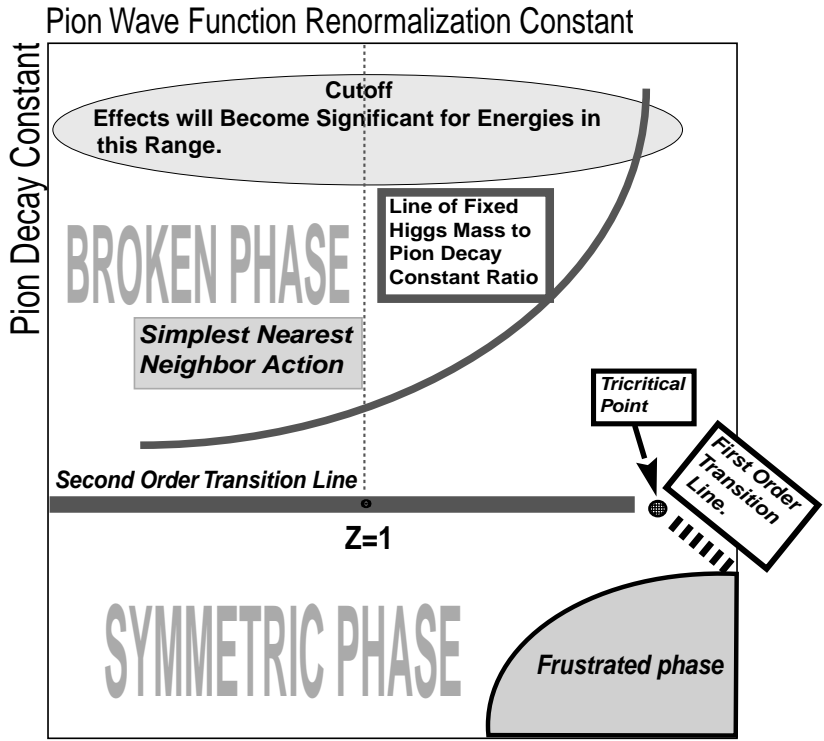
\def\captionthreetwo{Generic phase diagram containing the qualitative
features common to all models investigated.}

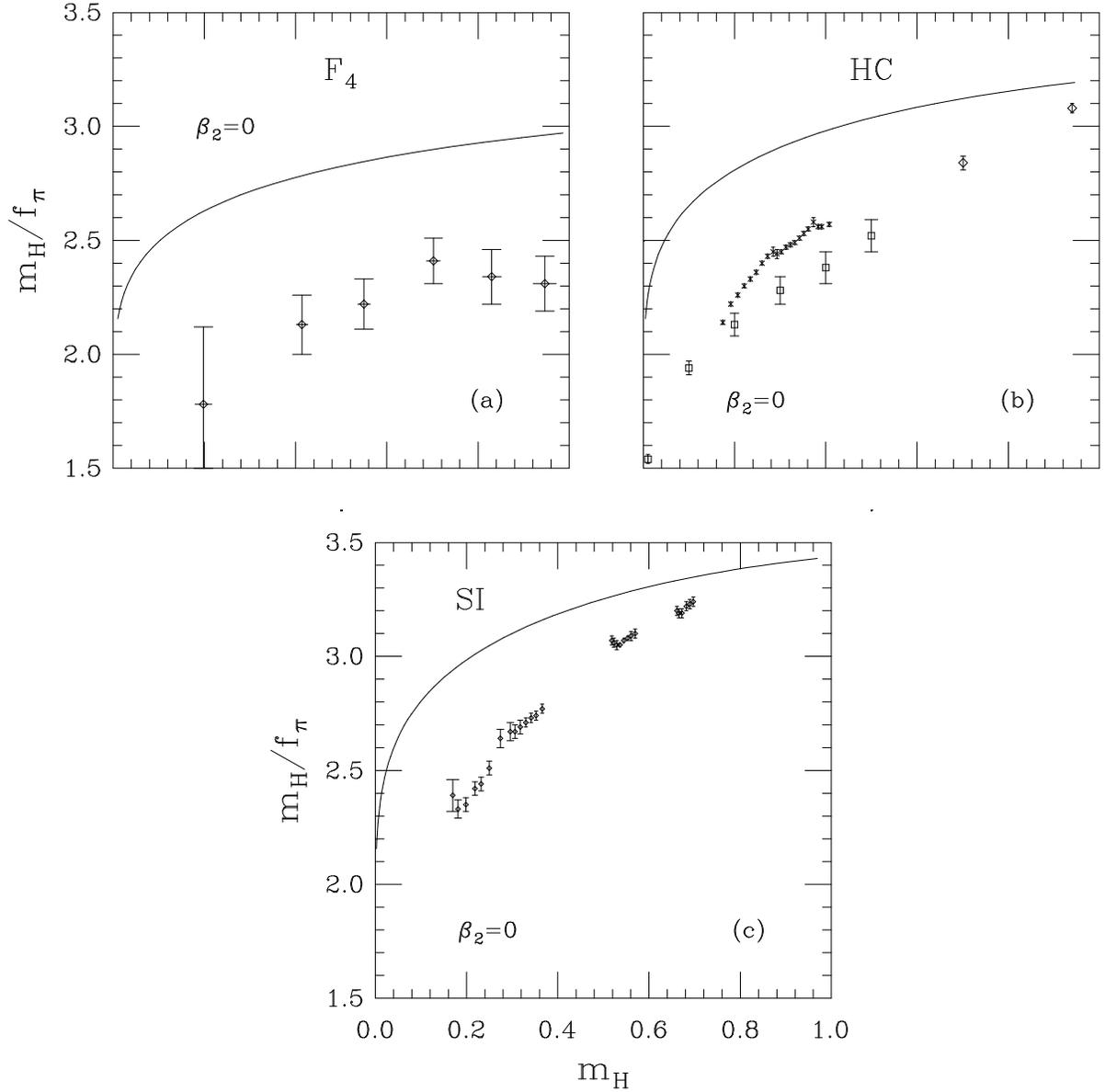
\def\captionsevenone{$m_H/f_\pi$ vs. $m_H$ for the na\"\i ve ($\beta_2=0$)
lattice actions. The solid line is the large $N$ result scaled to $N=4$.
The diamonds in fig.  (a) are numerical results from \cit{BBHN2}. In fig.
(b) the boxes are results from \cit{LW}, the diamonds are numerical results
from \cit{GOCK1}, and the crosses from \cit{GOCK2}. In fig (c) the diamonds
are numerical results from \cit{GOCK2}.}

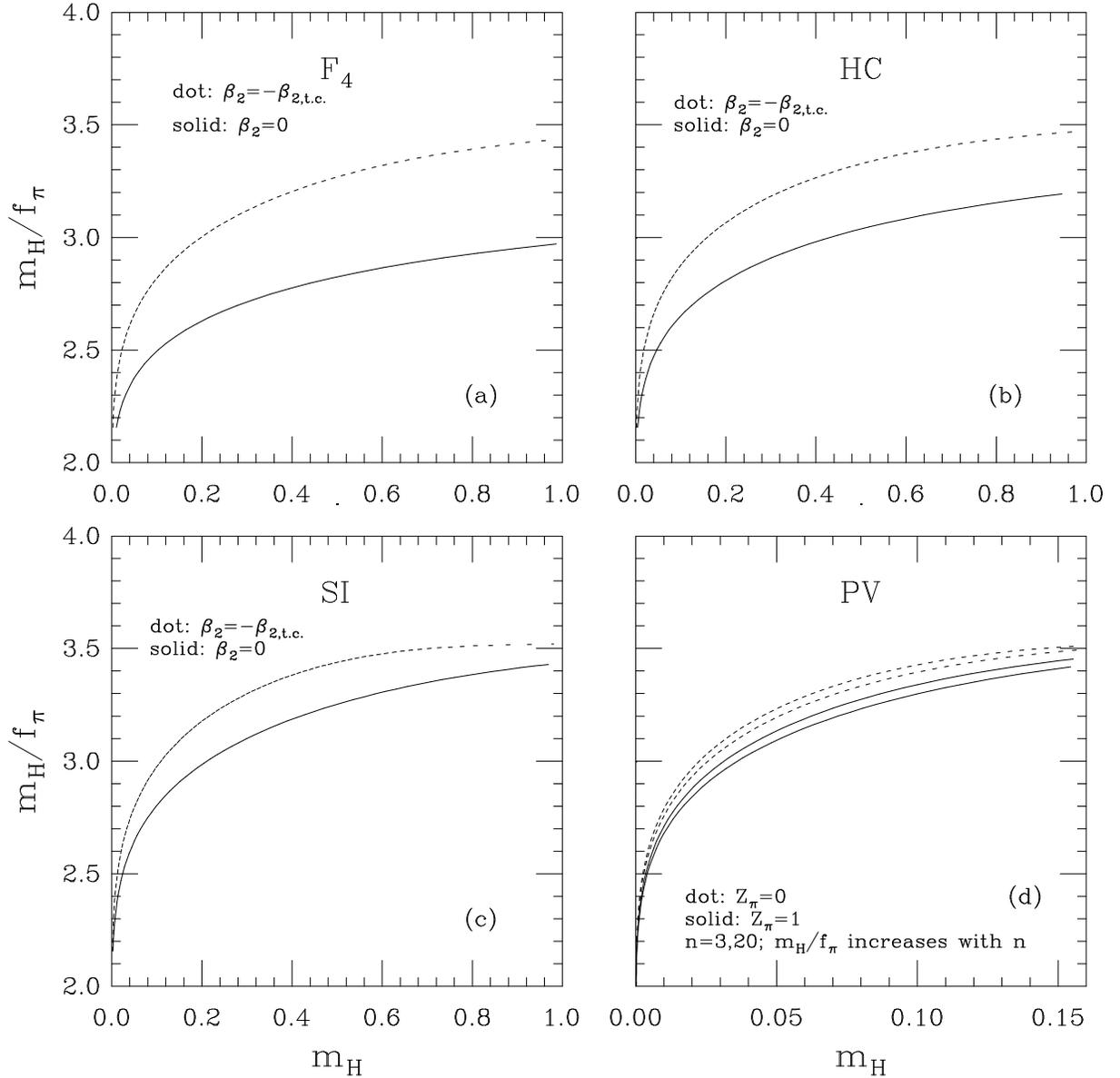
\def\captionseventwo{$m_H/f_\pi$ vs. $m_H$: The solid lines represent the
na\"\i ve actions ($\beta_2=0$ on the lattice and $Z_\pi=1$ for PV) the
dotted lines the actions with a four derivative term turned on to maximal
allowed strength ($\beta_2=-\beta_{2,t.c.}$ on the lattice and $Z_\pi=0$
for PV).  In figures (a)--(c) $m_H$ is measured in lattice units and in
fig. (d) it is measured in units of $\Lambda_s$.}

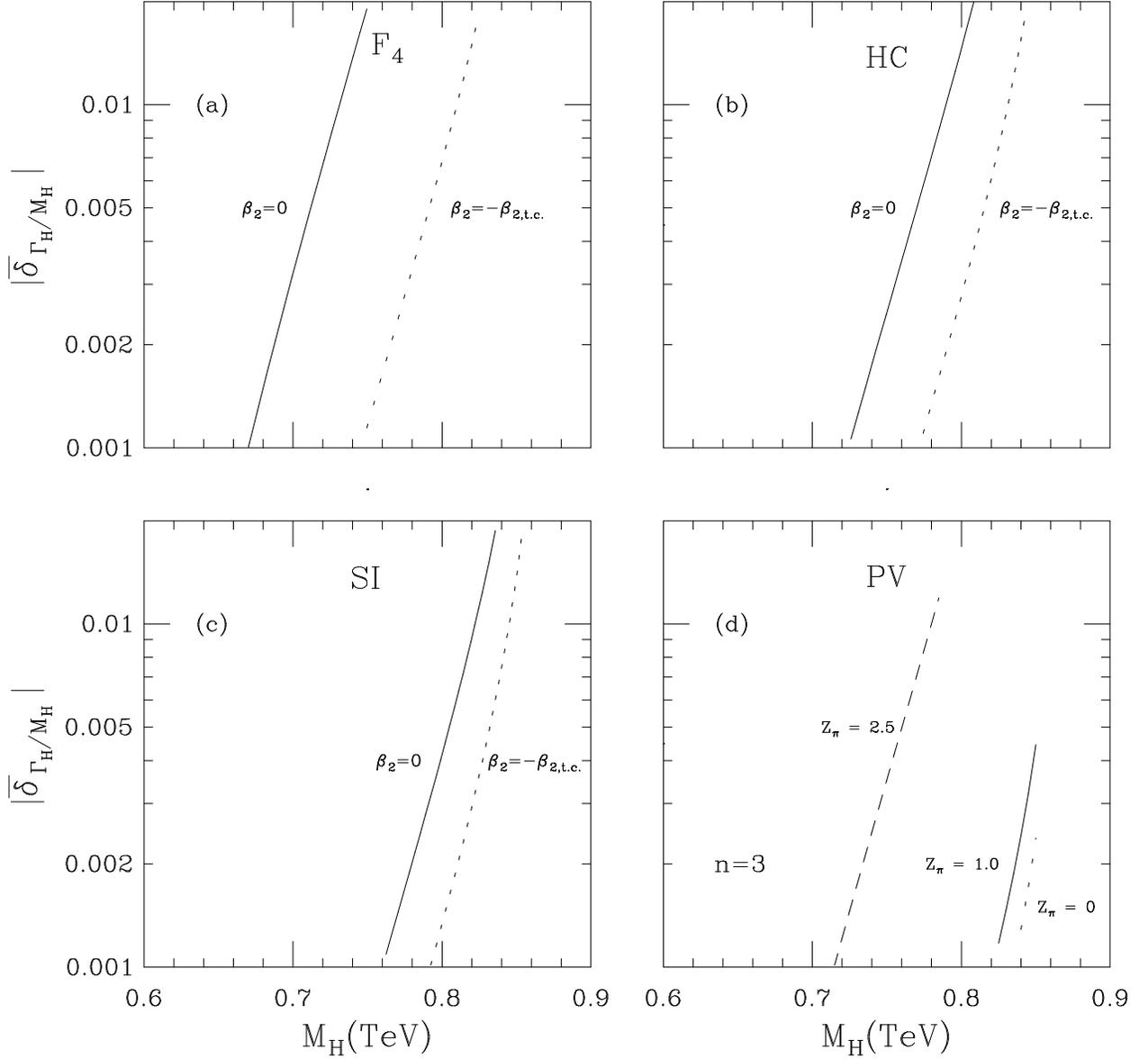
\def\captionseventhree{Leading order cutoff effects in the width to mass
ratio: The solid line represents the na\"\i ve actions  ($\beta_2=0$ on the
lattice and $Z_\pi=1$ for PV) and the dotted line the actions with a four
derivative term turned on to maximal allowed strength
($\beta_2=-\beta_{2,t.c.}$ on the lattice and $Z_\pi=0$ for  PV). The
dashed line in the PV case shows  an example with the ``wrong'' sign in
front of the four derivative term, corresponding to $Z_\pi =2.5$.}

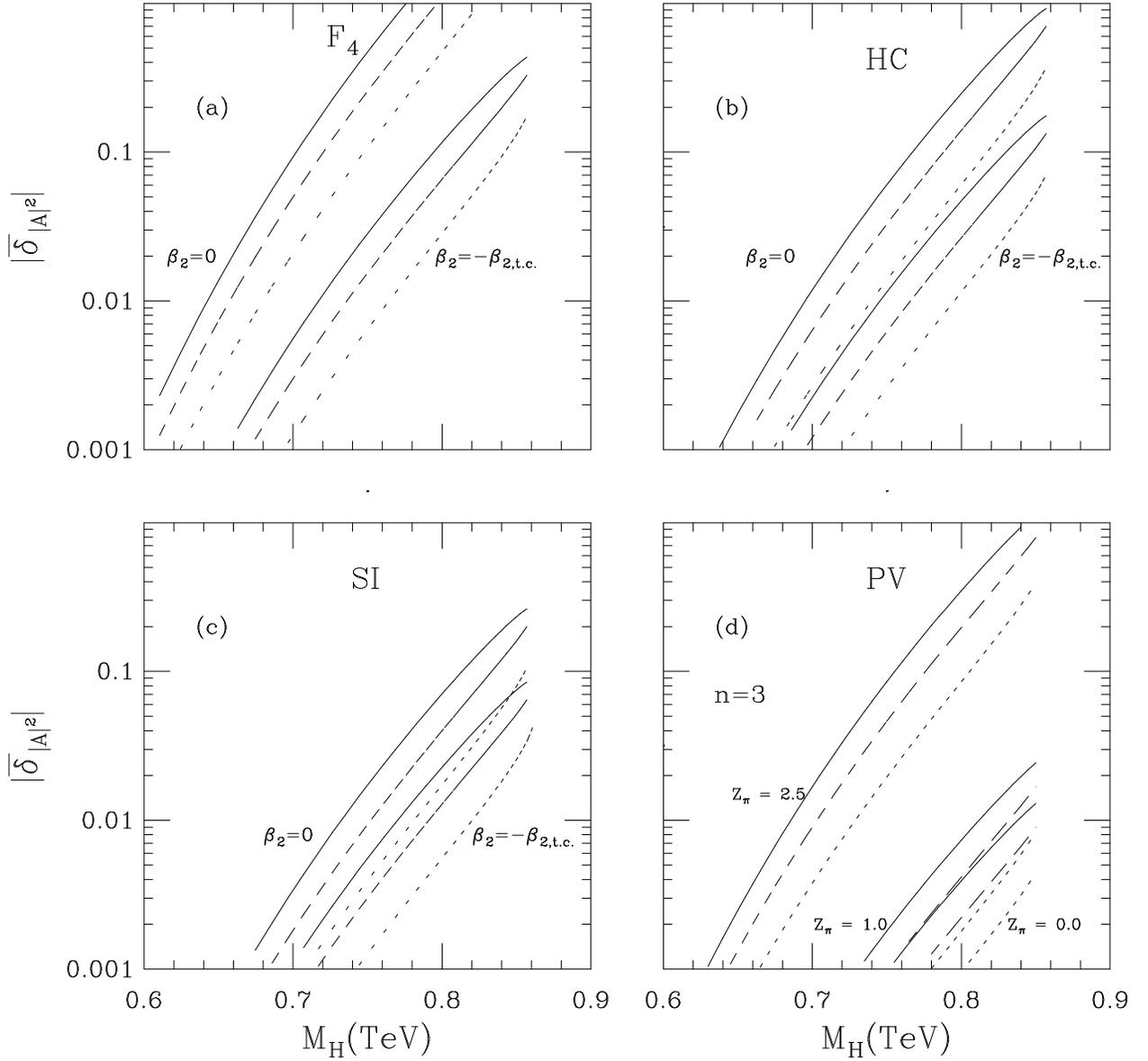
\def\captionsevenfour{Leading order cutoff effects in the  invariant
$\pi-\pi$ scattering amplitude at $90^0$ for the na\"\i ve actions
($\beta_2=0$ on the lattice and $Z_\pi=1$ for PV) and for the actions with
a four derivative term turned on to maximal allowed strength
($\beta_2=-\beta_{2,t.c.}$ on the  lattice and $Z_\pi=0$ for PV). The
dotted line represents center of mass energies $W=2M_H$, the dashed line
$W=3M_H$ and the solid line $W=4M_H$. In the  PV case we again show an
example with the ``wrong'' sign in front of the four derivative term
corresponding to $Z_\pi =2.5$.}

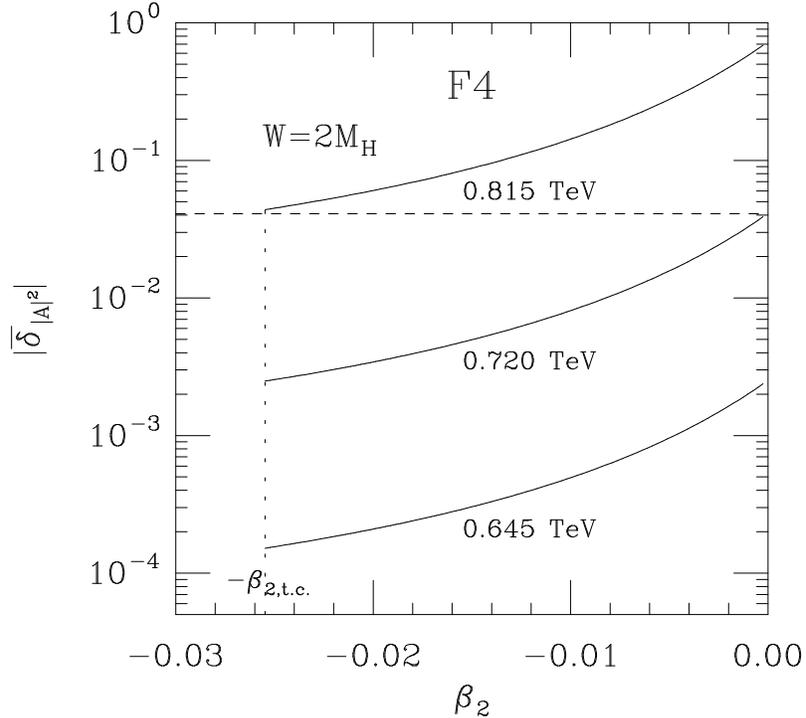
\def\captionsevenfive{Leading order cutoff effects in the invariant  $90^0$
$\pi-\pi$ scattering  amplitude vs. $\beta_2$ for three values of $M_H$ on
the $F_4$ lattice. The center of mass energy is set at $W=2M_H$. The cutoff
effects for a $0.815~TeV$ Higgs at $\beta_2=-\beta_{2,t.c.}$ and for a
$0.720~TeV$ Higgs at $\beta_2=0$ are equal in  magnitude.}

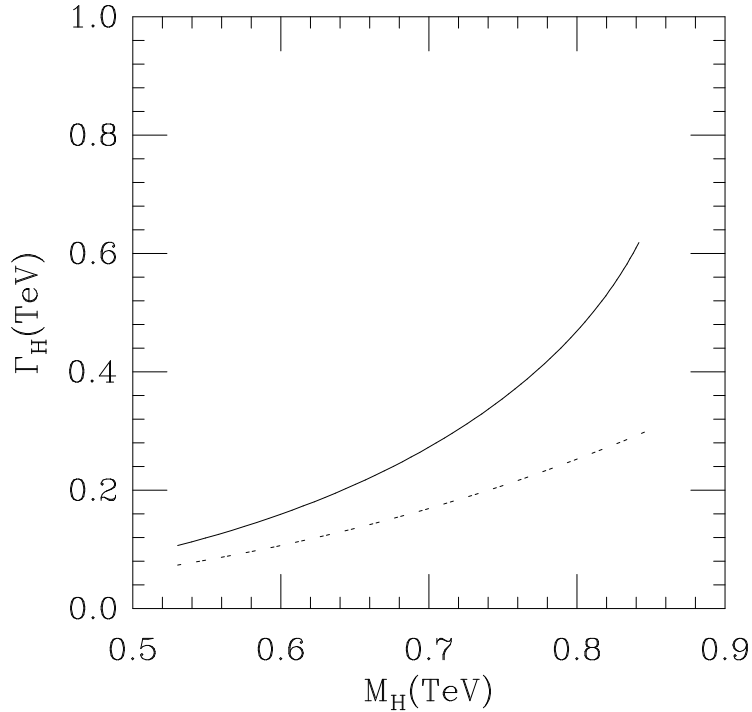
\def\captionsevensix{The regularization independent part of the width vs.
$M_H$.  The solid line displays the large $N$ result scaled to $N=4$ and
the dotted line  shows the leading order term in perturbation theory. For
large $M_H$,  the perturbative answer  severely underestimates the width.}

\if \draftversion Y


\fi

\def\today{\ifcase\month\or
	January\or February\or March\or April\or  May\or June\or
	July\or August\or September\or October\or November\or
	December\fi \space\number\day, \number\year}

\if \preprint Y \twelvepoint\oneandathirdspace \fi

\if \draftversion Y \rightline{\today{}} \fi

\if \preprint Y \rightline{FSU-SCRI-92-99} \fi \if \preprint Y
\rightline{RU-92-18} \fi

\title Large $N$ analysis of the Higgs mass triviality bound

\vskip 1. cm

\author Urs M.~Heller${}^{a}$, Herbert~Neuberger${}^{b}$,
Pavlos~Vranas${}^{a}$

\affil
\vskip 1.cm
\centerline{\it ${}^{a}$ Supercomputer Computations Research Institute}
\centerline{\it The Florida State University}
\centerline{\it Tallahassee, FL 32306}
\vskip .2cm
\centerline{\it ${}^{b}$ Department of Physics and Astronomy}
\centerline{\it Rutgers University}
\centerline{\it Piscataway, NJ 08855--0849}
\vskip 1.5 cm

\goodbreak

\abstract We calculate the triviality bound on the Higgs mass in scalar
field theory models whose global symmetry group $SU(2)_L \times SU(2)_{\rm
custodial} \approx O(4)$ has been replaced by $O(N)$ and $N$ has been taken
to infinity.  Limits on observable cutoff effects at four percent in
several regularized models with tunable couplings in the bare action yield
triviality bounds displaying a large degree of universality. Extrapolating
from $N=\infty$ to $N=4$ we conservatively estimate that a Higgs particle
with mass up to $0.750~TeV$ and width up to $0.290~TeV$ is realizable
without large cutoff effects, indicating that strong scalar self
interactions in the standard model are not ruled out.

\endtitlepage

\if \preprint N \doublespace \fi

\head{1. Introduction.} \taghead{1.}

This paper examines the regularization scheme dependence  of the
triviality bound on the Higgs mass in a $O(N)$ symmetric scalar field
theory to leading order in $1/N$.  Our purpose in doing this is
two-fold: we wish to investigate the issue by explicit, analytical
calculations and we need the results both directly and indirectly to
complement numerical work at $N=4$.  The results are needed directly
for estimating cutoff effects on physical observables that are not
accessible by Monte Carlo and indirectly for guiding our search in the
space of lattice actions to the region where heavier Higgs particles
are possible.

A preliminary account of some of our results has been presented in
\cit{{PLBHNV},{LATT91},{ROME}}.  Our general reasons for suspecting
that the present numbers for the bound are too low in the context of
the minimal standard model have been explained before
\cit{{PLBHNV},{TALLA},{CAPRI}} and will not be repeated here.

The basic logic of our approach has also been explained before
\cit{{BBHN1},{CAPRI},{PLBF4}} but, in order to make this paper more or
less self contained, we shall briefly review the main points.  We are
working with a scalar field theory at a finite cutoff. The field has
$N$ components ($N=4$ for the minimal standard model) and the action is
$O(N)$ symmetric.  We are interested in the broken phase in which the
symmetry is spontaneously broken and wish to pick cutoff schemes that
preserve as many of the continuum space--time symmetries as possible.
Our cutoff scheme must also exclude unitarity violations at the
energies and momenta of interest which are small relative to the
cutoff.

The physical scale is set by $F_\pi$, the ``pion decay constant'', and
we are interested in $M_H$ the ``Higgs mass''. Because of triviality
the ultraviolet cutoff cannot be removed and therefore cutoff effects
are unavoidable.  We require those cutoff effects that are observable
to be bounded by a few percent in relative magnitude. This imposes a
limit on the ratio $M_H / F_\pi$.  Generically, the leading cutoff
effects are of order inverse cutoff squared.  It is possible in some
models to arrange that the leading cutoff effects be of order inverse
cutoff to the fourth power instead.  We regard such ``finely tuned''
models as overly contrived.  The cutoff effects of the low energy
effective theory are ultimately determined by the underlying full
theory and we believe it is highly unlikely that such a ``finely
tuned'' model would be produced without any particular reason.  We wish
to find a region in the space of actions where, without excessive
``fine tuning'',\footnote{*}{Except the one possibly needed to make
$F_\pi$ small relative to the cutoff in the first place.} $M_H /
F_\pi$ gets to be as large as possible.

Since the cutoff effects are assumed to be small and ``fine tuning''
has been excluded we conclude that we can use the RG to count the
dimensions of the space of actions we ought to look at. We have one
relevant operator of dimension 2, two operators of dimension 4 and four
operators of dimension 6. We do not need to consider higher dimension
operators because their effects are subleading and we have excluded the
fine tuning of the leading effects to zero. Since we are interested
only in physical effects, namely in cutoff effects in the S-matrix, we
have to eliminate the redundant operators induced by field
redefinitions. For dimensions up to six there are three such
operators.  Thus, we end up with a 7-3=4 dimensional space. One
dimension corresponds to the relevant direction and sets the scale,
another to the marginally irrelevant parameter and the two remaining
parameters span all the possible leading order observable cutoff
effects.  It is preferable to think only about dimensionless physical
observables with the natural physical quantity obtained by
multiplication by the appropriate power of $F_\pi$. All momenta are
also measured in units of $F_\pi$ and so is $M_H$; the marginally
relevant direction can be thought of as parametrized by $g=3M_H^2
/F_\pi^2 $.

Intuitively, we would expect to increase $g$ by approaching the
nonlinear limit where the field $\vec \phi$ has fixed length. Hence,
for mass bound studies one can restrict one's attention to nonlinear
models. This is known to be true with na\"\i ve lattice actions and we
have checked it for some other actions at $N=\infty$.  Hence we shall
concentrate on nonlinear actions. At the most na\"{\i}ve level one can
think of a nonlinear action as obtained from a linear one by taking the
bare four point coupling to infinity and adjusting the other couplings
so that the model has a limit. From this point of view one of the four
free parameters has been eliminated; in other words, the effort to
maximize $g$ is assumed to guarantee the elimination of one of the four
parameters, leaving us with only three.  The nonlinear actions can also
be viewed as chiral effective actions in the sense defined by
Weinberg\refto{WEINBERG}.  The low momentum behavior of any broken
theory (even with a cutoff if the latter is sufficiently symmetric) can
be parametrized by such an effective action and, to first nontrivial
order, one needs three parameters.  This is exactly the number we came
up with after reducing by one the original dimensionality. Therefore we
know that the three parameter set of nonlinear actions we intend to
focus on is reasonably general, since it can at least reproduce all
``pion'' scatterings to first subleading order in the external momenta.
To be sure, we still expect the cutoff effects to depend on {\bf two}
parameters as before the restriction to nonlinear actions, because such
a constraint should not reduce the number of contributing operators in
the vicinity of the Gaussian fixed point.

Our further approximation will be to take $N$ to infinity. We shall see
that this effectively reduces the number of parameters from three to
two and, moreover, at $N=\infty$ the cutoff effects we shall be
interested in can now be parametrized by a single
parameter.\footnote{*}{Even a bare action that has only a single free
parameter, will, in principle, have observable cutoff effects that are
parameterizable by two coefficients, associated with the leading
``irrelevant'' operators after elimination of ``redundants''.} The
situation becomes as simple as it could get: we have one parameter that
sets the scale (``coarseness'' of grid in the lattice case) and another
that represents the entire freedom that is available at first
subleading order in the expansion of the S-matrix in inverse powers of
the cutoff. For fixed value of the second parameter one can imagine
that some extremization of $g$ has already been carried out by taking
the bare four point coupling to infinity. This simplification at
infinite $N$ makes it easy to relate very different cutoff schemes and
leads to a reasonably ``universal'' bound on $g$. These facts, derived
below, provide a basis confirming the general validity of the
$N=\infty$ results\footnote{\d}{With some small, inessential
corrections.} obtained previously by Einhorn \cit{{EINH},{EINHTALLA}}.

We shall work with a class of Pauli--Villars regularizations
parametrized in addition to the continuous parameters discussed above
by an integer $n\ge 3$, and with lattice regularizations on hypercubic
and $F_4$ lattices.  When the range of the additional couplings is
suitably restricted (but not ``finely tuned'') all these
regularizations give very consistent estimates for the $N=\infty$
bound. We present our results properly scaled to $N=4$ and obtain, with
$F_\pi = 0.246~TeV$, $M_H \le 0.820~TeV$ with rather stringent bounds
on the cutoff effects; changing the latter by a factor of 10 may affect
the bound by about $0.050~TeV$ in the obvious direction. We can try to
guess what correction on the bound might be due to $N$ being four
rather than infinity by looking at the available numerical data at
$N=4$.  This leads us to expect for $N=4$ a bound somewhere between
$0.750~TeV$ and $0.800~TeV$. While the relatively recent previous
estimates were not much lower\footnote{\b}{Very early
estimates\refto{{WHIT},{TSYP}} were actually around $0.800~TeV$ but this
was accidental; these authors also used the simplest possible lattice
action for which we have nowadays better results.} ($0.600~TeV$ to
$0.650~TeV$) the effect on the width, $\Gamma_H$, is more significant:
$\Gamma_H / M_H$ may reach $0.4$ when the bound is close to saturation
and this implies that strong scalar self--interactions without strong
observable cutoff effects are possible.

Since the previous bounds were obtained from a very restricted class of
actions, chosen just because they were simpler to analyze for technical
reasons, it was somewhat premature to put forward the $0.550~TeV$ to
$0.650~TeV$ range as {\bf the} lattice triviality bounds with lattice
spacings between $1/(5M_H)$ and $1/(2M_H)$ and imply direct relevance
to experiment\refto{KUTIPRL}. To be sure, we do not contest the
validity of the numbers within the particular regularization that they
were obtained in, and, in retrospect, they were not very far off even
in the general context. Our analysis shows that we are nowhere near a
ratio $M_H / F_\pi \sim 6$ that would be expected in a QCD like theory
and our Higgs is quite ``elementary''. We have no explanation for why
the range $3 \le M_H / F_\pi \le 5$ seems to be so difficult to attain
with ``an elementary'' Higgs; this looks like a stronger numerical
reflection of triviality than one might have originally suspected.

Throughout the paper we shall use a notation with capital letters,
$F_\pi$, $M_H$ and $\Gamma_H$, to denote the physical quantities,
measured in $TeV$, and with lower case letters, $f_\pi$, $m_H$ and
$\gamma_H$, to denote those quantities in units of the cutoff,
$\Lambda$, or $a^{-1}$ for lattice regularizations. In the large $N$
limit the renormalized vacuum expectation value (pion decay constant),
$f_\pi$, diverges as $\sqrt N$ but $m_H$ and $\gamma_H$ stay finite. We
therefore rescale $f_\pi$ and $F_\pi$ by $\sqrt N$ to make them also
finite at $N \to \infty$. From now on and until sections 7 and 8
$f_\pi$ and $F_\pi$ refer to the rescaled values.  Of course, when we
use the large $N$ results for $N = 4$, we undo this rescaling; when we
present our large $N$ results in physical units we take the rescaled
$F_\pi$ as $0.246/\sqrt{4}~TeV = 0.123~TeV$.

The outline of the paper is as follows: In the next section we shall
briefly analyze linear models with particular higher dimensional
operators to see explicitly that in order to estimate the bound we have
to look only at the nonlinear limit. We proceed to define the class of
nonlinear models we shall consider in the rest of the paper. In section
3 we work out the large $N$ phase diagram of the models in the
Pauli--Villars class. Section 4 deals with the cutoff effects in the
Pauli--Villars class of models, first in a simplified manner and then
in detail. Sections 5 and 6 repeat the analysis for the lattice models.
In section 7 we present the results of our large $N$ study of cutoff
effects for the Higgs width to mass ratio and for Goldstone pion
scattering and the implications on the Higgs mass bound.  Also, a
comparison between large $N$ and available numerical results at $N=4$
is carried out. This allows us to make inferences from the results
obtained in this paper to the $N=4$ case of actions that have not yet
been studied numerically. Our conclusions are presented in section 8.
Three appendices have been added to present some technical details and
expand on side issues.

\head{2. Linear versus nonlinear action.} \taghead{2.}

In this section we briefly discuss the linear case. We shall see why we
were led to consider only nonlinear actions in the sequel. Also, by
comparing the formulae derived later on for the nonlinear case to the
ones in this section, one can see explicitly that the nonlinearity of
the bare action has no observable effect in the critical  regime as
expected by general RG arguments. More precisely, by making physical
measurements in the critical regime and at low enough momenta one
cannot say whether the bare theory was linear or not. It is because of
this that simulations employing a nonlinear action have something to
say about the standard model as we know it.

We shall use the following notations: The partition function is
$\zpar$, the metric is Euclidean and the action is denoted by $S$.
$\zpar$ is given by $$ \zpar =\int \exp (-S) \cm \eqno(e2p1)$$ where
the fields have been suppressed in  the action and integration
measure.  For spatial ($x,y\cdots$) and momentum ($p,q,k\cdots$)
integration we use $$ \int_x\equiv\int d^4
x~,~~~~~~~~~~\int_p\equiv\int{{d^4 p}\over{(2\pi)^4}} \cm \eqno(e2p2)$$
and as usual $$ (\partial_\mu f)^2\equiv\sum_\mu (\partial_\mu f )^2
\fs \eqno(e2p3)$$ $\Lambda$ denotes the cutoff. We shall suppress
arguments whenever possible.

\subhead{2.1. The general linear action and the saddle point
equations.} \taghead{2.1.}

According to the discussion in the introduction the most natural set of
actions we should be studying is given by $$\eqalign{ &S= \int_{x_0}\cr
&\left [ {1\over 2} \fhi_0 K_0 (-\partial_0^2 )\fhi_0 + {{\mu_0^2}\over
2}\fhi_0^2  + {{\lambda} \over {4N}}(\fhi_0^2 )^2 +{{\eta_{1,0}}\over
{2N}} \fhi_0^2 (\partial_\mu \fhi_0 )^2
+{{\eta_{2,0}}\over{2N}}(\partial_\mu (\fhi_0^2 ))^2
+{{\eta_{3,0}}\over{6N^2 }}(\fhi_0^2 )^3 \right ] \fs \cr}
\eqno(e2p1p1)$$ The subscript ``0'' denotes dimensionful parameters.
The kinetic term is regulated in a yet unspecified manner, but it is
assumed that all subsequent potential ultraviolet divergences are cut
off at about $\Lambda$. Thus we write $$ K_0 (-\partial_0^2 )=\Lambda^2
K(-\partial_0^2 /\Lambda^2 )\fs \eqno(e2p1p2)$$ \(e2p1p2) contains
seven free parameters (two of them are implicit in $K_0
(-\partial_0^2)$), one for each operator of dimension $\le 6$. Later we
shall choose to fix some of these parameters to simplify our analysis.

We scale $\Lambda$ out of the problem by defining $$ x=\Lambda
x_0~,~~\fhi_0 = \Lambda \fhi~,~~\mu_0^2=\Lambda^2 \mu^2~,~~
\eta_{1,0}={{\eta_1}\over{\Lambda^2}}~,~~
\eta_{2,0}={{\eta_2}\over{\Lambda^2}}~,~~
\eta_{3,0}={{\eta_3}\over{\Lambda^2}}\eqno(e2p1p3)$$ and all our
variables become dimensionless $$ S=\int_{x} \left [ {1\over 2} \fhi K
(-\partial^2 )\fhi + {{\mu^2}\over 2}(\fhi^2 )^2 +{{\lambda} \over
{4N}}(\fhi^2 )^2 +{{\eta_{1}}\over {2N}} \fhi^2 (\partial_\mu \fhi )^2
+{{\eta_{2}}\over{2N}}(\partial_\mu (\fhi^2 ))^2 +{{\eta_{3}}\over{6N^2
}}(\fhi^2 )^3 \right ] .\eqno(e2p1p4)$$

Introducing into the functional integral $$ \prod_x
\int_{-\infty}^{\infty} d\sigma (x) {N\over{4\pi}}
\int_{-i\infty}^{i\infty} d\omega (x) \exp \left\{{N\over 2} \omega(x)
\left[ \sigma (x) - {{\fhi^2 (x)}\over N} \right] \right\} =1 \cm
\eqno(e2p1p5)$$ we obtain a new action with more fields $$ S_1 ={1\over
2} \int_x \fhi [ K-\eta_1 \partial_\mu \sigma \partial_\mu + \omega
]\fhi+ {N\over 2} \int_x [\eta_2 (\partial_\mu \sigma )^2 +\mu^2 \sigma
+{{\lambda}\over 2}\sigma^2 +{{\eta_3}\over
 3}\sigma^3 -\sigma\omega ]\fs \eqno(e2p1p6)$$

The zero mode of $\fhi$ needs special treatment requiring that we
separate it out explicitly $$ \fhi (x) = \sqrt{N} \vec v + \hat v H(x)
+ \vec \pi (x) ~,~~\hat v ={{\vec v}\over v}~,~~|\vec v |=v {}~,~~\hat
v \cdot \vec \pi =0~,~~\int_x H(x)=\int_x \pi^j (x) =0 \fs
\eqno(e2p1p7)$$

We integrate out the $H$ and $\vec \pi$ fields and, with $\hat K = K -
\eta_1 \partial_\mu \sigma \partial_\mu +\omega$, restricted to operate
in the space of non-constant functions, we obtain

$$ S_2 = {N\over 2} {\rm Tr} \log {\hat K} -{N\over 2} v^2 \int
{\omega}^{\prime} {\hat K}^{-1} {\omega}^{\prime} +{N\over 2} v^2
\int_x \omega +{N\over 2}\int_x [\eta_2 (\partial_\mu \sigma )^2 +\mu^2
\sigma +{{\lambda}\over 2}\sigma^2 +{{\eta_3}\over 3}\sigma^3
-\sigma\omega] \fs \eqno(e2p1p8)$$ Here ${\omega}^{\prime}$ is the
non-constant part of $\omega$ $$ \partial_\mu (\omega -
{\omega}^{\prime} )=0~,~~~\int_x {\omega}^{\prime} =0 \fs
\eqno(e2p1p9)$$ ${\omega}^{\prime}$ rather than $\omega$ comes in
because $\int_x \omega H =\int_x {\omega}^{\prime} H$ as a result of
the zero mode, $\vec v$ being separated out. In the functional
integral one is still left with a piece $\int_0^{\infty} v^{N-1} dv$
because $v$ hasn't been integrated over. In the infinite volume limit,
and assuming translational invariance, this piece can
be ignored.\footnote{*}{The large $N$ analysis can be
easily extended to finite volumes where
the probability distribution of $v$ can be calculated showing explicitly how
the infinite volume limit is approached.}
Taking $N$ to infinity we obtain the general saddle point equations $$
\eqalign{ &\int_p {1 \over {K(p^2 )+\eta_1 \sigma_s p^2 + \omega_s}} +
v^2 =\sigma_s\cr \eta_1 &\int_p  {{p^2} \over {K(p^2 )+\eta_1 \sigma_s
p^2 + \omega_s}}+\mu^2+ \lambda\sigma_s +\eta_3 \sigma_s^2 =\omega_s
\fs \cr}\eqno(e2p1p10)$$ We notice that $\eta_2$ has disappeared from
the saddle point equations because we assumed translational invariance
of the saddle.

If we keep $v$ as a free variable and plug in the solutions of the
saddle point equations into $S_2$ we obtain the effective potential.
Here, however, we just wish to fix $v$ at the vacuum expectation value of
$\fhi$. In the broken phase one then has $\omega_s =0$. For $\omega_s >
0 ~$ $v$ will vanish.

We are looking for the candidate transition surface. It is only
``candidate'' because it might be cut by other transition surfaces
rendering some portions of it metastable. However, any true second
order transition point we are interested in must be on this surface.
The condition for criticality fixes $\mu^2 =\mu_c^2 (\eta_1 , \eta_3 ,
\lambda)$.  $\mu_c^2$ is implicitly defined by $$ \eqalign{ &\int_p {1
\over {K(p^2 )+\eta_1 \sigma_s p^2 }} =\sigma_s\cr \eta_1 &\int_p
{{p^2} \over {K(p^2 )+\eta_1 \sigma_s p^2 }}+\mu_c^2+ \lambda\sigma_s
+\eta_3 \sigma_s^2 =0 \fs \cr}\eqno(e2p1p11)$$

For the time being we ignore questions regarding global stability.  We
take $\delta\mu^2 \equiv \mu^2 -\mu_c^2$ to be negative and small
relative to unity. This should place us in the broken phase close to
the transition.

\subhead{2.2. Small fluctuations around the saddle point in the broken
phase.} \taghead{2.2.}

We now wish to compute the propagators. We take $\hat v$ to point in
direction 1 in the internal space. Hence $\pi^1 =0$; we use latin
letters to label the $N-1$ nonvanishing components and view the
``pion'' field as made out of them only. The $N=\infty$ propagator is
immediately read off, in Fourier space, as $$ <\pi^a \pi^b >
={{\delta^{ab}}\over{K(p^2 )+\eta_1 \sigma_s p^2}} \equiv
{{\delta^{ab}} \over{\Delta_s (p^2 )}} \fs \eqno(e2p2p1)$$

To get the Higgs field propagator we integrate out the $\pi^a$ fields
$$ \eqalign{ S_3 =& ~ {{N-1}\over 2} {\rm Tr} \log {\hat K}+{1\over
2}\int_x H {\hat K} H+{N\over 2} v^2 \int_x \omega +\sqrt{N} v\int_x
\omega H +\cr & ~ {N\over 2}\int_x [\eta_2 (\partial_\mu \sigma )^2
+\mu^2 \sigma +{{\lambda}\over 2}\sigma^2 +{{\eta_3}\over 3}\sigma^3
-\sigma\omega] \fs \cr}\eqno(e2p2p2)$$ At large $N$ we expand around
the saddle point with $\delta \sigma =\sigma -\sigma_s ,~\delta \omega
=\omega ~{\rm and}~\delta H = H$. We get, neglecting order $1/N$
corrections, $$ \eqalign{ S_3^{(2)} =& ~ {N\over 2} \int_p [ {1 \over
2} \delta \sigma B_{\sigma \sigma} \delta \sigma + {1\over 2} \delta
\omega B_{\omega \omega} \delta \omega +\delta \sigma B_{\sigma \omega}
\delta \omega ] +{1\over 2}\int_p \delta H \Delta_s \delta H + \sqrt
{N} v \int_p \delta \omega \delta H +\cr & ~ {N\over 2}\int_x[\eta_2
(\partial_\mu \delta \sigma )^2  +{{\lambda}\over 2}(\delta \sigma )^2
+\eta_3 \sigma_s (\delta \sigma )^2
-\delta\sigma\delta\omega]\cr}\eqno(e2p2p3)$$ where $$ \eqalign{
B_{\sigma \sigma} (p^2 )=& ~ -\eta_1^2\int_q {{({1\over 4}p^2 -q^2 )^2
}\over {\Delta_s ({1\over 2}p-q) \Delta_s ({1\over 2}p +q)}}\cr
B_{\omega \omega} (p^2 )=& ~ -\int_q {1\over {\Delta_s ({1\over 2}p-q)
\Delta_s ({1\over 2}p +q)}}\cr B_{\omega \sigma} (p^2 )=& ~
-\eta_1\int_q {{{1\over 4}p^2 -q^2  }\over {\Delta_s ({1\over 2}p-q)
\Delta_s ({1\over 2}p +q)}} \fs \cr}\eqno(e2p2p4)$$ To find the
propagator of $H$ one has to invert a $3\times 3$ matrix whose entries
can be read off from \(e2p2p3,e2p2p4). Note that $\eta_2$ reappeared
and the Higgs mass will depend on it.

\subhead{2.3. Restriction to $\eta_1 =\eta_3 =0$.} \taghead{2.3.}

The problem simplifies considerably when we set  $\eta_1 =\eta_3 =0$
because $\eta_2$ doesn't appear in the saddle point equations and
affects only the Higgs mass. Note that using field redefinitions and
keeping terms up to order inverse cutoff square $\eta_1$ and $\eta_3$
can be eliminated from the action \(e2p1p4).  Intuitively $\eta_2$
looks like a parameter that may have a significant effect because in
\(e2p1p4) it gives some extra ``stiffness'' to the modulus of $\fhi$.
We wish to see whether keeping $\eta_2 > 0$ can lead to a situation
where the Higgs mass is no longer monotonically increasing with
$\lambda$ (since $\sigma$ is a real field [see \(e2p1p5)] $\eta_2$ must
now be non-negative to keep the action bounded from below).  We shall
see that any reasonable limitation on the cutoff effects in $\pi$ $\pi$
scattering prohibits this from happening. Hence, as far as the
triviality bound is concerned, we end up being driven to $\lambda =
\infty$ and the dependence on $\eta_2$ ultimately drops out from the
Higgs mass too. We shall assume later on that turning on the other
$\eta$ couplings would have had similar consequences, and that the
triviality bound would be independent of them too.

Our action in dimensionless variable is now given by $$ S=\int_{x}
\left [ {1\over 2} \fhi K (-\partial^2 )\fhi + {{\mu^2}\over 2} \fhi^2
+{{\lambda} \over {4N}}(\fhi^2 )^2 +{{\eta_{2}}\over{2N}}(\partial_\mu
(\fhi^2 ))^2 \right ] \fs \eqno(e2p3p1)$$ The saddle point equations
become $$ v^2 +\int_p {1\over {K(p^2 )+\omega_s}}
=\sigma_s~,~~~~\mu^2+\lambda\sigma_s = \omega_s \fs  \eqno(e2p3p2)$$
One can show that there are no competing saddles here and hence, once
$\eta_1 =\eta_3 =0$, questions about global stability do not arise.

The critical line is given by $$ \mu_c^2=-\lambda\int_p {1\over {K(p^2
)}}\eqno(e2p3p3)$$ and the broken phase is where $$ \delta \mu^2 =\mu^2
-\mu_c^2 < 0 \fs \eqno(e2p3p4)$$ There, $\omega_s =0$ and the vacuum
expectation value $v$ is given by $$ v^2={{-\delta\mu^2}\over{\lambda}}
\fs \eqno(e2p3p5)$$

Denoting the fields $\delta H,~\delta\omega$ and $\delta\sigma$ by
$\psi_1,~\psi_2$ and $\psi_3$ respectively, we obtain the following
propagators in Fourier space $$ <\psi_A \psi_B >=
(M^{-1})_{AB}\eqno(e2p3p6)$$ where $$ M=\left ( \matrix{
K(p^2)&\sqrt{-{N\over{\lambda}}\delta\mu^2}&0\cr
\sqrt{-{N\over{\lambda}}\delta\mu^2}&-{N\over 2}I(p^2)&-{N\over 2}\cr
0&-{N\over 2}&{N\over 2}(\lambda + 2\eta_2 p^2)\cr} \right
)\eqno(e2p3p7)$$ and $$ I(p^2)=\int_q {1\over{K({1\over 2}p-q)K({1\over
2}p+q)}} \fs \eqno(e2p3p8)$$

Instead of looking at the Higgs propagator we look at $\pi$ $\pi$
scattering.  The invariant amplitude $\cal M$ for the process $\pi^a
(1) +\pi^b (2) \rightarrow \pi^c (3) + \pi^d (4)$ is $$\eqalign{
_{out}<cd|{\cal M}|ab>_{in}~ &=A(s,t,u)\delta_{ab}\delta_{cd}+
A(t,s,u)\delta_{ac}\delta_{bd}+A(u,t,s)\delta_{ad}\delta_{bc}\cr s=&~
(p_1 +p_2 )^2~,~~~t=(p_1 -p_2 )^2~,~~~u=(p_1 -p_4 )^2~,~~~p_j^2=0\cr }
\eqno(e2p3p9)$$ in standard Minkowski space notation. At $N=\infty$,
$A(s,t,u)$ depends only on $s$ and, in our restricted model, is
proportional to the $\delta\omega$ propagator at $p^2=-s$. Hence, by
looking at the $\delta\omega$ propagator we can get both the Higgs mass
and estimate the cutoff effects in $\pi$ $\pi$ scattering and on the
width to mass ratio for the Higgs.  {}From \(e2p3p6,e2p3p7) we obtain
$$ <\delta\omega \delta\omega > =-{2\over N} {{K(p^2 )} \over {2v^2
+K(p^2 )[I(p^2 )+{1\over{\lambda+2\eta_2 p^2}}]}} \fs \eqno(e2p3p10)$$

We now make our choice for $K(p^2 )$:  $$ K(p^2 )=p^2+(p^2)^2+\epsilon
(p^2 )^3 \fs \eqno(e2p3p11)$$

By setting the coefficient of the $(p^2)^2$ term to one we have
restricted another parameter, leaving us with 3 free parameters. With a
rescaling of the cutoff and the fields\footnote{*}{The rescaling of the
fields, by itself, would lead to unphysical cutoff effects only, but
the rescaling of the cutoff does induce a physically observable change
in the action.} one can easily adapt the analysis to the case where the
coefficient of  the $(p^2)^2$ term is different from one, as long as it
stays positive. We don't expect small changes in this coefficient to
affect our final conclusion in any way, but have not checked explicitly
whether our conclusion still holds when the changes are allowed to
become large.

To make $\mu_c^2$ finite we need $\epsilon > 0$. We introduce an
$\epsilon$ dependence in $\mu^2$ such that in the limit $\epsilon
\rightarrow 0^+$ $\mu^2$ diverges and $\delta\mu^2$ stays finite.
Everywhere else only $\delta\mu^2$ appears and $\epsilon$ can be set to
zero.\footnote{\d}{Note that when the coefficient of the $(p^2 )^2$
term is negative $\epsilon$ cannot be taken to zero; nearest neighbor
lattice actions have a negative coefficient for $(p^2 )^2$ and
therefore lattice actions ought to be analyzed separately. An example
of such an analysis is presented in Appendix C.} Thus, in the
definition of $I(p^2 )$ we take $$ {1\over{K(p^2 )}}={1\over
{p^2}}-{1\over{p^2 + 1}}\eqno(e2p3p12)$$ which gives us $$ I(p^2 )=
{1\over{8\pi^2 }} \left[ -{1\over 2}\log ( p^2 ) + (1+{1\over{p^2
}})\log (1+ p^2)-\sqrt{1+{{4}\over{p^2}}} {}~{\rm arcsinh}\left (
\sqrt{{{p^2}\over{4}}}\right ) \right ] \fs \eqno(e2p3p13)$$ It is easy
to check that the local stability requirement $I(p^2 ) > 0$ of the
saddle holds for all Euclidean momenta; it is known \refto{COLEMAN}
that if one extrapolates the leading asymptotic expression for $I(p^2
)$ valid for small momenta to large ones, the stability requirement
appears to be violated; this gives rise to the ``tachyon'' problem in
the renormalized expressions and signals that for high momenta one
cannot remove the cutoff dependence from the theory.

When we analytically continue to the physical regime for $\pi$ $\pi$
scattering we replace $p^2$ by $-w-i 0^+$ with positive $w$. We obtain
$$ \eqalign{ I(-w-i0^+ )=& ~ {i\over{16 \pi}} -\cr & ~ {1\over{8\pi^2
}}\left[ {1\over 2} \log (w) +({1\over w}-1)\log(1-w)+\sqrt{{{4}\over
{w}}-1}~{\rm arcsin}\left ( \sqrt{{w\over{4}}}\right ) \right ]\fs \cr}
\eqno(e2p3p14)$$

\subhead{2.4. Higgs mass in the restricted model.} \taghead{2.4.}

As a measure of the Higgs mass we choose the more accessible quantity,
$m_R^2$, defined as the smallest positive root of ${\rm Re}
[<\delta\omega\delta\omega >^{-1} (w)]$. It makes physical sense to
only allow center of mass energies for which ghosts cannot be produced.
On the other hand, the range of center of mass energies must be allowed
to surpass $m_R$ by a factor of  2--4. A pair of ghosts can be created
with $w\ge 4$. We therefore restrict $m_R$ by $m_R \le 0.5$. Later on
we shall sharpen this requirement somewhat.

It is clear that a zero will develop in $<\delta\omega\delta\omega >$
for $w =\lambda/\xi$, where we introduced $2\eta_2 =\xi >0$ (see
eq.~\(e2p3p10)).  This is a cutoff effect of large relative magnitude
that has to be forbidden at least for energies as low as $w\le 1$ and
maybe up to $w=4$. The unwanted zero is pushed to sufficiently high
energies if we impose the requirement $\lambda/\xi > 1$ or maybe even
$\lambda/\xi > 4$. We shall be able to derive our main conclusion even
with the milder restriction $\lambda/\xi > 1$ and therefore we shall
stick to it from now on.

Defining $x=m_R^2$ and $G=m_R^2/(2v^2 )$ we have to solve the equation
$$ {{8\pi^2 }\over G}= (1-x)\left[{{8\pi^2}\over{\lambda-\xi x}}-\phi
(x) \right ]\equiv{{\phi^\prime (x)}\over x}\eqno(e2p4p1)$$ with $$
\eqalign{ \phi(x)=& ~ {1\over 2}\log (x) +{{1-x}\over x}\log (1-x)
+\sqrt {{4\over x}-1} {}~{\rm arcsin}\left ( \sqrt{{x\over 4}}\right )
\cr \approx & ~ {1\over 2}\log x
+{5\over{12}}x+{{19}\over{120}}x^2+{{23}\over{280}}x^3+{{251}\over{5040}}x^4
+\cdots \fs \cr} \eqno(e2p4p2)$$ Note the absence of a constant term in
\(e2p4p2).  When $x$ is very small compared to unity we get $$
{{8\pi^2}\over G} \approx {{8\pi^2}\over{\lambda}}- {1\over 2}\log x
\cm \eqno(e2p4p3)$$ and clearly the maximal $G$ is obtained at
$\lambda=\infty$. This conclusion has been obtained in the perturbative
regime  and applies only when $G$ is quite small even at its largest
value.  We want to show that the conclusion holds irrespectively of the
magnitude of $G$.

Observe that $\phi^\prime (x)$ will have a  maximum, $0< \bar x
(\lambda , \xi ) < 1$ and a solution $x$ exists for a given $v^2$ only
if
 $v^2$ satisfies $\phi^\prime (\bar x ) \ge 16\pi^2 v^2$.  We are
interested in the smallest positive root of \(e2p4p1); it satisfies
$x<\bar x$.  A short analysis shows that $\bar x (\lambda , \xi ) \ge
\bar x (\infty , 0) \approx 0.18$ and that $\phi^\prime (x)/x$
decreases monotonically for $x\in (0,  {\rm C} )$.  $\rm C$ is some
number that can be shown to be larger than $\bar x (\infty , 0) $.

The cutoff effects can be chosen to be characterized by $m_R$.  We now
decide to limit the cutoff effects by the bound $m_R < \sqrt{p}$. We
pick some positive $p$ satisfying $p \le \bar x (\infty , 0) \approx
0.18$.  We shall see later on that this restriction on $p$ is not at
all severe (we already argued above for a $p\le 0.25$).  The smaller
$p$ is chosen to be, the more stringently the cutoff effects are
limited.

One can view $v^2,~\lambda,~{\rm and}~\xi$ as free positive parameters
restricted only by $\lambda/\xi > 1$. Indeed, any desired value for
$v^2$ can be attained by tuning $\mu^2$ in \(e2p3p5).  We now look for
the largest possible coupling $G$ that satisfies \(e2p4p1) with some
$x\le p$.  More explicitly,  we are looking for the set $\{ v^2
,\lambda ,\xi \} $ for which the equation $16\pi^2 v^2 =\phi^\prime
(x)$ (see eq.~\(e2p4p1)) gives an $x$ that leads to the largest $G$
possible while $x$ is not permitted to exceed  $p$.  Suppose we have
found this set and its associated $x$ and $G$. The monotonic decrease
of $\phi^\prime (x)/x$ and eq.~\(e2p4p1) imply $$ {{8\pi^2}\over G} \ge
{{\phi^\prime (p)}\over p } \ge  -(1-p)\phi (p) \fs \eqno(e2p4p4)$$ For
the largest $G$ possible both inequalities in eq.~\(e2p4p4) ought to
become equalities. Note that $\phi$ does not depend on $\lambda$ and
$\xi$.  The first inequality becomes an equality simply by setting
$x=p$, which shows explicitly that indeed the coupling is maximized
when the cutoff effects are as large as they are allowed to get. This
shows that, as expected, in order to make the coupling as large as
possible one has to allow the cutoff effects to grow up to the bound
one has declared from the beginning; in other words, the cutoff effects
as measured by the mass to cutoff ratio and the coupling, defined by
the mass to $v$ ratio, are monotonically related.  The second
inequality in eq.~\(e2p4p4) can be made into an equality only when
$\lambda \rightarrow \infty$;  $\xi$ is not restricted as long as it is
assumed to always obey $0\le\xi\le\lambda$.  With any finite $\lambda
\ge \xi \ge 0$ the coupling $G$ would have been smaller for any $x$,
including the ``best'' value, namely $x=p$.  Therefore the Higgs mass
bound is obtained in the limit $\lambda \to \infty$ and does not depend
on $\xi$ and hence the coupling $\eta_2$.

One can compute the true pole location and one gets, for example, with
$M_R/\Lambda = \sqrt x = \sqrt p = 0.397$ $M_H /\Lambda =0.188$ and
$\Gamma_H / \Lambda =0.198$ showing that our restriction on $p$ did
allow for sufficiently wide (and hence heavy) Higgs particles. Note
that $M_H/\Lambda$ and $m_R$ are very different numbers for such large
couplings; still, in the whole range they are monotonically related so
our main conclusion is unaffected.

\subhead{2.5. Nonlinear actions.} \taghead{2.5.}

In the introduction we counted the number of parameters our space of
actions should depend on. In this section we saw that just having the right
number of parameters is not sufficient; for example, with $\eta_1 = \eta_3
= 0$,  varying $\eta_2$ in the physical range had no effect on the bound,
which is obtained in the limit $\lambda \to \infty$. We also considered a
linear model on an $F_4$ lattice having the equivalent of the couplings
$\eta_2$ and $\eta_3$ and found again that the Higgs mass bound is obtained
in the limit $\lambda \to \infty$. This analysis is sketched in Appendix C,
since it relies heavily on methods developed in subsequent sections. We
believe that the same conclusion would be found if $\eta_1$, $\eta_2$ and
$\eta_3$ were allowed to vary simultaneously. When $\lambda \rightarrow
\infty$ with a properly adjusted $\mu^2 \rightarrow -\infty$ we get a
nonlinear action and the $\eta_{1,2,3}$ either drop out or can be absorbed
in $K$.  In the nonlinear limit, power counting reduces to derivative
counting. The leading cutoff effects are now parametrized by ``dimension''
four operators (the leading nonlinear term is of ``dimension'' two), and we
therefore need to add four-derivative terms to the non-linear action. There
are three such terms, but one is redundant and can be eliminated by a
suitable field redefinition.

We now have to address the question whether this time there will be
some dependence on the couplings when they vary in reasonable
intervals. By ``reasonable'' we mean intervals of order unity for the
dimensionless couplings; we expect all sorts of effects to limit the
range in which these couplings can vary but it is almost certain that
when everything is taken into account the allowed intervals will be of
order unity. We now argue that indeed, with a nonlinear action, sizable
variations can be induced in physical observables by variations of
order one in the couplings.

For this purpose it is useful to view the nonlinear action as a chiral
effective Lagrangian rather than a cutoff version of the usual
renormalizable scalar field theory. Let us do some numerology: From
previous work we know that  for the bound we have $M_H / (2F_\pi )\sim
2.5$\footnote{*}{Recall that we are working with a rescaled $F_\pi$.}
and $\Lambda$, the maximal allowed ``momentum'', is about $2\pi M_H$.
Thus $\Lambda/(4\pi (2 F_\pi )) \sim 1.25$ and loop effects will be of
relative order one for observables depending on external momenta
smaller than $\Lambda$\refto{GEORGI}.\footnote{\d}{There are arguments
that would lead us to consider, in our normalization, $\Lambda/(4\pi
({\sqrt{2}}F_\pi) \approx 1.76$ instead  \cit{CHIVU}.} Since loop
effects are more or less of the same order as tree level effects, we
can reasonably expect that order one variations in the bare couplings
will have measurable effects. In short, unlike in the linear case,
investigating the dependence of the bound on the additional terms won't
be a waste of time.

All our actions have a derivative expansion of the form $$ \eqalign{
S_c ~ &=\cr &\int_x ~ \left [{1\over 2} \vec \phi (-\partial^2 +2 b_0
(-\partial^2)^2 )\vec \phi - {b_1 \over {2N}} (\partial_\mu \vec \phi
\cdot \partial_\mu \vec \phi )^2 - {b_2 \over {2N}} (\partial_\mu \vec
\phi \cdot \partial_\nu \vec \phi - {1\over 4} \delta_{\mu , \nu }
\partial_\sigma \vec \phi \cdot \partial_\sigma \vec \phi )^2
\right]\cr} \eqno(e2p5p1)$$ where $\fhi^2 = N\beta$. The parameter $b_0
$ can be absorbed in $b_1$ and $b_2$ by a field redefinition $$ \fhi
\rightarrow {{\fhi - b_0 \partial^2 \fhi }\over{\sqrt{\fhi^2 + b_0^2
(\partial^2 \fhi )^2 -2b_0 \fhi\partial^2 \fhi }}}\sqrt{N\beta} \fs
\eqno(e2p5p2)$$ Out of the four parameters $\beta$, $b_0$, $b_1$ and
$b_2$ only three combinations affect the leading and subleading terms
in the expansion of pion scattering amplitudes in the external momenta.
With Pauli--Villars cutoff we shall simply set $b_0=0$, but on the
lattice it is usually more convenient to stick with the particular
value for $b_0$ that comes from the expansion of the lattice kinetic
energy term at small momenta. This has to be taken into account when a
bare action regulated by Pauli--Villars propagators is compared to one
regulated by a lattice. At infinite $N$ the $b_2$ term won't contribute
to the saddle point equations because the $O(N)$ invariant bilinear
that appears in it cannot acquire a vacuum expectation value as long as
Lorentz invariance is preserved. More analysis will show that this term
doesn't contribute to the leading cutoff effects on $\pi$ $\pi$
scattering and the Higgs width. Hence, up to questions of global
stability $b_2$ can be ignored and simply set to zero. This fact will
be used as an argument to simplify the kind of lattice actions we are
going to consider explicitly.

The class of actions regulated by Pauli--Villars terms we choose to
investigate is, in dimensionless units, given by $$ \eqalign{ S ~ &=\cr
\int_x ~ & \left [{1\over 2} \vec \phi [-\partial^2 +(- \partial^2
)^{n+1} ]\vec \phi - {1 \over {2Ng_1 }} (\partial_\mu \vec \phi \cdot
\partial_\mu \vec \phi )^2 - {1 \over {2Ng_2 }} (\partial_\mu \vec \phi
\cdot \partial_\nu \vec \phi - {1\over 4} \delta_{\mu , \nu }
\partial_\sigma \vec \phi \cdot \partial_\sigma \vec \phi )^2
\right]\cr} \eqno(e2p5p3)$$ where $\fhi^2=N\beta$ and $n\ge 3$ is an
integer. When $n\rightarrow \infty $ we approach a sharp momentum
cutoff. For such a cutoff the parameter counting presented in the
introduction doesn't work because even at first subleading order in the
inverse cutoff nonlocal higher dimensional operators make their
appearance. We are interested in large $n$ values, however, because
there one expects a greater similarity to the lattice, in the sense
that arbitrarily high momentum excitations are almost totally
suppressed.

We shall also study actions regularized on the lattice because only the
latter can be solved nonperturbatively by Monte Carlo. We shall
consider the $F_4$ lattice and the hypercubic lattice.  On a hypercubic
lattice the operator counting is slightly wrong because of the
existence of a Lorentz breaking dimension six operator which has no
counterpart in any conceivable extension of the minimal standard
model.  This has no effect on the particular observables that we will
consider at $N=\infty$, but would lead to measurable effects at order
inverse cutoff square in other observables. Such a problem does not
arise for the $F_4$ lattice.

The most concise action on the $F_4$ lattice that reduces, up to higher
order terms, in the continuum limit to the form \(e2p5p1) is $$
\eqalign{ S= & ~ -2N\beta_0 \sum_{<x,x'>} \Fhi (x) \cdot \Fhi (x')
-N\beta_1 \sum_{<x,x'>} [ \Fhi (x) \cdot \Fhi (x')  ]^2 \cr & ~
-N~{{\beta_2}\over{8}} \sum_x \sum_{{<ll'>}\atop {l,l' \cap  x \ne
\emptyset ,~ l\cap x' \ne\emptyset ,~ l'\cap x'' \ne\emptyset ,~
x,x',x'' {}~{\rm all~ n.n.}} } \left[ \left( \Fhi (x) \cdot \Fhi (x')
\right) \left( \Fhi (x) \cdot \Fhi (x'') \right) \right] \cr}
\eqno(e2p5p4)$$ where $x, x',x''$ denote sites and $<x,x'>,l,l'$ links.
Here the field is constrained by ${\Fhi}^2 (x) =1$. Geometrically, the
first two terms couple two fields connected by a nearest neighbor bond
and the last term couples three fields that live at the corners of an
equilateral triangle whose sides are nearest neighbor bonds. The
existence of elementary triangular ``plaquettes'' is a special feature
of the $F_4$ lattice and provides here for the desirable feature that
\(e2p5p4) couples only nearest neighbors.

This feature is desired for Monte Carlo simulations because these do
not yield direct measurements of physical cutoff effects and one
usually estimates the latter by looking at the range of interactions in the
ultraviolet expressed in terms of the measured estimates for the Higgs
mass. This range is simply the lattice spacing when the action couples
only nearest neighbors. It is independent of the values of the
couplings in the action. If the action contained both nearest neighbor
and next nearest neighbor couplings it would be less clear what to view
as the range and it would be quite reasonable to assume that the range
does depend this time on the couplings in the action. In the pure
nearest neighbor case we still don't know, from Monte Carlo alone, how
large the observable cutoff effects really are, but we can expect with
reasonable confidence that variations of the bare couplings affect the
cutoff effects only through variations in the Higgs mass as measured in
lattice units. However, when there is a mixture of nearest neighbor and
next nearest neighbor terms in the action one may suspect that what one
calls the ``lattice unit'' changes when the couplings are varied.

To take the continuum limit we rescale the field $\fhi = \sqrt
{6N(\beta_0 +\beta_1 +\beta_2)} \Fhi$ (we only consider the region
$\beta_0 +\beta_1 +\beta_2 > 0$).  The combination ${1\over{36}}
{{\beta_1}\over {(\beta_0 +\beta_1 +\beta_2 )^2}} $ is then analoguous
to $b_2$ and the combination ${1\over{48}} {{\beta_1 +\beta_2} \over
{(\beta_0 +\beta_1 +\beta_2 )^2}} $ analoguous to $b_1$. The first term
in \(e2p5p4) gives the usual kinetic term, $g(p)$, on the $F_4$
lattice, $$ g(p) = {1 \over 6} \sum_{\mu \ne \nu} \left[ 2 - \cos(p_\mu
+ p_\nu) - \cos(p_\mu - p_\nu) \right] = p^2 - {1 \over 12} (p^2)^2 +
O(p^6) \fs \eqno(e2p5p5)$$ The $p^4$ part corresponds to a negative
$b_0$ in \(e2p5p1) which, with the field redefinition \(e2p5p2),
generates a positive contribution to $b_1$.  Thus the na\"\i ve
nonlinear nearest neighbor $F_4$ lattice action corresponds effectively
to a continuum action with a positive $b_1$-type term.

For hypercubic lattices the inclusion of next nearest neighbor
couplings is unavoidable if one wishes to obtain \(e2p5p1) in the long
wavelength limit.  The argument against next nearest neighbor terms
holds only within the limitations of Monte Carlo methods and is
therefore relevant to the physical case of $N=4$. However, at
$N=\infty$ ``mixed'' actions are as useful as ``pure'' ones because we
have at our disposal means to directly evaluate the observable cutoff
effects. Therefore we shall also consider ``mixed'' lattice actions.

As a matter of fact the large $N$ analysis of the model \(e2p5p4) is
quite complicated while certain ``mixed'' models are easier. The reason
is that at large $N$ one has to introduce auxiliary fields to decouple
the terms in the action that are quartic in the fields. One would need
an auxiliary field for each of the ``positive'' bonds emanating from a
site (12 on an $F_4$ lattice) and additional auxiliaries for enforcing
the nonlinear constraint. This large number of coupled fields makes an
analytical treatment cumbersome on the $F_4$ lattice. In Pauli-Villars
regularization one also has quite a few auxiliary fields but most can
be easily decoupled exploiting Euclidean $O(4)$ invariance. On a
lattice the symmetry is only a discrete subgroup of $O(4)$, and while
this might be sufficient, we chose to avoid carrying out this exercise.
The Pauli--Villars analysis will teach us that all the extra
auxiliaries are irrelevant to the order in inverse cutoff that we are
interested in. But to this order, the $F_4$ lattice is essentially
$O(4)$ invariant so the same conclusion should hold on it too.
Therefore it seems somewhat a waste of effort to struggle to fully
solve (to infinite order in the inverse cutoff) the pure nearest
neighbor action.  Moreover, this ``exact solution'' only goes as far as
giving us expressions involving some lattice momentum integrals, and,
in practice, one still has to evaluate the latter by numerical means
making it again necessary to approximate by ignoring terms that have no
effect to leading order in the inverse cutoff. To be sure, in Monte
Carlo simulations, at $N=4$, we are advocating the use of the
``triangle'' action \(e2p5p4).

We have argued earlier that for Pauli--Villars regularization the $b_2$
term plays no role to the order in the inverse cutoff considered.  We
assume that this would also happen on the lattice and set $\beta_1$ to
$0$. We have also explained that we would like to avoid having to deal
with the $\beta_2$ term in \(e2p5p4).  To preserve the freedom of
varying the strength of the important couplings in the long wavelength
limit we need to replace this term by another one that has a similar
effect at long wavelength but which can be easily ``decoupled''
preferably employing one auxiliary field only.  We choose an action
that does this:  $$ S=-2N\beta_0 \sum_{<x,x'>} \Fhi (x) \cdot \Fhi (x')
-N{{\beta_2}\over{16 \gamma}} \sum_x \left[\sum_{{l\cap x \ne
\emptyset}\atop {l=<x,x'>} } \Fhi (x) \cdot \Fhi (x') \right]^2 \fs
\eqno(e2p5p6)$$ The parameter $\gamma$ has been introduced because the
same formula will work for the hypercubic lattice also; for the $F_4$
lattice $\gamma = 3$, but for the hypercubic lattice $\gamma=1$.
Geometrically, the new term corresponds to a coupling between any four
fields that live at the ends of two bonds that have a site in common.
Before, on the $F_4$ lattice, we only had such coupling if the two
bonds spanned an angle of sixty degrees at the common site. Now larger
angles are included and fields living on sites separated by more than a
single lattice spacing are coupled.

On a hypercubic lattice it is impossible to write down an action like
\(e2p5p4) involving only nearest neighbor terms. The second term in
\(e2p5p4), when considered on a hypercubic lattice, becomes in the
continuum limit proportional to $\sum_\mu (\partial_\mu \Fhi \cdot
\partial_\mu \Fhi )^2$ and thus breaks Lorentz invariance. We therefore
have an additional reason to ignore it as we did before for the $F_4$
lattice. The action \(e2p5p6) considered on a hypercubic lattice
reduces, after rescaling $ \fhi = \sqrt {2N(\beta_0 +\beta_2)} \Fhi$,
almost to the form \(e2p5p1) with $b_2 = 0$ and ${1 \over 32} {\beta_2
\over {(\beta_0 + \beta_2)^2}}$ analoguous to $b_1$. The difference is
that instead of the $b_0$ term in \(e2p5p1) we now have a Lorentz
invariance breaking term $\fhi \sum_\mu \partial_\mu^4 \fhi$. In view
of this we have no strong reason for considering this particular form
of the action, except that it includes, with $\beta_2$ set to 0, the
case most thoroughly investigated to date for $N=4$.

In na\"\i ve nonlinear models on the hypercubic lattice the Lorentz
invariance breaking term can be avoided by using a ``Symanzik
improved'' action. For $N=4$ the exact elimination of Lorentz
invariance breaking terms at order inverse cutoff square is not an
attractive proposition because it would necessitate an impracticable
amount of ``fine tuning''.  At infinite $N$ this problem is less severe
and we can do quite well with just generalizing the (tree level)
Symanzik improved action to also contain a $b_1$ like term with four
fields and four derivatives. We are thus led to also consider the
action $$ \eqalign{ S_{SI} = & ~ -2N\beta_0 \sum_{x,\mu} \left( {4
\over 3} \Fhi (x) \cdot \Fhi (x+\mu) - {1 \over 12} \Fhi (x) \cdot \Fhi
(x+2\mu) \right) \cr & ~ -N{{\beta_2}\over{20}} \sum_x \left[ \sum_{\pm
\mu} \left( {4 \over 3} \Fhi (x) \cdot \Fhi (x+\mu) - {1 \over 12} \Fhi
(x) \cdot \Fhi (x+2\mu) \right) \right]^2 \fs \cr} \eqno(e2p5p7)$$ As
will be seen later on, this choice of the transcription of the
four-derivative term exactly eliminates Lorentz breaking terms in the
full pion propagator at order inverse cutoff square.

\head{3. Phase diagram for the Pauli--Villars models.} \taghead{3.}

Similarly to the linear case we introduce auxiliary fields in \(e2p5p3)
to make the dependence on $\fhi$ bilinear. We obtain a new action $$
\eqalign{ S_1=& ~ \int_x \left [ {1\over 2}\fhi K \fhi +{1\over 2}
\lambda ( \partial_\mu \fhi )^2 +{1\over 2} \rho (\fhi^2 -N\beta )
+{1\over 2} \omega_{\mu\nu} [\partial_\mu \fhi \partial_\nu \fhi
-{1\over 4} \delta_{\mu \nu } ( \partial_\sigma \fhi )^2 ]\right ] +\cr
{1\over 8}& ~ \int_x  [g_1 N \lambda^2 + g_2 N \omega_{\mu \nu}
\omega_{\mu \nu} ]\cr} \eqno(e3p1)$$ where $\omega_{\mu\mu} =0$
(summation over repeated indices is implied) and
$\omega_{\mu\nu}=\omega_{\nu\mu}$ and hence the $\delta_{\mu\nu}$ terms
above can be replaced by zero. In more condensed notation we have $$
S_1 = {1\over 2} \int_x \fhi \hat K \fhi -{N\over 2}\int_x [\beta\rho
-{1\over 4} g_1 \lambda^2 - {1\over 4}g_2
\omega_{\mu\nu}\omega_{\mu\nu} ]\eqno(e3p2)$$ with $$ \hat K = K -
\partial_\mu \lambda \partial _\mu + \rho - \partial_\mu
\omega_{\mu\nu}  \partial_\nu \fs \eqno(e3p3)$$ We separate the zero
mode of $\fhi$ as in the linear case in \(e2p1p7) and integrate out
$\vec \pi$ and $H$. We obtain $$ S_2 = {N\over 2} \left [ {\rm Tr}\log
\hat K +v^2\int_x \rho - v^2 \int \rho^\prime {\hat K}^{-1} \rho^\prime
-\int_x (\beta\rho -{1\over 4} g_1 \lambda^2 -{1\over 4} g_2
\omega_{\mu\nu}\omega_{\mu\nu} )\right ]\eqno(e3p4)$$ with $$
\partial_\mu (\rho - \rho^\prime ) =0 ~,~~~~~\int_x \rho^\prime =0 \fs
\eqno(e3p5)$$

\subhead{3.1. Saddle point equations and dominating saddles.}
\taghead{3.1.}

The saddle point equations are $$ \eqalign{ &\int_p {1\over{K(p^2 )
+\lambda_s p^2 +\rho_s}} + v^2 =\beta\cr &\int_p {{p^2}\over{K(p^2 )
+\lambda_s p^2 +\rho_s}}=-{1\over 2} g_1 \lambda_s \fs \cr}
\eqno(e3p1p1)$$ As promised, the saddle point equations do not depend
on $g_2$.  Sometimes they may admit several solutions. In these cases
we need to find the dominating one.  Among the possible solutions some
may have broken symmetry and some unbroken symmetry all at the same
couplings. In such regions of the phase diagram the order--disorder
transition can become discontinuous. The analytic continuation of the
continuous transition into these domains yields metastable critical
regimes that have to be eliminated from our search for the Higss mass
bound. Therefore, a complete analysis of the competing saddles is
necessary. We shall ignore possible dominating ``end point''
contributions because we are pretty sure that they will not affect our
conclusions regarding the accessible critical regime. As will be clear
later on, the possibility of first order transitions cutting into the
critical regime is realized and has physical content teaching us
something about the dynamics that is both relevant and illuminating for
the mass bound issue.

{}From \(e3p4) it is clear that we wish to find the solution
$(\lambda_s , \rho_s )$ that minimizes the following function of
$\lambda$ and $\rho$:  $$ \psi(\lambda, \rho ) = \int_p \log {{p^2
(1+p^{2n}) +\lambda p^2 + \rho }\over{p^{2(n+1)}}} +v^2 \rho
-\beta\rho+{1\over 4} g_1 \lambda^2 \fs \eqno(e3p1p2)$$ The function is
completely defined when one adds that $v^2 > 0$ implies $\rho=0$ and
$\rho >0 $ implies $v^2 =0$. Both $\rho$ and $v^2$ are nonnegative.

To simplify the analysis we introduce some rescaled variables $$
u=(1+\lambda ) ( 16 \pi^2 \beta )^{{n\over {n-1}}}~,~~~~~t=\rho
(1+\lambda)^{-{{n+1}\over n}} \fs \eqno(e3p1p3)$$ It is also useful to
factor out a positive constant from $\psi$ and consider from now on
only the minimization of $\hat \psi$, defined by $$ \hat \psi (u,t)
=u^{{2\over n}} \int_0^{\infty} \xi d\xi  \log {{t+\xi +
\xi^{n+1}}\over {\xi^{n+1}}} +{{v^2} \over {\beta}} t u^{{{n+1}\over
n}} -t u^{{{n+1}\over n}} +{1\over 2} g^* (u-u^* )^2 \eqno(e3p1p4)$$
where $$ g_1 = 32\pi^2 \beta^2 g^*~,~~~~~~u^* =(16\pi^2 \beta
)^{{n\over {n-1}}} \fs \eqno(e3p1p5)$$ It is easy to check that the
saddle point equations are reproduced by setting the derivatives of
$\hat \psi$ with respect to $u$ and $t$ to zero. This of course had to
be true.

We now split the candidate saddles into two classes according to
whether the symmetry is broken at the saddle or not. In the symmetric
phase $v^2=0$ and the equation $\partial {\hat \psi}/ \partial t =0$
can be used to define a function $t(u)$ for $0\le u \le u_0 $ by $$
\int_0^{\infty} {{\xi d\xi}\over {\xi^{n+1}+\xi +t(u)}} =
u^{{{n-1}\over n}}~,~~~~~~u_0 = \left ( \int_0^{\infty} {{d\xi}\over
{1+\xi^n }} \right )^{{n\over{n-1}}}= \left ( {{\pi}\over {n\sin
({{\pi}\over n})}} \right )^{{n\over{n-1}}} \fs \eqno(e3p1p6)$$ $t(u)$
varies between zero and positive infinity when $u$ goes from $u_0$ to
$0$.

Let $u_s$ and $t_s$ be the coordinates of a saddle point in the
symmetric phase. Starting from $$ \hat \psi (u_s , t_s )
=\int_{u_0}^{u_s} {{\partial \hat \psi }\over {\partial u}} (u,t(u)) +
\hat \psi ( u_0 ,0)\eqno(e3p1p7)$$ we derive $$ \hat \psi (u_s , t_s )
=\int_{u_0}^{u_s} [ G(u) +g^* (u-u^* )] + \hat \psi ( u_0
,0)\eqno(e3p1p8)$$ where $$ G(u) = u^{{2-n}\over n} \int_0^{\infty}
{{\xi^2 d\xi}\over {\xi^{n+1} +\xi +t(u)}} \fs \eqno(e3p1p9)$$ $G$ is
defined in the interval $(0, u_0 )$ and is monotonically decreasing
there.

In the broken phase the equation $$ \left ( {   {\partial \hat \psi }
\over {\partial t } } \right )_{t=0} = 0\eqno(e3p1p10)$$ together with
the requirement $v^2 \ge 0$ yields the restriction $u_s \ge u_0$.  The
function $\hat \psi$ at the saddle can be written as $$ \hat \psi (u_s
,0) = {n\over 2} u^{2\over n} \int_{0}^{\infty}  {{\xi d\xi }\over
{1+\xi^n }} +{1\over 2} g^* (u-u^* )^2 \fs \eqno(e3p1p11)$$ If we
extend the range of the function $G$ from $u\in (0, u_0 )$ to the
segment $u\in [u_0, \infty)$ with $$ G(u)= u^{{2-n}\over n}
\int_0^{\infty} {{\xi d\xi }\over {1+\xi^n }}= u^{{2-n}\over n}
{{\pi}\over {n\sin ({{2\pi}\over n})}}~~~~{\rm for} ~~u >
u_0\eqno(e3p1p12)$$ we can rewrite \(e3p1p11) as $$ \hat \psi (u_s ,0 )
=\int_{u_0}^{u_s} [ G(u) +g^* (u-u^* )] + \hat \psi ( u_0, 0) \fs
\eqno(e3p1p13)$$ The advantage of these manipulations is that equations
\(e3p1p8) and \(e3p1p13) have identical forms and can be geometrically
interpreted.

$G(u)$ is somewhat complicated but independent of the continuously
varying couplings -- hence it can be computed once $n$ is given. The
dependence on the couplings comes in through the parameterization of
the straight line $-g^* (u-u^* )$; different couplings correspond to
different intercepts and slopes of the straight line. All candidate
saddles are found at intersections between the straight line and the
``universal'' function $G$. The quantity to be minimized is the signed
area bounded by $G$, the straight line, the line $u=u_0$ and by the
particular intersection point under investigation. When the straight
line $-g^* (u-u^* )$ intersects $G$ three times and the two areas
between the consecutive intersection points are of equal magnitude we
are at a symmetry breaking transition point of first order. When the
straight line intersects $G$ only once at the point $(u_0 , G(u_0 ))$
the transition is second order if the straight line is steeper than the
line representing the tricritical case. The tricritical case
corresponds to a straight line that goes through $(u_0 , G(u_0 ))$ and,
in addition, is tangent to $G$ there. These cases are illustrated in
Fig.~(3.1) for $n=3$. When the straight line does not intersect $G$
there is no translationally invariant saddle and we are in a frustrated
phase.

\figure{3.1}{\captionthreeone}{4.8}

In equations the conditions for criticality are as follows: The
tricritical point has parameters given by $$ g_{t.c.}^*
={{n-2}\over{2\pi}}\tan \left ( {{\pi}\over n} \right )~,~~~~~
u_{t.c.}^* =2{{n-1}\over{n-2}}u_0 \eqno(e3p1p14)$$ and the second
order line is described by $$ u^* =u_0 \left ( 1 +
{{n}\over{n-2}}{{g_{t.c.}^*}\over{g^*}}\right
)~,~~~~{{g_{t.c.}^*}\over{g^*}} < 1 \fs \eqno(e3p1p15)$$

\subhead{3.2. Physical properties of the phase diagram.} \taghead{3.2.}

The details of the complete phase diagram are not essential; the
schematic structure is presented in Fig.~(3.2). The ``frustrated
region'' corresponds to regions where translational invariance breaks
spontaneously and regular saddle points are either not competitive or
do not exist.

\figure{3.2}{\captionthreetwo}{4.8}

It is important to understand the source of the tricritical point: As
already mentioned, the action can also be viewed as an effective chiral
Lagrangian with couplings of order one. The single stable particles are
the pions. Two pions will attract if they are in a relative zero
angular momentum and total isospin singlet state. This is easy to
understand: If two field configurations corresponding to an
approximately localized pion are placed one on top of the other and the
isospin indices match appropriately, the net state will be closer to
the vacuum than a state where the pions are far apart. The index
matching will be right when the total isospin is zero. Hence, soft
pions attract in the $I=0$, $J=0$ state, a well known fact. One can
change the interaction between the pions only at subleading order in
their momenta, $p^4$ (even the logarithmic part $p^4 \log p^2$ is fixed
by current algebra), and this is the main physical effect produced by
varying the couplings (of course the value of $F_\pi /\Lambda$ is also
changed by the variation of the couplings, but this is an ``unphysical''
effect). If one introduces extra attraction, it is possible that two
pions of some nonvanishing momentum each (in units of the cutoff), can
bind to a zero mass state which will be stable and condense. We believe
that this is what is happening at the tricritical point and we shall
present some evidence later on. As far as the Higgs mass bound goes, we
want to make the resonance replacing the above bound state as heavy as
possible. This will be achieved if we introduce as much repulsion as
possible between the pions, thus delaying the formation of the
resonance to higher momenta of the ``constituent'' pions. We shall see
that this reasoning is born out by explicit calculations.

To complete our investigation of the region of the phase diagram we are
interested in, we must also ascertain the local stability of our saddle
points (global stability was checked in section (3.1) but we ignored
until now the question of local stability). To do this part of the
analysis we need to compute the small fluctuations around the saddle
points. Since we are interested only in a particular region, our
computations will be restricted to that part of the phase diagram. We
should also remember that some of the field variables are auxiliaries.
The integration contour for auxiliary fields that are unphysical can be
deformed in the stable directions, so stability isn't really an issue
there. However, we are not allowed to deform the integration contours
for the ``physical''  fields, because if we did that we would easily
loose the approximate unitarity we have at low energies. Among the
auxiliaries only $\rho$ is ``unphysical'' as its role was just to
impose the fixed length constraint; however, $\omega_{\mu\nu}$ and
$\lambda$ are simply related to bilinears of field gradients and should
stay real fields. In short, we must make sure that the pion, Higgs,
$\omega$ and $\lambda$ propagators come out to be positive in Euclidean
momentum space.

\subhead{3.3. Pion propagator and ghosts.} \taghead{3.3.}

{}From now on we shall always assume that the couplings are chosen in
such a way that we are somewhere in the broken phase close to the
second order transition.

We again separate out the zero mode from the field $\fhi$ and
parametrize the remainder by $H$ and $\vec \pi$  (see \(e2p1p7)).
Expanding around the saddle we introduce the shifted fields $$ \delta
\lambda = \lambda -\lambda_s~,~~\delta H=H~,~~\delta\vec \pi = \vec
\pi~,~~\delta \omega_{\mu\nu}=\omega_{\mu\nu} ~,~~\delta\rho =\rho \fs
\eqno(e3p3p1)$$
 From \(e3p3) we read off the pion propagator in Fourier space $$
 <\delta  \pi^a \delta  \pi^b > ={{\delta^{ab}}\over{K(p^2 )+\lambda_s
p^2}}={{\delta^{ab}}\over{(1+\lambda_s )p^2 +(p^2 )^{n+1}}} \fs
\eqno(e3p3p2)$$
 There are ghosts in this propagator (poles with residues that are not
 real positive numbers). Their
 presence reflects the fact that Lorentz invariant Pauli--Villars
regularization is achieved at the
 expense of exact unitarity in Minkowski space. The ghost poles are
 located on a circle
 in the complex $p^2$ plane $$ |p_{ghost}^2 |=(1+\lambda_s )^{1\over n}
 \fs \eqno(e3p3p3)$$
 The condition $\lambda_s \ge -1$ is always satisfied. When the
 energies in a process approach
 $(1+\lambda_s )^{1\over {2 n}} \Lambda \equiv \Lambda_s$ one expects
 violent cutoff effects to set
 in. Thus, the  ``physical'' cutoff scale is coupling dependent and
 different from the ``bare'' cutoff
 $\Lambda$. In retrospect we see that it makes more sense to measure
 our dimensionful
 quantities not in terms of $\Lambda$ as we did until now but rather in
 terms of $\Lambda_s$. When
 we do the lattice analysis we shall see that violations of Euclidean
rotational invariance occur
 in the pion propagator at a distance of the order of the lattice
 spacing and that there is no dependence
 on the couplings. In our case the correct analogue of the inverse
 lattice spacing is $\Lambda_s$, not $\Lambda$, and rescaling by
$\Lambda_s$ here is analogous to the standard practice of setting the
lattice spacing equal to unity in lattice work.

 We rescale all our fields  (including the auxiliaries) and couplings,
 but keep the old notation,
 $$ \eqalign{ \partial_{\mu} ,~p ~~~ \rightarrow& ~~~ \partial_{\mu}
 ,~p ~~\times (1+\lambda_s)^{1\over{2n}}\cr
 \partial_{p} ,~x ~~~ \rightarrow& ~~~ \partial_{p} ,~x ~~\times
(1+\lambda_s)^{-{1\over{2n}}}\cr \fhi ,~H,~\vec \pi ~~~ \rightarrow&
{}~~~ \fhi ,~H,~\vec \pi   ~~\times (1+\lambda_s)^{{1\over{2n}}}\cr
\lambda ~~~ \rightarrow& ~~~ \lambda\cr \rho ~~~ \rightarrow& ~~~ \rho
{}~~\times (1+\lambda_s)^{{1\over{n}}}\cr \omega ~~~ \rightarrow& ~~~
\omega\cr v ~~~ \rightarrow& ~~~ v~~\times
(1+\lambda_s)^{{1\over{2n}}}\cr g_{1,2} ~~~ \rightarrow& ~~~ g_{1,2}
{}~~\times (1+\lambda_s)^{{2\over{n}}} \fs \cr} \eqno(e3p3p4)$$ The
original constraint $\fhi^2 =N\beta$ now becomes $\fhi^2 = (1+\lambda_s
)^{-{1\over n}} N\beta$ and the pion propagator becomes $$
 <\delta  \pi^a \delta  \pi^b > ={{\delta^{ab}}\over{(1+\lambda_s )[p^2
 +(p^2 )^{n+1}]}}={{\delta^{ab}}\over{(1+\lambda_s )K(p^2 )}} \fs
\eqno(e3p3p5)$$
 With standard conventions we define the pion wave function
 renormalization constant $Z_\pi$
 by $$ Z_\pi = {1 \over {1+\lambda_s}} \fs \eqno(e3p3p6)$$
 This constant is fixed by the couplings through the saddle point
 equations.  {}From our analysis
 of the phase diagram we deduce that the range in which $Z_\pi$ is
 allowed to vary is
 $$ 0\le Z_\pi \le {{2(n-1)}\over {n-2}} \fs \eqno(e3p3p7)$$ $Z_\pi$
 depends only on $g_1$ and it is useful to invert the relationship
viewing $Z_\pi$ as the
 external control parameter restricted only by \(e3p3p7) $$ {1\over
 {g_1}}={1\over {2\int_k {{k^2}\over{K(k)}}}} \left ( {1\over{Z_\pi
}}-{1\over{Z_\pi^2}}\right ) \fs
 \eqno(e3p3p8)$$ \(e3p3p8) is a rewriting of the second saddle point
 equation and fixes $\lambda_s$. The other saddle point equation fixes
$v$.
 We know that ultimately we shall be more interested in the quantity
 $v^2 / Z_\pi$ because it
 is equal to the pion decay constant. So we choose to write the first
 saddle point equation in the form
 $$ {{v^2}\over {Z_\pi}} =\beta Z_\pi^{{1-n}\over n} -\int_k {1\over
 {K(k^2 )}} \fs \eqno(e3p3p9)$$

\subhead{3.4. Small fluctuations in the broken phase.} \taghead{3.4.}

After integrating out the pions from the action \(e3p2) and taking into
account the rescalings \(e3p3p4), we expand around the saddle point. To
quadratic order in the small fluctuations we obtain
 $$ \eqalign { S_2^{(2)}= {1\over {2 Z_\pi}} \int_k& ~ \delta H K
 \delta H +\sqrt{N} v \int_k \delta\rho\delta H +{N\over 8}
 \int_k (g_1 \delta\lambda^2 + g_2 \delta \omega_{\mu\nu}^2 ) +\cr
 {{N-1}\over 2} Z_\pi^2 \int_k & ~ \left [ {1\over 2} \delta \lambda
 B^{\lambda\lambda} \delta \lambda +\delta\lambda
B_{\mu\nu}^{\omega\lambda} \delta\omega_{\mu\nu} +{1\over 2}\delta
\omega_{\mu\nu} B_{\mu\nu ,\mu^\prime \nu^\prime}^{\omega\omega}
\delta\omega_{\mu^\prime \nu^\prime}\right ] + \cr
 {{N-1}\over 2} Z_\pi^2 \int_k & ~ \left [ {1\over 2}\delta\rho
 B^{\rho\rho} \delta\rho +\delta\rho
B^{\rho\lambda}\delta\lambda+\delta\rho
B_{\mu\nu}^{\omega\rho}\delta\omega_{\mu\nu} \right ]
\cr}\eqno(e3p4p1)$$ where $$ \eqalign { B^{\rho\rho}(p^2 )=& ~ -\int_k
{1 \over {K ({1\over 2}p +k ) K ({1\over 2}p-k)} }\cr
B^{\lambda\lambda} (p^2 )=& ~ -\int_k {{({1\over 4}p^2
-k^2)^2}\over{K({1\over 2}p +k ) K({1\over 2}p-k)}}\cr
B^{\rho\lambda}(p^2 )=& ~ \int_k {{{1\over 4}p^2 -k^2}\over{K({1\over
2}p +k ) K({1\over 2}p-k)}}\cr B_{\mu\nu}^{\omega\rho}(p^2 )=& ~ \int_k
{{ {1\over 4}p_\mu p_\nu -k_\mu k_\nu -{1\over 4} \delta_{\mu\nu}
({1\over 4}p^2 -k^2)} \over{K({1\over 2}p +k ) K({1\over2}p-k)} } \cr
B_{\mu\nu}^{\omega\lambda}(p^2 )=& ~ -\int_k {{[{1\over 4}p_\mu p_\nu
-k_\mu k_\nu -{1\over 4} \delta_{\mu\nu} ({1\over 4} p^2 -k^2)]
({1\over 4} p^2 -k^2)}\over{K({1\over 2}p +k ) K({1\over2}p-k)}} \cr
B_{\mu\nu , \mu^\prime \nu^\prime }^{\omega\omega}(p^2 )=&\cr -\int_k&
{}~ {{ [{1\over 4}p_\mu p_\nu -k_\mu k_\nu -{1\over 4} \delta_{\mu\nu}
({1\over 4}p^2 -k^2)] [{1\over 4}p_{\mu^\prime} p_{\nu^\prime}
-k_{\mu^\prime} k_{\nu^\prime} -{1\over 4} \delta_{{\mu^\prime}
{\nu^\prime}}  ({1\over 4}p^2 -k^2)]} \over{K({1\over 2}p +k )
K({1\over2}p-k)} } \fs \cr} \eqno(e3p4p2)$$

To compute the various propagators we would have to invert a $12 \times
12$ matrix. To simplify this task we exploit rotational invariance to
block diagonalize the matrix. We decompose the field
$\delta\omega_{\mu\nu}$ in a spin--zero, spin--one and spin--two
field.  To do this we introduce Euclidean polarization vectors $W_\mu^j
(p) $, where $j=1,2,3$, $$ W_\mu^j (p) W_\mu^k (p)
=\delta^{jk}~,~~~~~p_\mu W_\mu^j (p) =0 \fs \eqno(e3p4p3)$$ We now
define new $\omega$ fields $$ \eqalign{ \delta\omega_{\mu\nu} (p) ~ &
=\cr W_\mu^j (p) W_\nu^k (p) \delta\omega_{jk} (p) ~ & +{1\over {2|p|}}
[p_\mu W_\nu^j (p) +p_\nu W_\mu^j (p) ]\delta\omega_j (p)+{1\over
{p^2}} [p_\mu p_\nu -{1\over 4} \delta_{\mu\nu} p^2 ] \delta\omega (p)
\fs \cr}  \eqno(e3p4p4)$$
The spin--two field $\delta\omega_{jk}$ is
symmetric and traceless. Because of Lorentz invariance the $12 \times
12$ matrix must now split up into three blocks: a $4\times 4$ block
involving the scalar fields $\delta H$, $\delta\lambda$, $\delta\rho$
and $\delta\omega$, a diagonal $3\times 3$ block for the vector field
$\delta \omega_j$ and a diagonal $5\times 5$ block for the tensor field
$\delta \omega_{jk}$. This decomposition is exact in Pauli--Villars
regularization because of the preservation of rotational invariance; on
a lattice this decoupling would not be exact, but, as long as one is
ultimately going to expand in the inverse cutoff only to leading and
subleading order, a similar simplification should occur for $F_4$
lattice actions that don't break the lattice symmetries. In terms of
the new $\delta \omega$ fields the quadratic action becomes, ignoring
order $1/N$ corrections,
 $$ \eqalign { &S_2^{(2)}= {1\over {2 Z_\pi}} \int_k \delta H K \delta
 H +\sqrt{N} v \int_k \delta\rho\delta H +
 {{N}\over 2} Z_\pi^2  \cr\int_k  & \left [
 \delta \lambda ({1\over 2} B^{\lambda\lambda} +{1\over {4Z_\pi^2 }}
 g_1 ) \delta \lambda +\delta\lambda B^{\omega\lambda} \delta\omega +
{1\over 2} \delta\rho B^{\rho\rho} \delta\rho +\delta\rho
B^{\rho\lambda}\delta\lambda + \delta\rho B^{\omega\rho}\delta\omega
 \right ]+{{N}\over 2} Z_\pi^2  \cr \int_k & \left [
 \delta \omega ( {1\over 2} B^{\omega\omega} +{3\over {16Z_\pi^2}} g_2
 ) \delta\omega +
 \delta \omega_j  ( {1\over 2} B_V^{\omega\omega} +{1\over {8Z_\pi^2}}
 g_2 ) \delta\omega_{j}  +
 \delta \omega_{jk}  ( {1\over 2} B_T ^{\omega\omega} +{1\over
 {4Z_\pi^2}} g_2 )
 \delta \omega_{jk} \right  ] \fs \cr} \eqno(e3p4p5)$$
 The $B$ ``bubbles'' in \(e3p4p5) are projections of the $B$ bubbles in
\(e3p4p2) via \(e3p4p4). Their explicit forms can be read off equation
\(+3) below.  Introducing the four component fields $\psi_A$,
representing $\delta H$, $\delta\rho$, $\delta\lambda$ and
$\delta\omega$ for $A=1,2,3,4$, respectively, we rewrite eq.~\(e3p4p5)
as $$ S_2^{(2)}={1\over 2} \int_k \psi_A M_{AB} \psi_B +{1\over
2}\int_k [ M_V\sum_j (\delta\omega_j  )^2 + M_T\sum_{jk}
(\delta\omega_{jk} )^2 ] \fs \eqno(e3p4p6)$$ The variables $M$ are
functions of the couplings and momentum square; they can all be
expressed in terms of six elementary ``bubble'' integrals
$I_{n,m}(p^2)$, $m+n \le 2$:  $$ I_{n,m} (p^2) =\int_k {{(k^2)^n
(p\cdot k)^{2m}}\over {(p^2)^m K({1\over 2}p +k ) K({1\over2}p-k)}} \fs
\eqno(e3p4p7)$$ $M_{AB}$ is symmetric with non-vanishing entries $$
\eqalign{ M_{11} =& ~ {K\over {Z_\pi}}\cr M_{12}=& ~ \sqrt {N} v\cr
M_{22}=& ~ - \half N Z_\pi^2 I_{0,0}\cr M_{23}=& ~ \half N Z_\pi^2
\left ( {{p^2}\over 4} I_{0,0}-I_{1,0} \right)\cr M_{24}=& ~ \half N
Z_\pi^2 \left ( {{3p^2}\over {16}} I_{0,0}-I_{0,1}+{1\over 4} I_{1,0}
\right )\cr M_{33}=& ~ \half N Z_\pi^2 \left ({{g_1}\over {2 Z_\pi^2 }}
- {{(p^2 )^2}\over {16}} I_{0,0}+{{p^2}\over 2} I_{1,0} -I_{2,0}
\right)\cr M_{34}=& ~ \half N Z_\pi^2 \left( - {{3(p^2 )^2}\over {64}}
I_{0,0}+ {{p^2}\over 4} I_{0,1}+{{p^2}\over 8} I_{1,0}  -I_{1,1}+
	{1\over 4}I_{2,0}\right )\cr M_{44}=& ~ \half N Z_\pi^2 \left
({{3g_2}\over {8 Z_\pi^2 }}
	- {{9(p^2 )^2}\over {256}} I_{0,0}+{{3p^2}\over 8} I_{0,1}  -
	I_{0,2} - {{3 p^2}\over 32} I_{1,0}+{1\over 2}I_{1,1}  -
	{1\over 16} I_{2,0} \right )\cr M_V=& ~ \half N Z_\pi^2 \left
({{g_2}\over {4 Z_\pi^2 }}
	+{1\over 3} I_{0,2} -{1\over 3}I_{1,1} \right )\cr M_T=& ~
\half N Z_\pi^2 \left ({{g_2}\over {2 Z_\pi^2 }}
	-{2\over {15}} I_{0,2} +{4\over {15}} I_{1,1} -{2\over {15}}
	I_{2,0} \right) \fs \cr} \eqno(e3p4p8)$$

\subhead{3.5. Spontaneous breakdown of space--time invariances and
local stability.} \taghead{3.5.}

We are finally in position to check local stability for the vector and
tensor fields. The corresponding matrix entries $M_T$ and $M_V$ first
vanish at zero momentum when $g_2$ decreases to a critical value
$g_{2c}$ given by $$ g_{2c}= {{Z_\pi^2}\over 6} \int_k {{(k^2 )^2
}\over {K^2 (k)}} \fs \eqno(e3p5p1)$$ At $g_2 =g_{2c}$ also the entry
$M_{44}$ vanishes at zero momentum. In all our subsequent calculations
we shall assume $g_2$ to be safely larger than this critical value so
that even for negative (but small in absolute magnitude relative to
unity) values of $p^2$ these entries are of order unity and positive.
Physically, we wish to keep the masses of the vector and the tensor of
the order of the cutoff. Note that both fields are isoscalars.

On a lattice a related phenomenon would be that a non-translational
invariant saddle takes over. In Pauli--Villars regularization when
translations get broken so do rotations and we obtain both a massless
vector and a massless tensor.  On a lattice rotations aren't a
continuous symmetry and we would not expect a ``tensor'' like particle
to also become massless.

Thus, although $g_2$ doesn't appear in the phase diagram there is some
bound on it too. Once this bound is satisfied we have no evidence for
any other source of local instability and we believe that the
$4\times4$ matrix $M$ will be positive definite for all Euclidean
momenta. We have not tested this fully, but from the behavior at low
momenta, which we did test (see below), it seems very safe to assume
that no local instability is hiding at high momenta and that all
critical regions that we shall henceforth be interested in are indeed
accessible.

\subhead{3.6. Spontaneous breakdown of scale invariance at the
tricritical point.} \taghead{3.6.}

The limitation on $Z_\pi$ in \(e3p3p7) is turned by eq.~\(e3p3p8) into
a limitation on $g_1$ $$ -\infty < {1\over g_1 } < {{2\pi n^2 (n-2)
\sin {{2\pi}\over n}}\over {(n-1)^2}} \fs \eqno(e3p6p1)$$ The
tricritical point is at $g_1 =g_{1,t.c.}$ where $$ g_{1,t.c.}
={{(n-1)^2}\over {2\pi n^2 (n-2) \sin {{2\pi}\over n}}} \fs
\eqno(e3p6p2)$$ Note that the rescalings \(e3p3p4) have changed the
definition of $g_1$ so that \(e3p6p2) differs from the combined effect
of \(e3p1p5), \(e3p1p6) and \(e3p1p14) by a factor of $Z_\pi^{{2\over
n}}$.

When $g_1$ approaches $g_{1,t.c.}$ it is easy to check that the matrix
element $M_{33}$ from \(e3p4p8) vanishes at $p^2=0$. This implies that
$\det M (p^2 =0) =0$, because, at $p^2=0$ also $M_{11}$ and $M_{34}$
vanish and then the first and third row of $M$ are proportional to each
other. The zero eigenvector is a particular linear combination of
$\delta H$ and $\delta\lambda$.  The field $\lambda$ is related by the
equations of motion to $(\partial_{\mu} \vec \phi )^2$; this is obvious
from eq.~\(e3p1). A pole at zero momentum in $\delta\lambda$ is
therefore likely to be interpretable as a dilaton. Therefore, at $g_1
=g_{1,c}$ the Higgs particle becomes massless and also plays the role
of a dilaton. This behavior is similar to the one discovered by
Bardeen, Moshe and Bander in three dimensions with a bare action that
was renormalizable\refto{BMB}.

\head{4. Higgs mass bound and cutoff effects with Pauli--Villars
regularization.} \taghead{4.}

We are now ready to address the main problem: How large can one make
the ratio $m_H / f_\pi$ while keeping cutoff effects small? We shall
carry out our analysis in two stages. At the first stage an approximate
but simple calculation will tell us what to expect. At the second stage
we shall perform a complete analysis.

First we need to make an observation that will simplify matters for
both stages: The Higgs mass will be obtained from the matrix $M_{AB}$
of \(e3p4p8). We are only interested in small values of $f_\pi$ ({\it
i.e.,} $F_\pi$ in units of $\Lambda_s$) and for such values the Higgs
resonance will appear at small complex value of $p^2$ (also in units of
$\Lambda_s$) because the physical coupling, even if large, is a finite
number. Therefore, it is consistent to expand the entries $M_{AB} (p^2)$
in powers and logarithms of $p^2$. In ordinary, renormalized continuum
field theory (i.e.  when calculating only universal quantities), we
would stop the expansion at leading order. Here we shall be going to
one order higher. The observation we wish to make is that to first
subleading order the scalar $\delta\omega$ decouples and the Higgs mass
and width can be obtained from the $3\times 3$ upper left corner of the
$4\times 4$ scalar block of $M$. We shall denote this submatrix by
$M^r$.

In addition, $\delta\omega$ contributes to $\pi$ $\pi$ scattering,
given by the exchange of the scalar auxiliary fields, only at orders
that we shall be neglecting.  Almost all the information needed for
calculating $\pi$ $\pi$ scattering to first subleading order in the
inverse cutoff is contained in $M^r$.  Fluctuations in the
approximately decoupled low momentum components of $\delta\omega$ are
under control because, as we have seen, local stability requirements
for the vector and tensor low momentum excitations stabilize also the
scalar $\delta\omega$. Thus, finally, the dependence on the precise
value of $g_2$ also disappears from all the processes we shall be
interested in.

Once we decided to go only to first subleading order we have to focus
on $M^r$ and from \(e3p4p7) and \(e3p4p8) we see that we only need 3
out of the 6 ``bubble'' integrals in \(e3p4p7), namely $I_{n,0}$ for
$n\le 2$. For arbitrary $n$ it would be very inconvenient to try to
work out closed forms for the complete integrals as we did in the
linear case. Moreover, unlike in the linear case, we have already
dropped some higher order terms when we reduced our problem to $M^r$ so
we should be consistent and keep only the leading and subleading terms
in the external momentum for the three bubbles that we need. Some of
the technical details of the relevant computations are sketched in
Appendix A. The results are $$ \eqalign{ I_{0,0} (p^2 ) =& ~ {1\over
{16\pi^2 }} [-\log p^2 +1-{1\over n} -(1-{1\over {n^2}}) {{\pi}\over
{12 \sin {{\pi}\over {n}}}} p^2]+\cdots \cr I_{1,0} (p^2 ) =& ~ {1\over
{16\pi^2 }} [ (1-{1\over n}){{\pi}\over {n \sin {{\pi}\over
{n}}}}+{1\over 4} p^2 \log p^2  + {{3-6n- n^2}\over {12n}} p^2 ]+
\cdots\cr I_{2,0} (p^2 ) =& ~ {1\over {16\pi^2 }}  [ {{(n-2)\pi} \over
{n^2 \sin {{2\pi}\over {n}}}} -~{{(n-1)(n+4)\pi}\over {12 n^2 \sin
{{\pi}\over {n}}}}p^2]+\cdots \fs \cr}\eqno(e4p1)$$

Note that the limit $n\rightarrow \infty$ can be taken on the leading
terms but not on the subleading ones. This reflects the fact that with
a sharp momentum cutoff the first correction is suppressed by $|p|$ and
not by $p^2$ which means that non-local operators come into play. Also,
there is a slight error in the equation for the simplest bubble,
$I_{0,0}$, in \cit{{EINH},{EINHTALLA}}: at infinite $n$ we get to
leading order in $p^2$, $I_{0,0} (p^2 ) = {1\over {16\pi^2 }} [-\log
p^2 +1]$, while in \cit{{EINH},{EINHTALLA}} $I_{0,0} (p^2 ) = {1\over
{16\pi^2 }} [-\log p^2 +2]$ is used. This has no significant effect on
the physical numbers obtained in these papers. As a matter of fact,
many workers adopt a convention where the ``physical'' cutoff
$\Lambda_L$ (the ``Landau pole'') is defined as the point where the
asymptotic expansion of $I_{0,0}$, $I_{as}={1\over {16\pi^2 }} [-\log
p^2 +c]$ obviously breaks down, namely, $I_{as}(\Lambda_L^2 )=0$. With
such a convention, the particular value of the constant $c$ does not
need to be known, and all dependence on the cutoff scheme disappears.
This convention is of course quite arbitrary but not totally
unreasonable; it does explain to a certain extent the effective
``universality'' of the triviality bound.

\subhead{4.1. Approximate calculation.} \taghead{4.1.}

It is much easier to do calculations for weak couplings. Ultimately we
wish to find out how strong the physical coupling can become without
distorting the theory too much. This is achieved by limiting the cutoff
effects. If we make this limitation extremely stringent the cutoff is
very high and because of triviality we are forced to weak physical
couplings where it is easy to calculate. Even for very stringent
limitations on the cutoff effects there will be a dependence on the
bare couplings and in some region of the space of bare couplings the
physical coupling will be allowed to be larger (while still very small
in absolute magnitude) than in other regions. It is very reasonable to
assume that there is some smoothness in the dependence on the higher
dimensional operators that is induced by the variation in the bare
couplings and, therefore, the region we shall be interested in is the
one where we expect the largest possible coupling even when the cutoff
effects are less stringently limited. In particular, we already know
that in practice the physical coupling cannot be made large in a sense
that would invalidate perturbation theory completely. The major result
to date is that even for relatively small physical couplings the cutoff
effects become sizable disallowing further increases. Therefore, our
decision to first work where the physical coupling is small is not
expected to lead us astray.

When the physical coupling is small the width of the Higgs particle is
small too and, to simplify the formulae, we shall ignore width effects
and concentrate on the quantity $m_R$ defined by $$ {\rm Re} [\det M^r
(p^2)]_{p^2 = -m_R^2} =0 \fs \eqno(e4p1p1)$$ $m_R$ is related to, but
not equal to, $m_H$.  It can become quite different from $m_H$ even for
moderate physical couplings but the relationship between the two is
monotonic and, for the purpose of identifying the right region in the
action space, $m_R$ is perfectly adequate.  Similar considerations were
presented in our study of the linear models.

In our approximate analysis all the subleading effects are small but
still hard to calculate. Since they are small, they do not affect the
numerical value of the leading order answer.\footnote{*}{A quantitative
feeling may be obtained from the analogous situation in the linear
case: see the expansion in eq.~\(e2p4p2).} We could do without
calculating them explicitly if we had an independent means to tell when
a given case will have larger cutoff effects than another case. The
easiest is to use the pion propagator, because it has a simple explicit
form to all orders in the inverse cutoff (see eq.~\(e3p3p5)). In units
of $\Lambda_s$ we know that ghosts will appear for energies of order
one and this is independent of the bare couplings. Therefore, making
$m_R$ in units of $\Lambda_s$ as small as possible while keeping the
ratio $m_R / f_\pi$ constant will diminish the cutoff effects and
identify the right region in the space of bare actions where a larger
triviality bound ought to be found. To be sure, our approximation isn't
quite consistent logically because the cutoff effects on the pion
propagator are suppressed by more than one power of $p^2$ and hence do
not reflect the dimension six operators.  We still think that our
approximation is useful because, if one does not ``fine tune'' in the
sense explained before, once the cutoff effects induced by the
dimension 6 operators become sizable (of the order of 10 percent) all
the higher dimensional operators also quickly become important.
Therefore, the pion propagator does sense the point where cutoff
effects turn on in an overwhelming way. Our argument is also helped by
the fact that our more complete analysis confirms the findings we shall
present here. We should point out that most lattice work was
essentially equivalent to viewing the inverse lattice spacing as $\sim
\Lambda_s/\pi$. This point of view is not logically superior to the one
we adopt in this section but nevertheless produces quite reasonable
results when compared to the outcome of a more sophisticated analysis.

By general arguments we know that $$ {m_R^2 } \approx C^2 (Z_\pi ) \exp
[ - 96\pi^2  / g_R  ]\eqno(e4p1p2)$$ where $$ g_R =3 {{m_R^2}\over
{f_\pi^2}} ~,~~~~~~ f_\pi^2 = {{v^2}\over {Z_\pi}} \cm \eqno(e4p1p3)$$
and we have already used the fact that any desired value of $f_\pi$ can
be obtained for any $Z_\pi$ by tuning $\beta$. Therefore the constant
$C$ depends only on $Z_\pi$ (which has been traded for $g_1$ in
\(e3p3p8)).  All that is left to compute is this constant and for this
we only need the leading terms in eq.~\(e4p1). We plug in the needed
results from \(e4p1) in \(e3p4p8) and then in \(e4p1p1) ending up with
$$\eqalign{ C(Z_\pi ) = & ~ \exp \left[ {{n-1}\over{2n}} \left( 1+
{\pi\over{n \tan (\pi /n)}}\zeta(Z_\pi ) \right) \right]\cr \zeta
(Z_\pi ) =& ~ {{Z_\pi -1} \over {1- Z_\pi (n-2)/(2(n-1)) }} \fs \cr
}\eqno(e4p1p4)$$

Since, as explained above, we want to find the range of $Z_\pi$ where,
for a fixed ratio $m_R / f_\pi$, $m_R$ is minimal, we need to minimize
$C$.  We must remember that $Z_\pi$ is restricted by \(e3p3p7). The
fact that $Z_\pi$ has to be positive is no surprise and we now see
that, at the other end of the interval, where $Z_\pi = 2(n-1)/(n-2)$,
$C$ diverges showing that the critical region shrinks to zero there.
This is where the tricritical point is located and our analysis
confirms our physical arguments that indicated that the bound would be
largest when one stays as far as possible from the tricritical point.

As far as the numerical validity of the approximation is concerned we
can get some feeling by looking at the higher order (neglected here)
terms in \(e4p1) or at the expansion in \(e2p4p2) for an analogous
problem. An accurate evaluation will be given later on in subsection
(4.6). The numerical values one gets from \(e4p1p2) are, for example,
good to one percent if $m_R $ corresponds to a physical Higgs mass no
larger than $0.800~TeV$, $Z_\pi \le 1$ and $n\le 100$.  For larger values
of $n$ the accuracy deteriorates somewhat.

\subhead{4.2. Some numbers.} \taghead{4.2.}

To get a feeling for the numbers involved let us take $N=4$ and ask
what change in $C$ will induce a noticeable change in
${{m_R}\over{2f_\pi}}$ away from the value ${{m_R}\over{2f_\pi}}
={{\pi}\over{2^{1/3}}}\approx 2.5$ corresponding more or less to the
present bound.  Maintaining the cutoff effects at a fixed magnitude
with $\delta(m_R ) =0$ we get $$ -\delta \log C= {{8\pi^2}\over
{({{m_R}\over{2f_\pi}})^3}} \delta ({ {{m_R}\over{2f_\pi}}}) =
{{16}\over{\pi}}\delta ({{{m_R}\over{2f_\pi}}}) \approx 5~ \delta
({{{m_R}\over{2f_\pi}}})=20\delta M_R~[{\rm in}~TeV] \fs
\eqno(e4p2p1)$$ To increase $M_R$ by $0.100~TeV$ we need to decrease $C$
by a factor of $e^2 \approx 7.4$ Between $Z_\pi =0$ and $Z_\pi =1$
little variation is induced, but when $Z_\pi$ approaches
$2{{n-1}\over{n-2}}$, $\ratio$ will decrease sharply. In summary, for
almost all large enough $n$'s the region $0\le Z_\pi \le 1$ seems to
give reasonable estimates for the bound. In terms of $g_1$ this region
corresponds to $g_1 \le 0$, showing that the parameterization of the
space of actions we chose is a reasonable one and lending support to
our physical understanding of the mechanism affecting the bound.

When $Z_\pi =1$ ${1\over {g_1}} =0$ by \(e3p3p8) and the model is the
simplest nonlinear model possible. While this is not the most optimal
place to look for the bound, it is reasonably close to it. Therefore,
even the simplest model, and even setting $n=\infty$, as we discussed
above, will give us reasonable and quite high estimates for the bound.
On the lattice the situation is not exactly the same, however, and this
is the main reason for present estimates being quite significantly lower.
A lattice action of the na\"\i ve type has generically, in the notation
of eq.~\(e2p5p1), a nonvanishing negative parameter $b_0$ in the
derivative expansion. To approximately translate this into a
Pauli--Villars regularization one must redefine the fields according to
\(e2p5p2) and this induces an effective $g_1$ coupling (and some $g_2$
coupling also, but this has no effect) with a positive sign.  Hence
na\"\i ve lattice actions are somewhat outside the region where
reasonable estimates for the bound can be obtained and, because of
this, the lattice bound is expected to increase. We shall see that the
increase is not negligible, but not very dramatic either.

\subhead{4.3. Computing leading cutoff effects in $1/ N$.}
\taghead{4.3.}

We have reached the point that we wish to do a complete calculation. In
this subsection we explain in detail the calculation conceptually.

The physical scale is set by the unitless $f_\pi$. Therefore, we
consider $f_\pi$ as an exact function of the bare parameters. By
definition there are no ``cutoff effects'' in $f_\pi$. Similarly we
define another ``exact'' quantity, the coupling $g = 3 m_H^2 / f_\pi^2
$. $m_H$ is the exact real part of the pole of the amplitude for $I=0$,
$J=0$ $\pi$-- $\pi$ scattering when continued to the second sheet below
the physical cut. $m_H$ is a function of the bare couplings with no
``cutoff effects'' by definition.

Consider now some new physical quantity. We make it dimensionless by
extracting the appropriate power of $f_\pi$ and denote it by $P$. $P$
may depend on momenta $q_i$ which we measure also in units of $f_\pi$:
$q_i =r_i  f_\pi$. Let the bare action depend on $n$ bare parameters.
We imagine changing variables to $f_\pi$, $g$ and $n-2$ remaining
parameters $p$ and consider $P$ as a function of them: $P = P(r_1 ,r_2
,\dots ; f_\pi, g ; p) $. The change of variables holds in some
neighborhood of a particular point in the broken phase that we wish to
investigate. We keep $p$ fixed and carry out a double expansion of $P$
in $f_\pi^2$ and $g$ without paying attention to whether the change of
variables is one to one also in the region where this expansion is
sensible. Note that the dependence on $f_\pi$ is both explicit and
implicit via the momenta.

If we expand $P$ in $g$ at fixed $r_i$, $f_\pi$ and $p$
renormalizability and universality tell us that to any finite order in
$g$ the limit $f_\pi \rightarrow 0$ exists and is independent of $p$.
The summation of all orders in $g$ of the $f_\pi =0$ terms is ambiguous
in the full theory because the series is badly behaved.  However, in
our case we expand $P$ in $1/ N$ after the appropriate rescaling and
redefinition of $g$. To leading order in $1/ N$ $P$ has a nontrivial
limit as $f_\pi \rightarrow 0$ even without expanding in $g$. It is
unlikely that this continues to be the case at subleading orders in $1/
N$ because one cannot replace the leading order (in $1/ N$) propagators
in higher order (in $1/ N$) terms by their leading asymptotic
expressions in $f_\pi$ as this will induce divergences in the momentum
integration at the positions of the ``Landau poles''. But at leading
order we are lucky and this simplifies our task considerably.

The above may look like a contradiction to triviality, because we just
argued that at leading order in $1/ N$ we can obtain a nontrivial
continuum limit. The catch is that the change of variables from the
bare parameters to $f_\pi$, $g$ and $p$ excludes precisely the region
where the continuum limit is taken. There are no values of the bare
parameters that are physically acceptable and can maintain a
nonvanishing coupling $g$ while $f_\pi$ is taken to zero.\footnote{*}
{This can be seen as follows: For small $g$ and fixed $Z_\pi$, $m_H$
satisfies an equation that is similar to \(e4p1p2) and for all
acceptable $Z_\pi$ one has $C(Z_\pi )\ge C_{min} >0$. Replacing $m_H$
on the left hand side of the equation by $f_\pi$ and $g$ we see that if
$g$ is fixed $f_\pi$ is bounded from below away from zero.} However, we
can carry out our change of variables away from the critical line in
the broken phase and work out the expansion by analytically continuing
(even without noticing) into the unaccessible
region.\footnote{\d}{There are claims that if this is really done with
care one will discover ambiguities and these are a direct reflection of
the singularity structure of $P$ in the Borel variable conjugate to
$g$\refto{BERGERDAVID}.} In this way we can, at least at leading order
in $1/N$, disentangle the issue of triviality from the issue of
summability of the renormalized perturbation theory.

Once we have the $f_\pi \rightarrow 0$ limit, $P_{N=\infty} (r_1 ,r_2
,\dots ;0,g)$, we can expand in the inverse cutoff $$ \eqalign{
P_{N=\infty}& (r_1 ,r_2 ,\dots ; f_\pi^2 , g ; p ) - P_{N=\infty} (r_1
,r_2 ,\dots ;0,g)=\cr &\Delta P_{N=\infty} (r_1 ,r_2 ,\dots ; f_\pi^2 ,
g ; p ) + O ( f_\pi^4 \log^\Upsilon \!\! f_\pi)\cr} \eqno(e4p3p1)$$
where $\Upsilon$ is some finite number. We define the leading cutoff
effect by $$ \delta_P (r_1 ,r_2 ,\dots ; f_\pi^2 , g ; p )={{\Delta
P_{N=\infty} (r_1 ,r_2 ,\dots ; f_\pi^2 , g ; p ) } \over {P_{N=\infty}
(r_1 ,r_2 ,\dots ;0,g)}} \fs \eqno(e4p3p2)$$

In practice we always work on a one dimensional line connecting the
given point in the broken phase where we wish to calculate the cutoff
effects to some point on the critical line. We change variables in our
parameter space (and this change is globally well defined in a large
region that includes the critical manifold) from the original $n$
parameters to $n-1$ parameters $\bar p$ and one parameter $\xi$. $\xi$
can be traded in a one to one fashion for $g$. Varying $\bar p$ and $g$
we span the whole region we are interested in. We organize our work by
fixing $\bar p$ and varying $g$ along the line.  Our purpose is to find
the region in $\bar p$ where $g$ can be maximized under the restriction
that the cutoff effects on some physical observable(s) do not exceed a
given amount. Along the line $f_\pi$ is monotonically dependent on $g$
and the cutoff effect is measured by $\bar \delta$ where $$ \bar
\delta_P ( r_1 ,r_2 ,\dots ;  g ; \bar p )\equiv \delta_P (r_1 ,r_2
,\dots ; f_\pi^2 (g) , g ; p ) \fs \eqno(e4p3p3)$$

In practice, we cannot compute the function $g$ exactly in terms of the
bare parameters and cannot carry out the change of variables
explicitly. But, we can compute $\bar \delta$ exactly, because errors
of order $f_\pi^4 \log^\Upsilon \!\! f_\pi$ or higher in $g$ have no
effect and it is practical to compute $f_\pi (g)$ neglecting order
$f_\pi^4 \log^\Upsilon \!\! f_\pi$ contributions. In plots we shall
always show $\bar \delta$ as a function of $M_H \equiv \sqrt{g /3}
\times 0.123~TeV$ in order to be able to immediately read off the
Higgs mass bound in $TeV$.

There is a fundamental difference between our way of calculating cutoff
effects here and the one employed previously in the triviality context.
Here, the single approximation in computing $\bar \delta$ is the
expansion in $1/N$.  In other computations, at $N=4$, one uses the loop
expansion. In terms of the bare coupling, the loop expansion is a
finite rearrangement of the perturbative expansion. An answer correct
to $l$ loops will be also correct to order $l$ in the bare coupling.
This is still true when the series is reexpressed in terms of an
appropriately defined renormalized coupling, $g$. The series for $\bar
\delta$ in $g$ has the following general structure:  $$ \bar \delta =
g^\Upsilon f_\pi^2 \sum_{n=0}^{\infty} g^n \sum_{l=0}^n {\cal P}_{n,l}
(\log f_\pi^2 )\eqno(e4p3p4)$$ where ${\cal P}_{n,l}$ is a polynomial
of degree $l$, $l$ is the number of loops, and $\Upsilon$ is some
number.  Along a line in the space of bare actions that touches the
critical surface at its end we have $f_\pi \sim \exp [-48\pi^2 / g]$ (at
$N=\infty$) and one needs to go to an infinite number of loops to get
$\bar \delta$ to a reasonable accuracy, even when $g$ is small. Therefore,
estimates of cutoff effects obtained by truncating the loop expansion
are not under good control. In Appendix B we give an explicit example
of this problem.

\subhead{4.4. Leading corrections to the width to mass ratio.}
\taghead{4.4.}

 The Higgs resonance is a complex root of $\det M^r (p^2 )=0$ on the
 second sheet
 $$ \det M^r (-(m_H -{i\over 2} \gamma_H )^2 ) =0 \fs \eqno(e4p4p1)$$
 What is meant by ``second sheet'' is that the branch of the logarithm
 is chosen so that, for
 $0\le \gamma_H /m_H << 1 $, $$ \log (-(m_H -{i\over 2} \gamma_H )^2 )
 \approx \log (m_H^2 ) -i{{\gamma_H }\over {m_H}} -i\pi \fs
 \eqno(e4p4p2)$$ In practice it will turn out that we have to restrict
our attention to the region where the width does not exceed the mass by
much.

With rescaled bubble integrals $$ B_j (p^2) =16\pi^2 I_{j,0} (p^2 ) \cm
\eqno(e4p4p3)$$ the explicit form of eq.~\(e4p4p1), for $p^2 =-(m_H
-{i\over 2} \gamma_H )^2$ is $$ {{32 \pi^2 f_\pi^2}\over {K(p^2 ) }} =
-B_0 (p^2 ) -{{[{1\over 4}p^2 B_0 (p^2 ) -B_1 (p^2) ]^2}\over {{n \over
{(n-2)(Z_\pi -1 )}} B_2 (0) -{1\over {16}} p^4 B_0 (p^2 ) + {1\over
2}p^2 B_1 (p^2 ) - B_2 (p^2 )}} \fs \eqno(e4p4p4)$$ Our problem is to
solve this complex equation for the real unknowns $m_H$ and $\gamma_H$
as a function of $f_\pi^2$, to first subleading order (logarithms are
counted as order one) for small $f_\pi^2$.  We no longer assume weak
coupling, so the ratio $m_H /f _\pi$ is taken to be of order one too
(up to logarithms).

We are working along lines of constant $Z_\pi$ which plays the role of
the set of parameters $\bar p$ in section (4.3). It is useful to
introduce a convenient parameterization of the line which is chosen so
that the equation simplifies.  On the line we have three real unknowns,
all small relative to unity: $f_\pi$, $m_H$ and $\gamma_H$. \(e4p4p4)
gives us two relations among the unknowns and the remaining free
parameter describes the line. We choose the free parameter to be an
angle $\theta$ defined by $$ (m_H -{i\over 2} \gamma_H )^2 =\mu^2 \exp
[-i\theta ]\eqno(e4p4p5)$$ where $\mu^2$ is real and positive. We shall
be interested only in that portion of the line where the mapping
between $\theta$ and the coupling $$g=3 {{m_H^2 }\over { f_\pi^2}}
\eqno(e4p4p6)$$ is one to one (in practice this limits the magnitude of
$g$ but cutoff effects become large before this limit is reached). This
portion of the line includes the critical point at one of its ends.

The two real equations \(e4p4p4) have to be solved now for $\mu^2$ and
$g$ in terms of $\theta$. We shall obtain the solution by first solving
to leading order and then perturbing around it with the subleading
terms. The leading order solution is $$ \eqalign{ \mu_0 = & ~ C(Z_\pi )
\exp \left [ - {{\theta_0 + \pi }\over { 2\tan (\theta_0 )}} \right
]\cr g_0 =& ~ {{96\pi^2 \cos^2 ({{\theta_0 }\over 2}) \sin ({\theta_0
})}\over {(\theta_0  +\pi )}} \fs \cr}\eqno(e4p4p7)$$ It is easy to
check that, as $\theta_0 \rightarrow 0$ we get for $m_H$ the same
equation we wrote down for $m_R$  in \(e4p1p2). Note that
$g_0(\theta_0)$ is bounded in the interval of interest, $0 \le
\theta_0 \le \pi$.

The physical quantity, $P(;0,g)={{\gamma_H}\over{m_H}}$, we are
interested in is dimensionless, and has no external momenta (we
suppress the subscript $N=\infty$ that we used in section (4.3)). To
leading order in the cutoff we have
 $$ P(;0,g_0 )= 2 \tan ({{\theta_0 }\over2})\eqno(e4p4p8)$$ where
 $\theta_0 $ is given in terms of $g_0$ by the inverse of the second
 equation in \(e4p4p7).

 To calculate to next order we treat all three variables $\theta$, $g$
 and $\mu^2$ on equal footing. These variables are defined by \(e4p4p5)
 and \(e4p4p6). At leading order we had $\theta=\theta_0$, $\mu=\mu_0$
 and $g=g_0$ given by \(e4p4p7) with $\theta_0$ as a free parameter, a
 coordinate along the constant $Z_\pi$ line. If we treat all three
 parameters on an equal footing we ought to also allow for the
 subleading terms to change the coordinate along the line. We set $$
 \eqalign{ \mu^2 =& ~ \mu_0^2 + \delta_\mu \cr \theta =& ~ \theta_0 +
 \delta_\theta \cr g=& ~ g_0 +\delta_g \fs \cr } \eqno(e4p4p9)$$ The
 equations \(e4p4p4), when expanded to subleading order give us two
 relations among the three $\delta$'s above. The needed third relation
comes from noting that the quantity $P(;f_\pi^2 , g; Z_\pi )$ in
\(e4p3p1) is evaluated at the same $g$ as the quantity $P(;0,g)$ there.
Therefore we have $$ \delta_g =0 \fs \eqno(e4p4p10)$$ Carrying out the
algebra we obtain $$ \bar \delta_{{{\Gamma_H}\over{M_H}}} (g_0 ; Z_\pi
) = {{D(Z_\pi ,n) \mu_0^2}\over{ 1+(\pi + \theta_0 ) \left [ \tan \left
( {{\theta_0}\over 2} \right ) - \cot (\theta_0 ) \right ]
}}\eqno(e4p4p11)$$ where, on the right hand side, $\theta_0$ and
$\mu_0$ are functions of $g_0$ defined in \(e4p4p7). $g_0$ is a free
parameter and the subscript $0$  has no particular meaning any more.
The function $D(Z_\pi ,n)$ is given by
 $$ D(Z_\pi ,n)= {{\pi (n^2 -1)}\over {12n^2 \sin {{\pi}\over n}}}+
 {{(n+3) \cos {{\pi}\over n}}\over {6 }}\zeta (Z_\pi ,n) + {{\pi
(n-1)(n-2)\cos^2 {{\pi}\over n}}\over{12n^2 \sin {{\pi}\over n} }}
\zeta^2 (Z_\pi ,n) \fs
 \eqno(e4p4p12)$$ The function $\zeta (Z_\pi ,n)$ was defined in
 \(e4p1p4), only now we have written out explicitly its dependence on
 $n$.

 The overall structure of eq.~\(e4p4p11) is very simple. The correction
has factorized into a function that
 carries all the dependence on the coupling $g$ and is universal times
 a function that carries all the information about the cutoff, through
 the dependence on $Z_\pi$ and $n$, namely
 $C^2 (Z_\pi ,n) D(Z_\pi ,n) $. It is amusing to note that the
dependence on $Z_\pi$ always comes in through the combination that
appears in the definition of
 $\zeta$.

 This factorization is more powerful than what one would have expected
 on the basis of the Symanzik improvement formulae, as they are usually
 presented, in several respects. First, we seem to have only a single
 operator insertion because if we had a linear superposition of several
 operator insertions nothing special should be evident when one looks
 at a single momentum independent observable as we are. Second, the
 coefficient of the operator is independent of the renormalized
 coupling. Third, the logarithms in the subleading terms have been
 summed in closed form.  To be sure, we are looking at a physical
 operator, not an arbitrary correlation function, so the operator
 counting should be different. Also, we are working at $N=\infty$ where
 additional simplifications should occur.  Our result could be
explained if it were true that, at $N=\infty$, all cutoff effects in
on--shell dimensionless physical quantities (that are functions of
dimensionless momenta) are given by an effective renormalized action,
$$ S_{eff}=S_R + c \exp [-96\pi^2 / g ] {\cal O} \cm \eqno(e4p4p13)$$
where ${\cal O}$ is a renormalized operator, $c$ is a $g$ independent
free parameter containing all the non--universal information and $S_R$
is describing the usual universal part of physical observables with the
unit of energy set by $f_\pi =1$. In terms of general RG reasoning this
representation of the $S_{eff}$ is not unreasonable if one accepts that
at $N=\infty$ the number of available independent operators decreases.
We shall see that this would also explain our results for the
$\pi$--$\pi$ scattering amplitude.

$D(Z_\pi ,n)$ is a quadratic polynomial in the variable  $\zeta$ and
has two negative roots. It turns out that one of the roots can be
realized in the allowed range of $Z_\pi$. For $n=3$ the root is at
$Z_\pi \approx 0.52$ while when $n\rightarrow \infty$ the root is
realized by $Z_\pi \approx {6\over {\sqrt{n}}}$.  We see that for all
$n$ the roots are for $Z_\pi$ between $1$ and $0$. If we set $Z_\pi$ to
be at the appropriate root for the value of $n$ chosen we have achieved
``improvement'' in the sense of Symanzik in that the leading cutoff
correction has been made to vanish. This would be the type of fine
tuning that we excluded because, as far as we know, it is unreasonable
to expect the more complete theory in which the minimal standard model
is embedded (assuming that this is the situation in nature) to conspire
to achieve such a more refined decoupling. We are entitled however to
use a reasonable range, say $0\le Z_\pi \le 0.5$ and $n\le 100$  to
estimate the bound.

We find that for $M_H \le 0.820~TeV$ the cutoff effect on the width to
mass ratio is less than half a percent, as will be shown in Fig.~7.3d.
At such a high mass at infinite $N$ the system is strongly
interacting.

\subhead{4.5. Leading corrections to $\pi$--$\pi$ scattering}
\taghead{4.5.}

 We now repeat the analysis of the previous section for $\pi$--$\pi$
scattering. We
 are considering the process $\pi^a (1) +\pi^b (2) \rightarrow \pi^c
 (3) + \pi^d (4)$ where
 $a,b,c,d$ are isospin indices and (1--4) denote momenta. We work in
 the center of
 mass frame and wish to compute the invariant amplitude $A(s,t,u)$
 defined in \(e2p3p9)
 to first non--vanishing order  in ${1\over N}$.

To leading order in $1/N$ and up to and including subleading order in
the inverse cutoff, $A(s,t,u)$ is dominated by the exchange of
scalar--isoscalar fields and therefore is just a function of $s$,
$A(s)$. The differential cross--section with identical isospin indices
for the in-coming pions, in the center of mass frame, is isotropic and
given by
 $$ \left ( {{d\sigma}\over{d\Omega}} \right )_{CM} = {{N}\over{64\pi^2
 s}} |A(s)|^2\eqno(e4p5p1)$$
 where $s$ is the square of the center of mass energy, henceforth
 denoted by $W$, and
 we are working to leading order in $1/N$.

 In fact, keeping only the leading order in $1/N$ alone, at any finite
 cutoff, is enough to exclude the contribution of exchanges of the
 spin--one field, $\delta\omega_j$, by Bose symmetry.  However,
 the exchange of the spin--two field, $\delta\omega_{jk}$, does come
 in, only at one
 order higher in the inverse cutoff than the order we are calculating
 to. In the case
 of a linear action we had a similar simplification. An immediate
 consequence of this
 is that we shall see no higher spin resonance in our computations (a
$\rho$--like
 particle for example). To see an even spin resonance it might be
 sufficient to just keep
 all orders in the cutoff, but to see an odd spin resonance one must go
 to a higher
 order in $1/N$.

 To evaluate $A(s)$ and see where the observations in the previous
 paragraph come from, we need to write down the pion interactions. We
 go back to equation \(e3p2) and introduce the pion fields and the
Higgs field by eq.~\(e2p1p7):
 $$ S_1 = {1\over 2} \int_x  [ \vec \pi \hat K \vec \pi + H \hat K H]
 -{N\over 2}v^2 \int_x \rho {}~ -\sqrt{N} v \int_x \rho H -{N\over
2}\int_x [\beta\rho -{1\over 4} g_1 \lambda^2 - {1\over 4}g_2
\omega_{\mu\nu}\omega_{\mu\nu} ] \fs \eqno(e4p5p2)$$ Here $\hat K$ is
defined in \(e3p3) as $ \hat K = K - \partial_\mu \lambda \partial _\mu
+ \rho - \partial_\mu \omega_{\mu\nu}  \partial_\nu$. We expand around
the saddle point, carry out the rescalings \(e3p3p4), but do not
integrate out $\delta H$ and $\delta \vec \pi$ yet. There are three
types of vertices involving pions, all involve two pion fields with
isospin indices contracted and a third field, which is $\delta\rho$,
$\delta\lambda$ and $\delta\omega_{\mu\nu}$, respectively. At leading
order in $1/N$, $A(s)$ is dominated by diagrams containing two vertices
of the above type and the full propagators (to leading order in $1/N$)
of the $\psi$ fields (\(e3p4p6)) and of $\delta\omega_{jk}$
(\(e3p4p4)).

 The expansion in the inverse cutoff is an expansion in $f_\pi^2$
 where, up to logarithms, $m_H^2$ and all momenta $p_j^2$ are of order
 unity in $f_\pi^2$.
 We are still assuming that $g_2$ is safely larger than $g_{2c}$
 (\(e3p5p1)) and therefore
 the spin--two field has a ``mass'' of order unity. We also know that
ultimately we shall
 continue analytically to Minkowski space with on-shell external pions,
 $p_j^2 =0$.
 Under these conditions we first carry out an order of magnitude
 estimation of the
 vertices and propagators that can be read off \(e4p5p2). For this
 purpose it is better
 not to think in terms of the full propagator, but use multiple
 insertions of the two point vertices in \(e4p5p2), which amounts to
 the same thing. It is then easy
 to see that the only kind of diagrams that contribute to leading order
 have the external pions connected to two $\pi\pi\delta\rho$ vertices.
 The internal line
 has multiple insertions of  $\delta\rho\delta H$ vertices and of the
 various $\pi\pi$ bubbles. When we decide to also keep the first
 subleading terms we have to add diagrams where two of the external
 pions are still coupled by a $\pi\pi\rho$ vertex but the other two
 external pions couple to a
 $\pi\pi\lambda$ vertex. The internal lines in these new diagrams have
essentially
 the same structure as before. The conclusion is that, in terms of full
propagators,
 we have, with Euclidean $q=p_1 +p_2$, but keeping only terms that will
contribute
 to the on--shell amplitude, $$ A(q^2 )\propto
 [<\delta\rho\delta\rho>(q^2 ) -q^2 <\delta\rho\delta\lambda> (q^2 )]
 \fs \eqno(e4p5p3)$$ The full propagator $<\delta\rho\delta\rho>(q^2 )$
 is needed to leading and subleading
 order, while the full propagator $<\delta\rho\delta\lambda>(q^2 )$ is
 needed only to leading order. These propagators are obtained from the
 inverse of the matrix $M$ \(e3p4p6,e3p4p7,e3p4p8).
 Some more order of magnitude estimates show that the $4 \times 4$
 matrix $M$ can be replaced, to the order we are interested in, by the
 $3\times 3$ matrix $M^r$ defined at the beginning of section 4.

 In the notation of section (4.3), the dimensionless quantity we are
 interested in here
 is chosen to be $$ P(r;  f_\pi^2 ,g; Z_\pi )= |A \left (p^2
 \rightarrow -r^2 (f_\pi^2 +i 0^+ ) \right ) |^2
 \eqno(e4p5p4)$$ where $r = W/F_\pi$. One expects a bump in $P$ when
 $W=M_H$ and the width $\Gamma_H$ is small. It therefore makes more
 sense to replace $r$ by another dimensionless number, $R$, so that in
 $R$ the bump would be independent of $g$ when the width of the Higgs
 is small. We therefore define $$ R={{W^2}\over
 {4M_H^2}}\equiv{{3r^2}\over{4g}}\eqno(e4p5p5)$$ and will express the
 center of mass energy dependence via $R$. When computing the
 cutoff effect we are required to keep $r$ and $g$ fixed and this is
 equivalent to
 keeping $R$ and $g$ fixed.

 Using \(e4p4p7) we obtain, for the leading order, $$\eqalign{ P(R; 0,
 g_0 )=& ~ \big \vert A_0 (R,g_0 ) {\big \vert}^2\cr N A_0 (R,g_0 )=& ~
{   {4Rg_0 }\over {
 3\left [ 1+4R [-\cos^2 ({{\theta_0}\over 2}) \cos (\theta_0 )+{{\cos^2
({{\theta_0}\over 2}) \sin (\theta_0 )}\over {\theta_0 +\pi}}( \log(
4R\cos^2 ({{\theta_0}\over 2} ) ) -i\pi ) ] \right ]  }   } \fs
 \cr } \eqno(e4p5p6)$$ Including the subleading order, we obtain $$
 \eqalign { &P(R; f_\pi^2 ,g_0 ; Z_\pi ) = P(R; 0, g_0 )
  {\bigg\vert} 1+ {{NA_0 (R,g_0 )\mu_0^2 D(Z_\pi ,n)}\over
 {32\pi^2}}\cr& \left [ 4R\cos^2 ({{\theta_0}\over 2}) -\cos (\theta_0
)
 -\sin(\theta_0 ) {{(\pi+\theta_0 )(1+\cot (\theta_0 ) \tan
 ({{\theta_0}\over 2})) -\tan
 ({{\theta_0}\over 2} )}\over {1+(\pi+\theta_0 ) (\tan ({{\theta_0}
 \over 2 } ) -\cot (\theta_0 ) )}}
 \right ]
  {\bigg\vert}^2 \fs \cr} \eqno(e4p5p7)$$ The variable $g_0$ is free
  and expressible in terms of $\theta_0$ by the second equation in
 \(e4p4p7). The function $D(Z_\pi ,n)$ is given in \(e4p4p12). From
\(e4p5p7) we obtain the cutoff correction
  $$ \eqalign{ \bar \delta_{|A|^2} ~ & = {{\mu_0^2 D(Z_\pi ,n)}\over
 {16\pi^2}} ~ {\rm Re} [N A_0 (R,g_0 )]
  \cr&\left [ 4R\cos^2 ({{\theta_0}\over 2}) -\cos(\theta_0 )
 -\sin(\theta_0 ) {{(\pi+\theta_0 )(1+\cot (\theta_0 ) \tan
 ({{\theta_0}\over 2})) -\tan
 ({{\theta_0}\over 2} )}\over {1+(\pi+\theta_0 ) (\tan ({{\theta_0}
 \over 2 } ) -\cot (\theta_0 ) )}} \right ] \fs \cr }\eqno(e4p5p8)$$
 Its structure is similar to the one seen in the calculation of the
 width to mass ratio.
 Note that again the dependence on $Z_\pi$ and $n$ factorizes and that
 it comes in through the same coefficient as in \(e4p4p11). Here this
 is even more non--trivial because it holds for all center of mass
energies measured by $R$. This
 result provides additional evidence for \(e4p4p13) showing that for
 the cross-section we
 would be needing the same $c$ as for the width to mass ratio. All the
dependence on
 $g_0$ and on $R$ is coming from the operator ${\cal O}$ and the action
$S_R$. Another
 check for eq.~\(e4p4p13) will be provided by our lattice work where we
 shall see that the
 same dependence on $g_0$ and on $R$ enters, the whole difference being
 expressible by a different parameter $c$. In the present case, a
 particular observation we
 can already make is that,  had we decided to ``improve'' the width to
 mass ratio
 by fine tuning $Z_\pi$ so that $D(Z_\pi ,n)=0$, we would also have
automatically ``improved'' the scattering cross--section.

Some feeling for the numbers involved can be obtained from Fig. 7.4d,
showing $\bar \delta_{|A|^2}$ as a function of the Higgs mass in $TeV$
for several choices of $R$ and at several values of $Z_\pi$.

With $n\le 100$ and $0\le Z_\pi \le 0.5$, the cutoff
effect on the square of the invariant amplitude is less than 3 percent
for center of mass energies less than four times the Higgs mass, as long
as the Higgs mass does not exceed $0.820~TeV$. So one can have a
strongly interacting Higgs particle, at infinite $N$, with relatively
minor cutoff effects.

\subhead{4.6. Leading corrections to $m_R$.} \taghead{4.6.}

As promised in subsection (4.1) we present here the leading correction
to equation \(e4p1p2) there. We write $$ m_R^2=m_{R0}^2 (1+
\delta_{m_{R0}} m_{R0}^2) \fs \eqno(e4p6p1)$$ Here $m_{R0}^2$ is a
function of $g_{R0}$ of the same form as in \(e4p1p2).  By essentially
the same methods as before, only this time the coupling we keep
constant is $g_{R0}$ (as defined in \(e4p1p3)), we derive a formula for
$\delta_{m_{R0}}$. It reads $$ \delta_{m_{R0}} = D(Z_\pi ,n) +{{96\pi^2
\cos ({{\pi}\over n}) \zeta (Z_\pi ,n )}\over {g_{R0}}} \fs
\eqno(e4p6p2)$$

In addition to completing the calculations in section (4.1) this
formula shows us that an unphysical quantity will not necessarily have
its cutoff dependent corrections proportional to the function $D(Z_\pi
,n)$ only. The quantity $m_R$ is unphysical because its definition in
equation \(e4p1p1) involved ${\rm Re} [\det M^r (p^2 )]$. This matrix
$M^r$ has a rather complicated relation to the correlation functions,
and it is obvious that the zero of the real part of its determinant is
not an ``on--shell'' quantity. Even if we used fine tuning  to
``improve'' all the physical observables by setting $D(Z_\pi ,n)$ to
zero the correction to $m_R$ would still be nonvanishing at leading
order in the inverse cutoff.  However, without  ``fine tuning'', the
correction to $m_R$ is quite similar numerically to the corrections to
the physical observables. Therefore, it made
practical sense to discuss $m_R$ first, exploiting its relative
simplicity, in order to get an indication for where in the space of
actions the mass bound is likely to be larger.

\head{5. Phase diagram for the lattice models.} \taghead{5.}

We are going to investigate the lattice models defined in \(e2p5p6) for
the $F_4$ and hypercubic lattices and in \(e2p5p7) for the hypercubic
Symanzik improved case. We shall introduce a common notation for all
three cases with the parameters and symbols having a slightly different
meaning in each case as explained below.

Using the constraint $\Fhi^2 = 1$ we rewrite the lattice actions, up to
additive constants, as $$ S = \eta N (\beta_0 + \beta_2) \half
\int_{x,y} \Fhi (x) g_{x,y} \Fhi (y) - N \beta_2 {\eta^2 \over {8
\epsilon} } \int_x \left[ \int_y \Fhi (x) g_{x,y} \Fhi (y) \right]^2
\fs \eqno(e5p1)$$ $g_{x,y}$ is the kinetic term with Fourier transform
$g(p) = p^2 + \dots$ with the higher order terms explicitly given below
for each case.  To obtain the continuum normalization the fields have
to be rescaled as $\fhi = \sqrt {\eta N(\beta_0 +\beta_2)} \Fhi$. We
therefore restrict ourselves to the region $\beta_0 + \beta_2 > 0$.

It is convenient to use a lattice type dependent Kronecker delta
function, $\tilde \delta_{x,y} = b \delta_{x,y}$, where the parameter
$b$ is the volume of the Brillouin zone on the particular lattice. On
$F_4$ $b= {1/2}$ while on the hypercubic lattice $b=1$.

In the $F_4$ case $\eta$, $\epsilon$, $\int_x$ and $g$ are given by $$
\eqalign{ \eta = 6 ~,~~~ \epsilon = & ~ 12 ~,~~~ \int_x = 2 \sum_x
{}~,~~~ \int_p = \int_{B^*} {d^4 p \over {(2 \pi)^2}} \cr g_{x,y} = & ~ 4
\tilde \delta_{x,y} - {1 \over 6} \sum_{{l\cap x \ne \emptyset}\atop
{l=<x,x'>}} \tilde \delta_{y,x'} \fs \cr} \eqno(e5p2)$$ $g(p)$ was
already given explicitly in \(e2p5p5) and $B^*$ is the Brillouin zone
of $F_4$\refto{BBHN1}.

For the hypercubic lattice we are considering two cases:

\noindent The first hypercubic case (HC) is without improvement and has
$$ \eqalign{ \eta = & ~ 2 ~,~~~ \epsilon = 8 ~,~~~ \int_x = \sum_x
{}~,~~~ \int_p = \int_{-\pi}^\pi {d^4 p \over {(2 \pi)^2}} \cr g_{x,y} =
& ~ 8 \tilde \delta_{x,y} - \sum_{\pm \mu} \tilde \delta_{y,x+\mu} \cr
g(p) = & ~ 2 \sum_\mu ( 1 -\cos(p_\mu) ) = p^2 - {1 \over 12} \sum_\mu
p_\mu^4 + O(p^6) \fs \cr} \eqno(e5p3)$$ The second hypercubic case (SI)
has Symanzik improvement, and the definitions are $$ \eqalign{ \eta = &
{}~ 2 ~,~~~ \epsilon = 10 ~,~~~ \int_x = \sum_x ~,~~~ \int_p =
\int_{-\pi}^\pi {d^4 p \over {(2 \pi)^2}} \cr g_{x,y} = & ~ 10 \tilde
\delta_{x,y} - \sum_{\pm \mu} \left[ {4 \over 3} \tilde
\delta_{y,x+\mu} - {1 \over 12 } \tilde \delta_{y,x+2\mu} \right] \cr
g(p) = & ~ \sum_\mu \left[ {4 \over 3} (1 -\cos(p_\mu) ) - {1 \over 12}
(1 -\cos(2 p_\mu) ) \right] = p^2 - O(p^6) \fs \cr} \eqno(e5p4)$$

With the above conventions we can treat all three cases simultaneously.
We first introduce auxiliary fields in \(e5p1) to make the dependence
on $\Fhi$ bilinear and relax the fixed length constraint.  $$ \eqalign{
S_1 = & ~ \eta N (\beta_0 + \beta_2) \half \int_{x,y} \Fhi (x) g_{x,y}
\Fhi (y) + \eta N \half \int_x \lambda (x) \left[ \int_y \Fhi (x)
g_{x,y} \Fhi (y) \right] + \cr & ~ \half N \int_x \rho (x) ( \Fhi^2 (x)
- 1 ) + \half N {\epsilon \over \beta_2} \int_x \lambda^2 (x) \cr = & ~
{N \over 2} \int_{x,y} \Fhi \hat K \Fhi + {N \over 2} \int_x \left[
{\epsilon \over \beta_2} \lambda^2 - \rho \right] \cr} \eqno(e5p5)$$
with $$ \hat K_{x,y} = \eta \left( \beta_0 + \beta_2 + \half
(\lambda(x) + \lambda(y)) \right)  g_{x,y} + \rho (x) \tilde
\delta_{x,y} \fs \eqno(e5p6)$$

We separate out the zero mode by $$ \Fhi (x) = \vec v + {1 \over \sqrt
N} \hat v H(x) + {1 \over \sqrt N} \vec \pi (x) ~,~~\hat v ={{\vec
v}\over v}~,~~|\vec v |= v {}~,~~\hat v \cdot \vec \pi =0~,~~\int_x
H(x)=\int_x \pi^j (x) = 0 \cm \eqno(e5p7)$$ and integrate out $\vec
\pi$ and $H$. Then we obtain $$ S_2 = {N \over 2} \left[ {\rm Tr}\log
\hat K + v^2 \int_x \rho - v^2 \int (\rho^\prime + {\eta \over 2}
\lambda^\prime g ) {\hat K}^{-1} (\rho^\prime + {\eta \over 2} g
\lambda^\prime ) + \int_x \left( {\epsilon \over \beta_2} \lambda^2 -
\rho \right) \right] \eqno(e5p8)$$ where $\lambda^\prime$ and
$\rho^\prime$ are, as usual, the non-constant parts of $\lambda$ and
$\rho$ and $\hat K$ now denotes the operator from \(e5p6) restricted to
the space of functions whose lattice average vanishes.

\subhead{5.1. Saddle point equations and dominating saddles.}
\taghead{5.1.}

The saddle point equations are $$ \eqalign{ & 1 - v^2 = {1 \over {\eta
(\beta_0 + \beta_2 + \lambda_s)}} \int_p {1 \over {g(p ) + \hat
\rho_s}} = {J_1(\hat \rho_s) \over {\eta (\beta_0 + \beta_2 +
\lambda_s)}} \cr & - {2 \epsilon \over \beta_2} \lambda_s = {1 \over
{(\beta_0 + \beta_2 + \lambda_s)}} \int_p { g(p) \over {g(p) + \hat
\rho_s}} = {1 \over {(\beta_0 + \beta_2 + \lambda_s)}} \left( b - \hat
\rho_s J_1(\hat \rho_s) \right) \fs \cr} \eqno(e5p1p2)$$ In this
equation we have defined $$ \rho = \eta (\beta_0 + \beta_2 + \lambda)
\hat \rho \fs \eqno(e5p1p1)$$ Also, $J_1$ is the lattice integral
introduced by \cit{LW} for the hypercubic and by \cit{BBHN1} for the
$F_4$ lattice. The definition for SI is the same in our notation. By
definition, one has $\int_p 1 = b$.

The symmetric phase is characterized by $v^2 = 0$, $\hat \rho_s > 0$
and the broken phase by $v^2 > 0$, $\hat \rho_s = 0$. A candidate
critical point has $v^2 = \hat \rho_s = 0$. As for the Pauli--Villars
models the saddle point equations may admit several competing
solutions. We shall find the dominating saddle by the method used in
section (3.1) for the Pauli--Villars models. Our presentation will be
much more sketchy here.

Defining $$ u = \eta (\beta_0 + \beta_2 + \lambda) ~,~~~ \beta^* = {2
\epsilon \over {\eta^2 \beta_2}} ~,~~~ u^* = \eta (\beta_0 + \beta_2)
\cm \eqno(e5p1p3)$$ we have to minimize the function $$ \hat \Psi (u,
\hat \rho) = \int_p \log [g(p) + \hat \rho] + b \log u + (v^2 -1) u
\hat \rho + \half \beta^* (u - u^*)^2 \fs \eqno(e5p1p4)$$

In the symmetric phase, with $v^2 = 0$, the equation $\partial \hat
\Psi / \partial \hat \rho = 0$ can be used to define a function $\hat
\rho (u)$ for $0 \le u \le u_0$ $$ J_1(\hat \rho(u) ) = u ~,~~~~~~ u_0
= J_1(0) = r_0 \cm \eqno(e5p1p5)$$ with the lattice constant $r_0$
given by $$ r_0 = 0.13823047 ~~~ {\rm for} ~~ F_4 ~,~~~~~ r_0 =
0.15493339 ~~~ {\rm for ~~ HC ~~~~~ and} ~~~~ r_0 = 0.12919024 ~~~ {\rm
for ~~ SI} \fs \eqno(e5p1p6)$$ The values for the HC and $F_4$ cases
can be found in \cit{{LW},{BBHN1}}, while the value for the SI case was
computed by similar methods.  We then derive $$ \eqalign{ \hat \Psi
(u_s, \hat \rho_s) = & ~ \int_{u_0}^{u_s} [ G(u) + \beta^* (u - u^*) ]
+ \hat \Psi (u_0, 0) , \cr G(u) = & ~ {b \over u} - J_1^{-1} (u)  ~~~~~
{\rm for} ~~~~ 0 \le u \le u_0 \fs \cr} \eqno(e5p1p7)$$

In the broken phase $(\partial \hat \Psi / \partial \hat \rho )_{ \hat
\rho = 0} = 0$ together with $v^2 \ge 0$ yields the restriction $u_s
\ge u_0$. Then, extending the range of the function $G$ from $u \in (0,
u_0 )$ to the segment $u\in [u_0, \infty)$ by $G(u) = b/u$, we can
write the function $\hat \Psi$ at the saddle as $$ \hat \Psi (u_s, 0)
=  \int_{u_0}^{u_s} [ G(u) + \beta^* (u - u^*) ] + \hat \Psi (u_0, 0)
\fs \eqno(e5p1p8)$$

Again we have identical forms in both phases and can introduce a
geometrical interpretation like in the Pauli--Villars case. We find a
tricritical point at $$ \beta^*_{t.c.} = {b \over u_0^2} = {b \over
r_0^2} ~,~~~~~ u^*_{t.c.} = 2 u_0 = 2 r_0 \fs \eqno(e5p1p9)$$ In terms
of the original couplings this corresponds to $$ \beta_{2,t.c.} = {2
\epsilon r_0^2 \over {b \eta^2}} ~,~~~~ \beta_{0,t.c.} = {2 r_0 \over
\eta} - \beta_{2,t.c.} \fs \eqno(e5p1p10)$$ The second order critical
line is described by $$ u^* = \left( {\beta^*_{t.c.} \over \beta^*} + 1
\right) u_0 ~,~~~~~ {\beta^*_{t.c.} \over \beta^*} < 1 \cm
\eqno(e5p1p11)$$ which translates into $$ \beta_{0,c} = \left( {
\beta_2 \over \beta_{2,t.c.}} + 1 \right) {r_0 \over \eta} - \beta_2
{}~,~~~~~~ \beta_2 < \beta_{2,t.c.} \fs \eqno(e5p1p12)$$ The restriction
$(\beta_0 + \beta_2) > 0$ leads on the critical line to the requirement
$$ \beta_2 > - \beta_{2,t.c.} = - {2 \epsilon r_0^2 \over {b \eta^2}}
\fs \eqno(e5p1p13)$$

To ascertain the local stability of our saddle points we need to
investigate the small fluctuations around the saddle points. This will
be done in the next section.

\subhead{5.2. Small fluctuations in the broken phase.} \taghead{5.2.}

Separating out the zero mode from the field $\Fhi$ as in \(e5p7) we
want to expand the action \(e5p5) around the saddle point. For this we
introduce the shifted fields, recalling that $\rho_s = 0$ in the broken
phase, $$ \delta \lambda = \lambda - \lambda_s ~,~~ \delta H = H ~,~~
\delta\vec \pi = \vec \pi ~,~~ \delta\rho =\rho \fs \eqno(e5p2p1)$$ We
can easily read of the pion propagator in Fourier space $$ < \delta
\pi^a \delta \pi^b > = {\delta^{a,b} \over { \eta ( \beta_0 + \beta_2 +
\lambda_s) g(p) }} \fs \eqno(e5p2p2)$$ {}From this we find the pion
wave function renormalization constant $$ Z_\pi = {1 \over { \eta (
\beta_0 + \beta_2 + \lambda_s) }} \fs \eqno(e5p2p3)$$ We see that the
renormalized pion propagator is independent of the couplings and hence
so are its cutoff corrections. We also see that in the SI case the
inverse pion propagator has no order $p^4$ contribution and thus is
Symanzik improved.

Using $Z_\pi$ given above and $v^2 = Z_\pi f^2_\pi$ we obtain from the
first saddle point equation in \(e5p1p2) ($\hat \rho_s = 0$ in the
broken phase) $$ Z_\pi = {1 \over {r_0 + f^2_\pi  }} \fs
\eqno(e5p2p4)$$ Then, also using the second saddle point equation, we
find $$ \beta_0 = {1 \over {\eta Z_\pi}} + \left( {b \eta Z_\pi \over
{2 \epsilon}} - 1 \right) \beta_2 = {r_0 + f^2_\pi \over \eta } +
\left( {b \eta \over {2 \epsilon (r_0 + f^2_\pi) }} - 1 \right) \beta_2
\fs \eqno(e5p2p5)$$ Therefore, at fixed $\beta_2$, we can trade the
coupling $\beta_0$ for $f^2_\pi$.

Integrating out the pions from the action \(e5p5), with the zero mode
separated out, we obtain, up to quadratic order in the fluctuations, $$
S_2^{(2)} = \half \int_p \psi_A M_{AB} \psi_B \fs \eqno(e5p2p6)$$ Here
the three component fields $\psi_A$ represent $\delta H$, $\delta\rho$
and $\delta\lambda$ for $A = 1,2,3$ respectively. The non-vanishing
entries of the symmetric matrix $M_{AB} (p)$ are, neglecting $1/N$
corrections, $$ \eqalign{ M_{11} = & ~ {g(p) \over {Z_\pi}} \cr M_{12}
= & ~ \sqrt {N} v \cr M_{13} = & ~ \half \sqrt {N} v \eta g(p) \cr
M_{22} = & ~  - \half N Z_\pi^2 \int_k {1 \over {g(k+ \half p) g(k-
\half p) }} \equiv - \half N Z_\pi^2 I(p) \cr M_{23} = & ~ - \half N
Z_\pi^2 \eta \int_k { \half [g(k+ \half p) + g(k- \half p)] \over {g(k+
\half p) g(k- \half p) }} = - \half N Z_\pi^2 \eta r_0 \cr M_{33} = & ~
- \half N Z_\pi^2 \eta^2 \int_k {{ {1 \over 4} [g(k+ \half p) + g(k-
\half p)]^2 } \over {g(k+ \half p) g(k- \half p) }} + {N \epsilon \over
\beta_2 } \equiv - \half N Z_\pi^2 \eta^2 \left( Q(p) - {2 \epsilon
\over {Z_\pi^2 \eta^2 \beta_2}} \right) \fs \cr} \eqno(e5p2p7)$$

We need two ``bubble" integrals, $I(p)$ and $Q(p)$. $I(p)$ has already
been computed in an expansion in the momentum $p$ in \cit{LW} for the
HC lattice and in \cit{BBHN1} for $F_4$. The computation for SI,
employing the same method, is straightforward. One finds for $F_4$ and
SI $$ I(p) = - {1 \over {16 \pi^2}} \log p^2 + c_1 - c_2 p^2 + O(p^4
\log p^2) \eqno(e5p2p8)$$ with the constants $c_i$ $$ \eqalign{ c_1 = &
{}~ 0.0466316 ~,~~~~ c_2 = - 5.4968 \cdot 10^{-4} ~~~~~{\rm for} ~~ F_4
\cr c_1 = & ~ 0.0283716 ~,~~~~ c_2 = - 3.5981 \cdot 10^{-4} ~~~~~{\rm
for ~~ SI} \fs \cr} \eqno(e5p2p9)$$ For HC the result is $$ I(p) = - {1
\over {16 \pi^2}} \log p^2 + c_1 - c_{2,1} p^2 - c_{2,2} \sum_\mu
p^4_\mu / p^2 + O(p^4 \log p^2) \eqno(e5p2p10)$$ with $$ c_1 =
0.0366783 ~,~~~~ c_{2,1} = - 7.524 \cdot 10^{-5}  ~,~~~~ c_{2,2} = -
2.6386 \cdot 10^{-4} \fs \eqno(e5p2p11)$$ Again, at order $1/
\Lambda^2$, the HC lattice has a Lorentz invariance breaking term.
However we shall need $I(p)$ only for `time' like, and thus on-axis,
momenta. Then the HC result reduces to the form \(e5p2p8) with $c_2 =
c_{2,1} + c_{2,2} = - 3.3910 \cdot 10^{-4}$. We see here explicitly how
the specific choices of observables that we made effectively hide the
Lorentz breaking effects of the HC case.

An expansion in $p$ for the integral $Q(p)$ is easily computed with the
result $$ Q(p) = b + \gamma p^2 + O(p^4) \eqno(e5p2p12)$$ with $$
\gamma = 0.00661524 ~~~ {\rm for} ~~ F_4 ~,~~~~~ \gamma = 0.01496670
{}~~~ {\rm for ~~ HC ~~~~~ and} ~~~~ \gamma = 0.01736345 ~~~ {\rm for ~~
SI} \fs \eqno(e5p2p13)$$

We are now in the position to discuss the local stability of our saddle
points. It is easy to see that on the critical line, {\sl i.e.,} for $Z_\pi
= r^{-1}_0$ (\(e5p2p4)), $M_{33}$ vanishes at $p=0$ when $\beta_2$
approaches $\beta_{2,t.c.}$ from below. At $p=0$ $M_{11}$ and $M_{13}$ also
vanish and hence at the tricritical point $\det M = 0$. Thus, as in the
Pauli--Villars models, the Higgs particle becomes massless at the
tricritical point and also plays the role of a dilaton. For $\beta_2 <
\beta_{2,t.c.}$ we find, at least in the neighborhood of the critical line,
that $\det M$ stays positive for Euclidean momenta. Thus the saddle points
in the critical regime that we are interested in are locally stable.

\head{6. Higgs mass bound and cutoff effects with lattice
regularizations.} \taghead{6.}

As in the Pauli--Villars case we will first do an approximate
calculation to see the basic trends. Next we follow up with an accurate
evaluation including the computation of the leading cutoff correction
to the width to mass ratio and to $\pi - \pi$ scattering.

\subhead{6.1. Approximate calculation.} \taghead{6.1.}

We will first study the quantity $m_R$, defined by $$ {\rm Re} [\det M
(p)]_{p = (i m_R, \vec 0)} =0 \fs \eqno(e6p1p1)$$ {}From \(e5p2p7) we
obtain $$ \eqalign{ & \det M(p) = {1 \over 4} N^2 Z^3_\pi \eta^2 \cr &
\left[ \left( Q(p) - {2 \epsilon \over {Z_\pi^2 \eta^2 \beta_2}}
\right) \left( g(p) I(p) + 2 f^2_\pi \right) - g(p) r^2_0 - 2 r_0
f^2_\pi g(p) + \half f^2_\pi g(p)^2 I(p) \right] \fs \cr }
\eqno(e6p1p2)$$

We use $Z^{-1}_\pi = r_0 + f^2_\pi$ and solve \(e6p1p1) for $m^2_R ,~
f^2_\pi << 1$, counting logs to be of order 1. To leading order we
find, as expected, the form \(e4p1p2,e4p1p3) with the proportionality
constant $C$ now depending on $\beta_2$ $$ C(\beta_2) = \exp \left\{ 8
\pi^2 c_1 - { 8 \pi^2 r^2_0 \over {b - {2 \epsilon r^2_0 \over \eta^2
\beta_2}} } \right\} \fs \eqno(e6p1p3)$$

We see that $C$ diverges, as for the Pauli--Villars models, when the
tricritical point is approached. Again we want to be as far away from
this point as possible.

When we decrease $\beta_2$, $C(\beta_2)$ and hence $m_R$ (in lattice
units) becomes smaller at constant $g_R$ (which means that the mass is
kept fixed in physical units although $m_R$ varies). Then the cutoff
effects become smaller and $g_R$ can be allowed to increase.  Thus we
expect to obtain the largest mass bound at the smallest acceptable
$\beta_2$, \(e5p1p13). The ratio of the $C$'s for the standard
non-linear action and for this extreme case is found to be $$ { C(0)
\over C(- \beta_{2,t.c.}) } = \exp \left\{ { 4 \pi^2 r^2_0 \over b }
\right\} \fs \eqno(e6p1p4)$$ This evaluates to 4.521, 2.580 and 1.933
for $F_4$, HC and SI respectively.  According to the rough estimates of
section (4.2) we would expect the change of $\beta_2$ from $0$ to
$-\beta_{2,tc}$ to induce increases of about
$0.075~TeV,~~0.047~TeV,~~0.033~TeV$ in the Higgs mass bound for the
three cases. In the next section we shall see that these rough
estimates for the increase of the Higgs mass are accurate within
a factor of two.

As in section (4.6) for the Pauli--Villars case we can compute the
leading correction to $m_R$ given by \(e4p1p2) and \(e6p1p3). In the
notation of \(e4p6p2) we find $$ \delta_{m_{R0}} = 16 \pi^2 \left( c_2
- { \gamma r^2_0 \eta^4 \beta^2_2 \over { (b \eta^2 \beta_2 - 2
\epsilon r^2_0 )^2 }} \right) + {96 \pi^2 \over {g_{R0}}} \left( \zeta
- { b r_0 \eta^4 \beta^2_2 \over { (b \eta^2 \beta_2 - 2 \epsilon r^2_0
)^2 }} \right) \eqno(e6p1p5)$$
where $\zeta$ is defined through $g(p) = p^2 - \zeta p^4 + O(p^6)$ for
on-axis momenta $p$. From equations \(e5p2,e5p3,e5p4) we find $\zeta =
1/12$ for $F_4$ and HC and $\zeta = 0$ for SI. This correction is less
than 15\% in the region of interest.

We can make the lattice result look even more like the Pauli--Villars
one by going to continuum normalization. We denote the corresponding
pion wavefunction renormalization constant by $\tilde Z_\pi$, $\tilde
Z_\pi = \eta (\beta_0 + \beta_2) Z_\pi$. Then, using eq.~\(e5p2p4) and
\(e5p2p5) we obtain $$ \eqalign{ \eta (\beta_0 + \beta_2) = & ~ \tilde
Z_\pi (r_0 + f^2_\pi) \cr {b \over {2 \epsilon}} {\beta_2 \over
{(\beta_0 + \beta_2)^2}} = & ~ {\tilde Z_\pi - 1 \over {\tilde Z^2_\pi
}} \fs \cr} \eqno(e6p1p6)$$ Therefore we can now trade the couplings
$\beta_0$ and $\beta_2$ for $f^2_\pi$ and $\tilde Z_\pi$. The critical
line is now given by $f^2_\pi = 0$ and $0 \le \tilde Z_\pi \le 2$. The
upper limit corresponds to the tricritical point. $\tilde Z_\pi = 1$
denotes the standard non-linear lattice action, without the term with
four derivatives and four fields.

In this normalization the term $Q(p) - {2 \epsilon \over {Z_\pi^2
\eta^2 \beta_2}}$ in $\det M$ of eq.~\(e6p1p2) has to be replaced by $b
{(\tilde Z_\pi - 2) \over {(\tilde Z_\pi -1)}} + Q(p) - b$. The
proportionality constant in eq~\(e4p1p2) now depends on $\tilde Z_\pi$
and becomes $$ C(\tilde Z_\pi) = \exp \left\{ 8 \pi^2 c_1 - { 8 \pi^2
r^2_0 \over b }~ {{\tilde Z_\pi - 1} \over {\tilde Z_\pi - 2 }}
\right\} \eqno(e6p1p7)$$ to be compared to eq.~\(e4p1p4) for the
Pauli--Villars models.

However, keeping $\tilde Z_\pi$ fixed, while varying $f^2_\pi$ does not
correspond to keeping $\beta_2$ fixed. In a numerical simulation of the
lattice models one cannot easily follow a line of fixed $\tilde Z_\pi$.
It is much more natural to carry out simulations at a fixed value of
$\beta_2$, approach the critical point along this line and subsequently
vary $\beta_2$.  Therefore we will return in what follows to the usual
lattice normalization.

\subhead{6.2. Leading correction to the width to mass ratio.}
\taghead{6.2.}

Again, the computation in this section closely follows that of the
Pauli--Villars model, section (4.4). The Higgs resonance is a complex
root on the second sheet of $\det M(p) = 0$ for $p = (i (m_H - {i \over
2} \gamma_H), \vec 0 )$. $\det M$ is given in \(e6p1p2) where we again
insert $Z^{-1}_\pi = r_0 + f^2_\pi$. The leading order solution, with
notation \(e4p4p5) and \(e4p4p6) is $$ \eqalign{ \mu_0 = & ~ C(\beta_2)
\exp \left[ - {{\theta_0 + \pi }\over { 2 \tan (\theta_0 )}} \right]
\cr g_0 = & ~ {{96\pi^2 \cos^2 ({\theta_0 \over 2}) \sin
(\theta_0)}\over {\theta_0  + \pi }} \cr} \eqno(e6p2p1)$$ with
$C(\beta_2)$ given in eq.~\(e6p1p3).

To compute the subleading terms we expand again as in \(e4p4p9) and set
$\delta_g = 0$ (see \(e4p4p10) and the explanation leading to it).
Straightforward algebra then leads to $$ \bar \delta_{{\Gamma_H \over
M_H}} (g_0 ; \beta_2 ) = { 16 \pi^2 \left( c_2 - { \gamma r^2_0 \eta^4
\beta^2_2 \over { (b \eta^2 \beta_2 - 2 \epsilon r^2_0 )^2 }} \right)
\mu_0^2 \over { 1+(\pi + \theta_0 ) \left[ \tan \left( {\theta_0 \over
2} \right) - \cot (\theta_0 ) \right] }} \fs \eqno(e6p2p2)$$

As for the Pauli--Villars case, the cutoff correction factorizes into a
universal $g$ dependent part and a part carrying all the information
about the cutoff through the dependence on $\beta_2$. The latter is
given here by $16 \pi^2 [ c_2 - { \gamma r^2_0 \eta^4 \beta^2_2 \over {
(b \eta^2 \beta_2 - 2 \epsilon r^2_0 )^2 }} ] C^2(\beta_2)$.  The
universal factor is identical to the one obtained in the Pauli--Villars
case, eq.~\(e4p4p11).

On the lattice we are also interested in comparing directly to
numerical data obtained at $N=4$. For this comparison we need some
``unphysical'' quantity, like $m_H$.  We therefore write down the
explicit formula to first subleading order in the cutoff $$ m^2_H =
m^2_{H0} ( 1 + \delta_{m_{H0}} \mu^2_0 + O( \mu^4_0 ) ) \eqno(e6p2p3)$$
and obtain $$ \eqalign{ m_{H0} = & ~ \mu_0 \cos \left( {\theta_0 \over
2} \right) \cr \delta_{m_{H0}} = & ~ 16 \pi^2 \left( c_2 - { \gamma
r^2_0 \eta^4 \beta^2_2 \over { (b \eta^2 \beta_2 - 2 \epsilon r^2_0 )^2
}} \right) \cr & \left[ \cos \theta_0 + \sin(\theta_0 ) {{(\pi +
\theta_0) (1 + \cot (\theta_0) \tan ({\theta_0 \over 2})) - \tan
({\theta_0 \over 2} )} \over { 1 + (\pi + \theta_0 ) (\tan ({\theta_0
\over 2 } ) - \cot (\theta_0 ) )}} \right] \cr + & ~ { \pi + \theta_0
\over {\sin \theta_0 }} \left[ \zeta - { b r_0 \eta^4 \beta^2_2 \over {
(b \eta^2 \beta_2 - 2 \epsilon r^2_0 )^2 }} \right] \fs \cr}
\eqno(e6p2p4)$$

\subhead{6.3. Leading correction to $\pi - \pi$ scattering.}
\taghead{6.3.}

Scattering on the lattice was discussed in \cit{LW} for the HC
lattice and generalized in \cit{BBHN1} to the $F_4$ lattice. It is easy
to see that at first non-vanishing order in $1/N$ the invariant
amplitude $A(\hat s, \hat t,\hat u)$, where $\hat s$, $\hat t$ and
$\hat u$ are the lattice versions of the Mandelstam variables $s$, $t$
and $u$,\refto{BBHN1} describing $\pi - \pi$ scattering in the center
of mass frame, is only a function of the lattice center of mass energy
squared $\hat s$. For our purposes we do not need the full expressions
for the variables with hats and it suffices to know that $\hat s = W^2
+ O(W^6 )$ where $W$ is the center of mass energy. The differential
cross-section with equal isospin indices on the incoming pions is given
by $$ \left( {d \sigma \over {d \Omega}} \right)_{CM} = {N \over
{64\pi^2 W^2 }} |A(\hat s)|^2 \fs \eqno(e6p3p1)$$ The kinematic
prefactor has no cutoff corrections of order $1 / \Lambda^2$.  For the
$F_4$ and SI models this is true for incoming and outgoing momenta in
any direction, while for the HC model, due to the Lorentz invariance
breaking at order $1 / \Lambda^2$, this property only holds for on-axis
momenta. For the $F_4$ and SI models anisotropies in the differential
cross-section only occur at order $1 / \Lambda^4$.

To evaluate $A(\hat s)$ we need to know the pion interactions. For this
we go back to the action \(e5p5) and introduce pion and Higgs fields
via eq.~\(e5p7). We then expand around the saddle but do not integrate
over $\delta H$ and $\delta \vec \pi$ yet. The part of the action
giving the pion interaction is then $$ S_1^{int} = {\eta \over 2}
\int_x \delta \lambda (x) \left[ \int_y \delta \vec \pi (x) g_{x,y}
\delta \vec \pi (y) \right] + \half \int_x \delta \rho (x) (\delta \vec
\pi)^2 (x) \fs \eqno(e6p3p2)$$ The $\pi \pi \rho$ and $\pi \pi \lambda$
vertices are now easily obtained, and one can see that the latter
vanishes for on-shell pions, since then $g(p) = 0$. Thus, the on-shell
amplitude with $q = p_1 + p_2$ is given by $$ A(q) \propto ~ < \delta
\rho \delta \rho >(q) \fs \eqno(e6p3p3)$$

Following closely section (4.5) where the Pauli--Villars case was dealt
with and using the same notation, we find the same leading order result
for the invariant scattering amplitude, eq.~\(e4p5p6). Including the
subleading order we obtain on the lattice $$ \eqalign{ \bar
\delta_{|A|^2} ~ & = \left( c_2 - { \gamma r^2_0 \eta^4 \beta^2_2 \over
{ (b \eta^2 \beta_2 - 2 \epsilon r^2_0 )^2 }} \right) \mu_0^2 {\rm Re}
[N A_0 (R,g_0 )] \cr & \left[ 4R \cos^2 ({\theta_0 \over 2}) - \cos
(\theta_0)  - \sin(\theta_0 ) {{(\pi + \theta_0) (1 + \cot (\theta_0)
\tan ({\theta_0 \over 2})) - \tan  ({\theta_0 \over 2} )} \over { 1 +
(\pi + \theta_0 ) (\tan ({\theta_0 \over 2 } ) - \cot (\theta_0 ) )}}
\right] \fs \cr } \eqno(e6p3p4)$$ This cutoff correction has again
factorized into a cutoff dependent part and a function of $g_0$  which
is the same as the one found for the Pauli--Villars models,
eq.~\(e4p5p7). The cutoff dependent part is identical to the cutoff
dependent factor we have obtained for the width to mass ratio  in
equation \(e6p2p2).

\head{7. From $N=\infty$ to $N=4$.} \taghead{7.}

In this section we shall extract quantitative information about the
Higgs mass bound from our large $N$ results. To do that we need to
first discuss how large the corrections to the $N=\infty$ limit are
when $N=4$. Our objective will be to argue that, in the region of
interest for the bound, the $N=\infty$ numbers should be expected to be
different from the $N=4$ values by about 25 percent ($1/N$). The
question of the difference between $N=\infty$ and $N=4$ has been
studied in some detail before by Lin, Kuti and Shen\cit{LINTALLA}. They
arrived at the conclusion that the difference between $N=\infty$ and
$N=4$ is very large and that $N=\infty$ results are qualitatively wrong
at $N=4$. We disagree with this conclusion in the region where the
bound is close to saturation, and this is the region of interest to us.
Some of the approximations used in \cit{LINTALLA}  for the lattice
``bubble'' diagram are unjustified and we suspect this to be the reason
for the discrepancy between our views on the usefulness of the $1/N$
expansion for the Higgs mass bound problem.

Throughout this and the next section we undo the rescalings of $F_\pi$
and $f_\pi$ by $\sqrt N$ so that, at $N=4$ $F_\pi = 0.246~TeV$.

\subhead{7.1. Simple error estimates.} \taghead{7.1.}

The quantity that we are extracting quantitative information from is
the $\pi$--$\pi$ scattering invariant amplitude $A(q^2 )$. From it we
extracted the Higgs mass and width, $m_H$ and $\gamma_H$, the quantity
$m_R$ and the cutoff correction $\bar \delta_{|A|^2} $.  Moreover, from
the behavior of the amplitude when $q^2 \rightarrow 0$ we can also
extract $f_\pi^2$. Therefore our basic question is how accurately is
the ``true'', $N=4$, scattering amplitude $A(q^2 )$ approximated by the
$N=\infty$ expression for complex $q^2$ with $|q^2 | \le 4m_H^2$.

The best approach would be to compute the $1/N$ correction explicitly but
this is a demanding calculation which we have not done. We are attempting
to guess what the right order of magnitude of the result of such a
calculation would be. The most na\"{\i}ve guess is that the leading
correction will be suppressed by $1/N$, meaning that our results are good
to 25{\%} when applied to $N=4$. Indeed, the most conspicuous omission in
the $N=\infty$ limit is the fact that the number of pion types is $N-1$
rather than $N$; the na\"{\i}ve error estimate then follows from the
observation that pairs of pions of identical type contribute more or less
additively. There is one example where this sort of pion counting argument
is believed to be exact: When one computes the finite volume leading
correction to the $<\fhi\cdot\fhi>$ correlation at zero momentum in the
$1/N$ expansion, the leading and subleading terms are related by exactly a
factor of $N$\refto{NEUBFPI}. Moreover, it is believed that all higher
order corrections in $1/N$ vanish if the pion wave function renormalization
constant is extracted \cit{HJJLLN}.

It is well known \refto{DASHN} that one cannot always expect the $1/N$
corrections to obey these na\"{\i}ve estimates; this point was also
discussed in ref. \cit{LINTALLA}.  Let us repeat the argument here, but
rephrased for the amplitude $A(q^2 )$: If the system is sufficiently
close to criticality, the low $q^2$ behavior of $A$ can be extracted
from an appropriate RG equation and is dominated by the noninteracting
fixed point. The function $\beta (g)$ appearing in this equation can be
approximated by its leading term in $g$ and that term is calculable
because it comes from one loop diagrams.  It is known that, to one
loop, the beta function for the properly rescaled coupling is
proportional to $1+8/N$ and, thus, the finite $N$ answer is for $N=4$
larger by a factor of $3$ than the $N=\infty$ answer. This will affect
the amplitude because the solution of the RG equation will simply say
that $A(q^2 )$ is well approximated by the tree level expression with
an effective coupling that is dependent on $q^2$ with a dependence
dictated by the $\beta$-function. If the coupling is also weak at
cutoff scales, the effective coupling can be related to the bare
coupling via the one loop $\beta$ function; in this way one obtains
$A(q^2 )$ in terms of the bare coupling. If we are very close to the
critical point the cutoff is very large (for example, when compared to
the Higgs mass) and the effective coupling becomes almost entirely
independent of the bare coupling; in that case the large error in the
$\beta$ function induces a large error in $A(q^2 )$ and we can't trust
the numbers obtained at $N=\infty$ when $N=4$.  Note that the
difficulty arises due to the existence of two very disparate scales in
the problem:  One scale is the cutoff and the other is set by the pion
decay constant.  Two very different scales exist only when the system
is very close to a critical point because $f_\pi$ vanishes there.

For the Higgs mass bound problem we are mostly interested in cases
where $4\pi F_\pi \approx \Lambda$; in these cases the usefulness of
the RG equations is very limited because we do not have to connect two
very different scales.  Therefore, for those actions that lead to the
approximate saturation of the bound we do not need to worry about the
$8/N$ error in the $\beta$--function and can hope for accuracies of
order 25{\%} at $N=4$.  In the next section we shall use available
numerical data as evidence that this hope is realized in the broken
phase of the models we are interested in.

In the symmetric phase more is known about the $1/N$ series. For the
simplest nearest neighbor action on the hypercubic lattice one is able
to get into the region where the triviality bound on the self--coupling
is saturated (at a correlation length $\xi$ of about $2$ lattice
spacings) with an accuracy on the coupling of the order of 2--3{\%} if
one keeps three orders in $1/N$ at $N=4$ \cit{FLYV}.  This is
compatible with the assumption that the $1/N$ series for the coupling
has a radius of convergence of order unity when $\xi\approx 2$.

Also, regarding the value of the bound on the coupling in the symmetric
and broken phases, one can compare the results obtained for the single
component model ($N=1$) to those obtained for $N=4$; in spite of the
fact that there are no Goldstone bosons at $N=1$ the upper bound on the
coupling is relatively stable when expressed as a fraction of the tree
level unitarity bound \cit{{LW1},{LW2},{LW}}. Again the $1/N$ series
looks reasonably well behaved.

\subhead{7.2. Numerical test of error estimates.} \taghead{7.2.}

In Figures 7.1a to 7.1c we compare the large $N$ predictions with $N=4$
results for the simplest lattice models investigated to date by Monte
Carlo and independent non--perturbative means. For the $F_4$ lattice
(Fig. 7.1a) the numerical data follows the large $N$ prediction quite
consistently over the entire range of interest, $0.3 < m_H < 0.8$, with
the large $N$ numbers exceeding the $N=4$ numbers by 20--30{\%}. For
the hypercubic lattice the excess is slightly smaller, 15--20{\%} in
the region of interest, and for the hypercubic improved less than
15\%.

\figure{7.1}{\captionsevenone}{6.7}

More importantly we see that the $N=\infty$ {\bf difference} in $M_H/F_\pi$
between the various lattices in the region of interest  underestimates the
$N=4$ difference by up to a factor of two. Therefore, if we knew the $N=4$
results for only one of the lattices we could predict the results for the
others, using infinite $N$, to an overall relative accuracy of about
7-9{\%}, which is of the same order of magnitude as the statistical error
in a  typical Monte--Carlo simulation. One of our objectives is to predict
the results for the coupling at $N=4$ for new lattice actions. Being
interested in actions that are not fundamentally different from the ones
that have already been investigated by other means, we can expect our
predictions based on large $N$ to have an accuracy of the order of several
percent. Preliminary results in the $F_4$ case provide further support to
this claim \cit{LATT91}. This is very useful because one  does not want to
invest resources in simulating actions that end up not giving a higher
bound.   We believe that quite reliable order of magnitude estimates (and
definitely the right sign) for the effects of different terms in the action
on the bound can be obtained from the $N=\infty$ results.

Until now we only tested the large $N$ predictions against results that
could be obtained at $N=4$ by other non--perturbative means.  But the
estimates for the bound also need non--perturbative evaluations of the
cutoff effects contributing to physical observables.  There do not
exist any tested, practical, non-perturbative methods yet that make it
feasible to obtain numerical evaluations of physically observable
cutoff effects at $N=4$. It is possible that such methods will be
developed but it is doubtful that one will be able to reach reasonable
accuracies for estimating such small corrections by Monte Carlo
methods. Therefore we are left with only two
options for evaluating the cutoff effects:
the loop expansion and the $1/N$
expansion.  It is gratifying that the order of magnitude of the results
at tree level order for nearest neighbor hypercubic and $F_4$ actions
are in agreement with the large $N$ estimates when the coupling
constants are close to the upper bound. We could have worried about the
tree level estimates because, as explained in Appendix B, higher loop
effects are not negligible and a method for resumming the ``leading
log'' terms contributing to order $1/\Lambda^2$ terms is not yet
available. However, it seems that this does not have a major effect
when the coupling is sufficiently large, probably for the same physical
reasons that the $\beta$--function problem was relatively harmless for
large couplings. We are further helped by the fact that the bound
depends weakly on the magnitude of the cutoff effects:  Changes of one
order of magnitude in the latter have an effect on the bound of the
order of only a few percent. If we assume that, when the coupling is
sufficiently strong, the accuracy of the $N=\infty$ answer is of order
25{\%} for the cutoff effects too, we can use the large $N$ estimates
for the leading cutoff effects to estimate the Higgs mass bound in a
Monte Carlo calculation.  We have not devised a method for computing
the cutoff effects in the loop expansion for the more complicated
actions that were the subject of this paper because we feel that the
large $N$ results are at least as reliable. Nevertheless, we think that
it would be interesting to see the explicit $N$ dependence emerging
from a perturbative computation.

\subhead{7.3. The large $N$ numbers.} \taghead{7.3.}

Let us now go over the quantitative results of our work. We start with
the kind of results that are the typical outcome of a Monte Carlo
simulation.  In Figures 7.2a-d we show plots of the ratio $m_H / f_\pi$
as a function of the quantity $m_H$. The latter is equal to $M_H
/\Lambda$ and contains an implicit understanding for what the cutoff is
taken to be. For all the lattice results we take the cutoff as the
inverse lattice spacing of the elementary hypercubic lattice on which
the action is defined (the $F_4$ lattice is viewed here as a hypercubic
lattice for which a choice of action was made so that no fields reside
on sites who have an odd sum of integer coordinates). In the
Pauli--Villars case the cutoff was defined as the absolute magnitude of
the distance, in the complex momentum square plane, to the ghost pole
in the pion propagator closest to the origin.

\figure{7.2}{\captionseventwo}{6.7}

For the three lattice cases we show two lines on each graph. The lower
one gives $m_H /f_\pi $ for the simplest action which is bilinear in
the fields. From the higher line one can read off the increase in $m_H
/f_\pi$ that can be induced, at the same value of $M_H / \Lambda$, by
turning on the four field terms to maximal acceptable strength in the
direction of increasing repulsion between the pions. We see that the
largest relative effect takes place in the $F_4$ case while the
smallest relative effect takes place in the hypercubic ``Symanzik tree
level improved'' case. Since the lines for the simple actions give
higher values of $m_H /f_\pi$ for exactly those cases for which the
effect of the four field terms is smaller, we see a tendency towards
``equalization'' when the four--field terms are turned on. This fact
supports some sort of approximate ``universality'' among the various
bounds. Of course we cannot expect anything rigorous to hold in this
respect, but it is impressive that the higher lines for all three
lattice actions are closer to each other than the lower lines are.

In the Pauli--Villars case we can compare different values of $n$ and
see similar effects. While the overall ``plateau'' attained by the
higher curves is quite similar to the lattice cases, a detailed
comparison cannot be made because there is no ``natural'' relationship
between the ``cutoffs'' in the two schemes.\footnote{*}{At $N=\infty$
one can define the ratio of the Pauli--Villars cutoff to a particular
lattice cutoff by requiring exact matching of the coefficient $c$ in
equation \(e4p4p13). This is equivalent to an exact match of the cutoff
effects at order $1/\Lambda^2$. However, this match cannot always be
realized since some Pauli--Villars actions can make $c$ vanish and this
is impossible to achieve with the lattice actions that we
investigated.  In the text we meant by ``natural'' the sort of
relationship one would guess by just looking at the bare action, not
the more sophisticated relation described above in this footnote.}

There is a simple explanation for the systematics we observed: Suppose
each one of the actions is evaluated for slowly varying fields and a
field rescaling is carried out in accordance with eq. \(e2p5p2) to
remove the four derivative term from the term in the Lagrangian that is
bilinear in the fields (this cannot be done for the simple hypercubic
action and we shall discuss this presently). We see that the ordering
of the lattices by $m_H /f_\pi$ at fixed $m_H$ and with na\"{\i}ve
actions follows the magnitude of the four field coupling induced by the
field redefinition of \(e2p5p2). For the hypercubic lattice there is no
field redefinition that can eliminate the four derivative term but it
is clear that this term is ``weaker'' than the corresponding term on
the $F_4$ lattice, and obviously ``stronger'' than the absent term in the
improved case. We obtain for the first time an explanation for why
$F_4$ results for the bound have turned out to be smaller than the
hypercubic ones \cit{BBHN2} and preliminary numerical results with
``Symanzik improvement'' seem to give bounds \refto{ZIMMER} that are
larger than the ones obtained on hypercubic lattices with the simplest
action.\footnote{\d}{Very recent work \refto{GOCK2} reaffirms the
trend; the newer data points have been included in Figures 7.1.}

Similar logic works when one tries to understand the effects of the
four field terms. The bottom line is that one has full support for the
``rule'' that strengthening the repulsion between the pions in the low
momentum approximation to the action, viewed as a chiral effective
Lagrangian, increases the bound. Still, the answer is not entirely
universal, and there are some finer differences, especially between the
Pauli--Villars cases and the lattice cases.

We now turn to the cutoff effects. In order for the features observed
above to really imply that higher Higgs masses are possible when the
action is changed, we have to show that the above trends are also
obeyed by the cutoff effects. This would free us from the somewhat
arbitrary choice of comparing different lattice actions at the same
$m_H$ with a specific choice for the ``cutoff'' in each case. Because
the mechanisms explaining the observed trends, even with this
arbitrariness unresolved, are of a physical origin it is to be expected
that the cutoff corrections will indeed follow these trends. Therefore
just for the purpose of knowing which kind of action one should try to
simulate at $N=4$ the indications obtained up to this point should be
taken seriously. However, to make well defined statements we shall need
to know the magnitude of the cutoff effects in greater detail anyhow,
even after Monte Carlo data replaces some of the large $N$ graphs in
Figures 7.2a-c. We therefore show in Figures 7.3a-d the cutoff effects
on the width to mass ratio for the four cases under consideration. The
numerical value of these effects is rather small, and this is in
agreement with tree level perturbation theory in the simplest lattice
action cases. We prefer to use the more stringent cutoff
effects on the differential cross section for $\pi$--$\pi$ scattering
at ninety degrees in the center of mass frame. We chose this observable
to be able to compare with existing perturbative results. The cutoff
effects on the invariant amplitude square are shown in Figures 7.4a-d.
Note that the trends are the same as in Fig 7.3a-d.

\figure{7.3}{\captionseventhree}{6.7}

\figure{7.4}{\captionsevenfour}{6.7}

We see that the trends already seen in Figures 7.2a-d are respected.
The amount of possible increase in the bound is smallest for the
hypercubic improved case and largest for the $F_4$ case. We can
immediately read off from the horizontal axis the mass in $TeV$ that is
permissible for the Higgs mass if one puts some restriction on the
cutoff effects. Again one sees a tendency toward approximate
``universality'' with a bound on the Higgs mass at $N=\infty$ of about
$0.820~TeV$.\footnote{*}
{The figure also shows that one does not need
to set $\beta_2$ at exactly
$-\beta_{2,t.c.}$ to obtain an effect of similar
magnitude.} This is achieved with the smallest cutoff effects in a
Pauli--Villars type of regularization. What seems to be special about
our PV regularizations is that the pion propagator has no order $p^4$
term; this is also true of the SI action and can also be achieved on
the $F_4$ lattice, but at the cost of introducing next nearest neighbor
interactions.\footnote{\d}{While it is true that order $p^4$ terms in
the bare action can be removed by field redefinitions, the nonlinear
relationship between the bare parameters in the action and the cutoff
effects implies that the elimination of the $p^4$ terms by field
redefinition is not exact.} The following is an ``improved'' action on
an $F_4$ lattice with  a tunable ``triangle term'' $$ \eqalign{ S= & ~
-2N(\beta_0 +\beta_1 + \beta_2 ) \left [ 2\sum_{<x,x'>} \Fhi (x) \cdot
\Fhi (x') -{1\over 2} \sum_{<x,x'''>} \Fhi (x) \cdot \Fhi (x''') \right
] \cr & ~ -N\beta_1 \sum_{<x,x'>} [ \Fhi (x) \cdot \Fhi (x')-1 ]^2 \cr
& ~ -N~{{\beta_2}\over{8}} \sum_x \sum_{{<ll'>}\atop {l,l' \cap  x \ne
\emptyset ,~ l\cap x' \ne\emptyset ,~ l'\cap x'' \ne\emptyset ,~
x,x',x'' {}~{\rm all~ n.n.}} } \left[ \left( \Fhi (x) \cdot \Fhi (x')
-1 \right) \left( \Fhi (x) \cdot \Fhi (x'')-1 \right) \right] \cr}
\eqno(e7p3p1)$$ where $x, x',x'',x'''$ denote sites, $<x,x'>,l,l'$
links, and  $<x,x'''> $ next nearest neighboring pairs. The field is
constrained by ${\Fhi}^2 (x) =1$. In \(e7p3p1) only sites separated at most
a distance of two, in units of the embedding hypercubic lattice, are
coupled.

\figure{7.5}{\captionsevenfive}{4.8}

To see more clearly how the bound changes with the action we show an
example in Figure 7.5 for the $F_4$ lattice: identical cutoff effects,
of 4{\%}, will be found at $N=\infty$ for a Higgs mass of $0.720~TeV$
with the simplest action, and for a Higgs mass of $0.815~TeV$ with an
action that has the maximal amount of four field interaction we have
allowed. It is important to understand that this difference is
substantial when the width is considered. In Figure 7.6 we show the
large $N$ Higgs width as a function of the Higgs mass (note that the
leading order weak coupling approximation severely underestimates the
width when the Higgs becomes heavier). In our example the width went
from $0.320~TeV$ to about $0.500~TeV$ and the heavier Higgs is
definitely strongly interacting. There is no doubt therefore that, at
least at infinite $N$, stopping the search for the Higgs mass bound at
the study of the simplest possible actions would have been misleading.

\figure{7.6}{\captionsevensix}{4.8}

\subhead{7.4. Predictions for $N=4$.} \taghead{7.4.}

We now turn to the most speculative, but quite important part of our
work, namely numerical predictions for the physically relevant case at
$N=4$. Our approach is as follows: For the simplest actions (which are
bilinear in the fields) numerical results at $N=4$ are already
available. We compare the $M_H / F_\pi$ ratios we calculated at
$N=\infty$, viewed as functions of $m_H$, to these numbers. We then use
the difference between the $N=\infty$ and $N=4$ numbers, at a given
$m_H$, as an estimate for the same difference for each one of the new
actions that have not yet been studied numerically.  In this way we are
able to make predictions for the new actions at $N=4$ based on our
$N=\infty$ results. We can test this method by assuming that we have
numerical results for only one of the simple actions and ``predict''
the numerical results for another simple action. We find that our
``predictions'' are good to a few percent in the region where cutoff
effects on the $\pi-\pi$ scattering cross section are of the order of a
few percent. Comparisons between the $N=\infty$ and $N=4$ results for
the simplest  lattice actions are given in figures 7.1a for $F_4$, 7.1b
for HC, and 7.1c for SI.\footnote{*}{Our plots include very recent data
that appeared in \cit{GOCK2}; this preprint became available during the
final stages of preparing our paper for submittal. We have not yet had
the time for a thorough study of \cit{GOCK2}.}

For the $F_4$ lattice, we see from figure 7.1a that for the range of
interest,\footnote{\d}{This is the range in which the bounds are
obtained when $-\beta_{2,t.c.} \le \beta_2 \le 0$.} $m_H > 0.5$ (with
$m_H$ the Higgs mass in lattice units of the large $N$ calculation at
$N=4$), the large N results are larger than the numerical ones by less
than $0.150~TeV$. For the HC lattice, figure 7.1b indicates that for
$m_H > 0.4$ the large $N$ results are larger by less than $0.100~TeV$.
Finally, for the SI case, figure 7.1c shows that for $m_H > 0.3$ the
large $N$ results are larger by less than $0.110~TeV$. From these
figures we also see that the large $N$ results for $M_H$ are
consistently larger than the numerical ones.  Therefore, given a large
$N$ estimate for $M_H$, we predict that the corresponding value,
calculated in a numerical simulation at $N=4$, will not exceed the
large $N$ estimate.  We also predict that it will not be lower by more
than the amounts given at the beginning of this paragraph, provided
that $m_H$ stays within the ranges of interest indicated there.

The largest Higgs mass is obtained when the four derivative term is
turned on to maximal allowed strength. Using the guidelines of the
previous paragraph and figures 7.3a -- 7.3c or 7.4a -- 7.4c
 we can estimate the Higgs mass bound for a given amount of allowed
cutoff effects. We decide to allow up to 4\% cutoff effects in the
cross section at $W \le 2M_H$ as computed in large $N$. For the $F_4$
case we find that the bound should be between $0.660$ and $0.810~TeV$;
this prediction is in agreement with our preliminary data \cit{LATT91}.
For the HC it should be between $0.740$ and $0.840~TeV$, and for the SI
between $0.750$ and $0.860~TeV$.  The second number in each case is the
large $N$ result ignoring the likely overestimates and hence is a very
conservative estimate for the bound.

Although we do not have any large $N$ results corresponding to the
action \(e7p3p1), we expect the bound to increase by about $0.050~TeV$
from the value given for the $F_4$ lattice in the previous paragraph,
to somewhere between $0.710$ and $0.860~TeV$.  This is expected because
the action \(e7p3p1) should be closer to the PV case, and from figures
7.4a and 7.4d we see that the Higgs mass bound given by the PV case is
larger than the one given by the $F_4$ lattice by at least
$0.050~TeV$.

We see that if we allow up to 4\% scaling violations in the cross
section at $W \le 2M_H$, we should expect, without being overly
conservative, a Higgs mass bound of about $0.750~TeV$.  Renormalized
perturbation theory suggests that such a heavy Higgs particle will have
a width of at least $0.210~TeV$.  The large $N$ results indicate an
even larger width, around $0.290~TeV$, already subtracting a possible
overestimation by about 25{\%}. It is unclear whether such a wide Higgs
particle can still be seen on the lattice without applying more
sophisticated techniques than mere direct measurement.

The approach of \cit{GOCK2} is not based on direct measurement of the
mass in the $I=0,~J=0$ channel but relies quite heavily on the
applicability of ordinary perturbation theory. We know from our large
$N$ computations that ordinary perturbation theory does not work that
well for $\Gamma_H / M_H$ when $M_H / F_\pi$ gets close to the bound.
However it is not ruled out that the particular quantities that are
calculated in perturbation theory in \cit{GOCK2} do have a ``better''
perturbative expansion. Using the expression for the variance of  the
magnetization squared in \cit{TALLA}\footnote{\b}{Equations (6) and
(7a-b) there; see also the related discussion.} (evaluated there also
for the purpose of finding alternative definitions for the scalar
self--coupling, definitions that lead to quantities that are more
easily accessible numerically and don't suffer from finite width
contamination) one could test the validity of perturbation theory in
the precise setting  of \cit{GOCK2}, at least at infinite $N$.

It is important to stress that all our predictions in this subsection
are relative to the presently generally accepted numerical values at
$N=4$. Thus, if there are systematic errors in the $N=4$ results, for
example due to ignoring finite width effects in the direct Monte Carlo
measurements and/or due to the application of perturbation theory in
\cit{GOCK2}, these errors will propagate into our predictions.

\head{8. Summary.}

In this paper we have studied the regularization scheme dependence of
the Higgs mass triviality bound in an $O(N)$ scalar field theory to
leading order in $1/N$. All possible leading cutoff effects can be
induced by adding only a few higher dimensional operators with
adjustable coefficients to the standard $\lambda \Phi^4$ action. We
have found that the highest Higgs masses are obtained in the non linear
limit, at infinite $\lambda$. This has led us to consider non linear
actions with the leading cutoff effects induced by four derivative
terms with tunable couplings.

We used one class of continuum regularization schemes, of Pauli Villars
type (PV), and three kinds of lattice regularizations: $F_4$,
Hypercubic (HC), and Symanzik Improved Hypercubic (SI).  For these
regularization schemes the action, when expanded for slowly varying
fields to order momentum to the fourth power, is of the form $$
\eqalign{ S_c&=\cr &\int_x ~ \left [{1\over 2} \vec \phi (-\partial^2
+2 b_0 \partial^4 )\vec \phi - {b_1 \over {2N}} (\partial_\mu \vec \phi
\cdot \partial_\mu \vec \phi )^2 - {b_2 \over {2N}} (\partial_\mu \vec
\phi \cdot \partial_\nu \vec \phi - {1\over 4} \delta_{\mu , \nu }
\partial_\sigma \vec \phi \cdot \partial_\sigma \vec \phi )^2
\right]\cr} \eqno(e2p5p1) $$ where $\fhi^2 = N\beta$.  There are four
control parameters in this action. One is redundant since to
this order it can be absorbed into the other parameters by a field
redefinition:  The parameter $b_0$ can be absorbed in $b_1$ and $b_2$
by $$ \fhi \rightarrow {{\fhi - b_0 \partial^2 \fhi }\over{\sqrt{\fhi^2
+ b_0^2 (\partial^2 \fhi )^2 -2b_0 \fhi\partial^2 \fhi }}}\sqrt{N\beta}
\fs \eqno(e2p5p2) $$ However, eliminating the dependence in $b_0$ to
order momentum to the fourth power in the bare action does not
necessarily imply that the dependence in $b_0$ has been eliminated to
leading order
in the inverse cutoff from physical observables. The vacuum fluctuations
are ``aware'' of the full bare action and will carry that information
to the non universal part of the physical observables.
For example the parameter $n$ of the PV case, although associated with
an operator of dimension larger than four, does appear in the
relation between bare and renormalized parameters and in the leading
cutoff corrections to the width and cross section. Similarly for the
lattice regularizations, some dependence on the full structure of the
lattice propagators comes in through the constants $r_0$, $c_1$, $c_2$,
and $\gamma$. Because of this, the effects of $b_0$, after the field
redefinition, are not completely absent from the leading cutoff
corrections to the physical observables.  Still, the effect of $b_0$ is
probably small and, as far as the value of the Higgs mass bound is
concerned, one may be able to cover the whole range of leading cutoff
effects by varying $b_1$ only. Our results indicate that this is
realized to a good extend, leading to an approximate universality of
the Higgs mass triviality bound.

For PV we simply set $b_0=0$, and for the lattice regularizations $b_0$
is set to the na\"\i ve value obtained in the expansion of the lattice
kinetic energy term.  We calculated the phase diagram for each
regularization and found a second order line that ends at a tricritical
point where a first order line begins.  We studied the physically
interesting region close to the second order line in the broken phase.
The parameter $\beta$ corresponds to the relevant direction and is
traded, as usual, for the pion decay constant $f_\pi$.  The parameters
$b_1$ and $b_2$ control the size of the leading order cutoff effects
for a given value of $f_\pi$.  Our analysis has shown that, to this
order, the quantities we considered do not depend on the parameter
$b_2$.  Therefore, a very simple situation emerged with the scale set
by $f_\pi$ and the leading cutoff effects parametrized by only one
parameter $b_1$.  This simplification at infinite $N$ makes it easy to
relate very different regularization schemes, and leads to a reasonably
``universal'' bound on the renormalized charge $g$.

We calculated the leading cutoff effects on two physical quantities,
namely the width of the Higgs particle and the $90$ degrees
$\pi-\pi$ scattering cross section. We found that they are given by
the product of a ``universal'' factor (identical for all
regularizations considered) which only depends on $g$ and the
dimensionless external momenta, and a non--universal factor that
depends on the parameter $b_1$ but not on $g$ or the momenta. The
universal factor associated with the width is different from the one
associated with the cross section, but the non--universal factors are
identical for a fixed regularization scheme.  These properties suggest
that, to leading order in $1/N$, all cutoff effects in on--shell
dimensionless physical quantities (viewed as functions of dimensionless
momenta) are given by an effective renormalized action, $$ S_{eff}=S_R
+ c \exp [-{{96\pi^2 }\over g} ] {\cal O} \cm \eqno(e4p4p13) $$ where
${\cal O}$ is a renormalized operator, $c$ is a $g$ independent free
parameter containing all the non--universal information (i.e. a
function of $b_1$ only), and $S_R$ is describing the usual universal
part of physical observables with the unit of energy set by $f_\pi =1$.
In the general RG framework this representation of $S_{eff}$ is not
unreasonable if one accepts that at $N=\infty$ the number of
independent operators that contribute to observables at order
$1/\Lambda^2$ decreases by one relatively to $N<\infty$. Because of the
nonlinear relationship between $c$ and the parameters in the bare
action, the range in which $c$ is allowed to vary depends on the type
of action chosen. Different actions that realize the same $c$ are
indistinguishable to order $1/\Lambda^2$; all actions have large
regions of total overlap but the triviality bounds are obtained from
the area near the edges of the ranges in which $c$ varies and therefore
are somewhat dependent on the particular regularization scheme. Still,
this dependence turns out to be weaker than what one would have been
inclined to believe on the basis of the few simulations that had no
continuous tuning ability for $c$ built in. Thus, some of the stronger
``universality'' evident in \(e4p4p13) also makes its way into the
triviality bound.

We developed an approximate physical picture associated with the models
we studied.  When the regularized model is nonlinear one has to think
about the Higgs resonance as a loose bound state of two pions in an
$I=0$, $J=0$ state. Pions in such a state attract because superposing
the field configurations corresponding to individual pions makes the
state look more like the vacuum and hence lowers the energy. The four
derivative term in the action can add or subtract to this attraction.
We found that the smallest cutoff effects are obtained when the
coupling $b_1$ of the four derivative term is set so that the term
induces the maximal possible repulsion between the pions, postponing
the appearance of the Higgs resonance to higher energies.

Our findings concerning the Higgs mass triviality bound are summarized
by figures 7.3a -- 7.3d and 7.4a -- 7.4d for the four different
regularization schemes considered. There the leading cutoff effects in
the width and cross section are plotted versus the Higgs mass in $TeV$
with the large $N$ results presented for $N=4$.  We decided to extract
the bound from figures 7.4a -- 7.4d since the cross section cutoff
effects provide a more stringent criterion. For all regularization
schemes we extracted the bound by restricting the allowed cutoff
effects in the cross section to 4\%, at center of mass energies up to
twice the Higgs mass.

To make predictions for the physically relevant case $N=4$ we used the
known differences between the $N=\infty$ and the $N=4$ numerical
results, for the simplest lattice actions, to extrapolate to the
actions with four derivative terms that have not yet been studied
numerically. We argued that the expansion in $1/N$ is expected to be
``well behaved'' in the region where the triviality bound is obtained,
and because of that this extrapolation is sensible.

Based on what we have learned, it seems that a more realistic and not
overly conservative estimate for the Higgs mass triviality bound is
$0.750~TeV$, and not $0.650~TeV$ as it is sometimes stated. At
$0.750~TeV$ the Higgs particle is expected to have a width of about
$0.290~TeV$ and is therefore quite strongly interacting.

\head{Acknowledgments.} This research was supported in part by the DOE
under grant \# DE-FG05-85ER\discretionary{}{}{}250000 (UMH and PV) and
under grant \# DE-FG05-90ER40559 (HN). UMH also acknowledges partial
support by the NSF under grant \# INT-8922411 and he would like to
thank F. Karsch and the other members of the Fakult\"at f\"ur Physik at
the University of Bielefeld for the kind hospitality while part of this
work was done.

\head{Appendix A.} \taghead{A.}

In this appendix we show one example of the calculations needed to
derive the asymptotic expansions of the ``bubble'' integrals in
\(e4p1). We shall work out explicitly only
the simplest bubble, $I_{0,0} (p^2 )$. The other bubbles are
calculated by similar means.
Our problem is to calculate the first two terms in the expansion in
$p^2$ of $$ \eqalign{ I_{0,0} (p^2) =& ~ \int_k {1\over {K({1\over 2}
p-k) K ({1\over 2} p+k)}}\cr {1\over {K(q)}} =& ~ {1\over {q^2
(1+q^{2n})}}={1\over {q^2}} -{1\over n} \sum_{j=1}^n
{1\over {q^2 +w_j}}\cr w_j =& ~ \exp [i\phi_j ] ~,~~~~~\phi_j
={{\pi}\over{n}} (2j-1-n)~~{\rm for}~j=1,2,\dots,n \fs \cr
} \eqno(Ap1)$$

 We rewrite the inverse propagator as $$ {1\over {K(q) }} =\sum_{j=0}^n
 {{C_j}\over {q^2+w_j }} ~~~~{\rm with}~~ w_0 \equiv 0 \fs \eqno(Ap2)$$
The values of the constants $C_j$ can be read off \(Ap1).  We now
analytically continue in all the $w_j$, $j=1,\dots,n$, moving them to
high positive imaginary parts.  The bubble integral is now evaluated
with this new propagator. We also analytically continue in the
dimensionality $d$ and first compute $I_{0,0} (p^2 )$ with the new
propagator in $d$ dimensions and then take the limit $d\rightarrow 4$
$$ B(p ) = \lim_{d\to 4^{-}} \sum_{0\le i,j \le n} C_i C_j \int {{d^d
k}\over {(2\pi )^d }} {1\over {[({1\over 2} p-k )^2 +w_j ] [({1\over 2}
p+k )^2 +w_i  ]}}\fs \eqno(Ap3)$$ At the end we intend to analytically
continue the $w_j$'s back to their values on the unit circle.

We intend to use Feynman's formula $$ {1\over {AB}} =\int_0^1
{{ds}\over {[sA+(1-s)B]^2}} \fs \eqno(Ap4)$$ This is allowed as long as
the straight line connecting the relevant points $A$ and $B$ never
passes through the origin. This is ensured by all the $w_j$'s being in
the upper half of the complex plane and thus making the segments
connecting $A$ and $B$ also stay in the upper half plane for all real
values of the momenta.  We introduce \(Ap4) into \(Ap3) and do the
angular part of the $k$ integration.  After a subsequent partial
integration over the magnitude of $k$ we obtain $$ B(p )= \lim_{d\to
4^{-}} {{d-2}\over {2^d \pi^{d\over 2} \Gamma ({d\over 2})}} \sum_{0\le
i,j \le n} C_i C_j \int_0^1 ds \int_0^{\infty} {{dk k^{d-3}}\over{k^2
+s(1-s)p^2+sw_i+(1-s)w_j}} \fs \eqno(Ap5)$$ Let $\Phi_s$ be the phase
of $z_s \equiv s(1-s)p^2+sw_i+(1-s)w_j $. It is clear that $0\le \Phi_s
< \pi$. We can deform the $k$ integral from the positive axis to a ray
starting from the origin at an angle $\Phi_s /2$. This ray will always
be in the first quadrant of the complex plane. The contribution from
the arc at infinity is seen to vanish. Along the ray we can easily do
the integral over $k$ and obtain $$
 B(p )= \lim_{d\to 4^{-}} {{d-2}\over {2^d \pi^{d\over 2} \Gamma
 ({d\over 2})}} \sum_{0\le i,j \le n} C_i C_j \int_0^1 ds
[s(1-s)p^2+sw_i+(1-s)w_j]^{{d-4}\over 2} \fs \eqno(Ap6)$$ The phase of
the integrand is ${{d-4}\over 2} \Phi_s$ so that the associated sheet
of the logarithm is the first sheet cut along the negative real axis in
the usual way.

We now take the limit on the dimension and obtain, with an
unambiguously defined logarithm, $$ \eqalign{ B(p )=& ~ \lim_{d\to
4^{-}} \left ({2\over {d-4}} -\gamma_E \right )
 ( \sum_{j=0}^n C_j )^2\cr -{1\over {16\pi^2}}& ~ \sum_{0\le i,j \le n}
C_i C_j \int_0^1 ds \int_0^{\infty} ds \log [ s(1-s)p^2+sw_i+(1-s)w_j
]\fs \cr} \eqno(Ap7)$$ The first term vanishes because $\sum_{j=0}^n
C_j =0$. We separate out the terms containing $w_0 $ $$ \eqalign{ B(p )
=& ~ J_1 +J_2 +J_3 \cr J_1 =& ~ -{1\over {16\pi^2}} \int_0^1 ds \log
[s(1-s)p^2 ]\cr J_2 =& ~ {1\over {8\pi^2 n}} \sum_{j=1}^n \int_0^1 \log
[s(1-s)p^2 +sw_j ]\cr J_3 =& ~ -{1\over{16\pi^2 n^2}} \sum_{i,j=1}^n
\int_0^1 ds \log [s(1-s)p^2 +sw_i +(1-s)w_j ] \fs \cr} \eqno(Ap8)$$
Since none of the integrals goes through the cut along the negative
real axis they can be expanded in $p^2$ quite easily leading to $$
\eqalign{ &J_1 = -{1\over {16\pi^2 }} \log p^2 +{1\over {8\pi^2 }}\cr
&J_2 = {1\over{8\pi^2 n}} \sum_{j=1}^n \left [ {{p^2}\over {2w_j}}
+\log w_j -1 \right ] + O((p^2 )^2 )\cr &J_3  = -{1\over {16 \pi^2
n^2}} \sum_{j=1}^n \left [ {{p^2}\over {6w_j}} +\log w_j  \right ]
-{1\over {16 \pi^2 n^2}}\cr \sum_{1\le i\ne j \le n} & \left [  {{w_i
\log w_i -w_j \log w_j}\over{w_i -w_j }} -1 +p^2 \left ( {1\over 2}
{{w_i +w_j }\over {(w_i -w_j )^2 }} - {{w_i w_j}\over {(w_i -w_j )^3}}
( \log w_i -\log w_j ) \right ) \right ]\cr
 &+  O((p^2 )^2 ) \fs \cr } \eqno(Ap9)$$

Finally we take the $w_j$'s back to their original values, always
staying on the first sheet, and never crossing the cut. We know then
that in the above equations $\log w_j = i\phi_j $ for $j\ge 1$ with the
angles $\phi_j$ given by \(Ap1). The summations can now be performed by
using $$ \sum_{j=1}^n {1\over {z+w_j}} ={{nz^{n-1}}\over {1+z^n}}
={d\over {dz}} \log [1+z^n ]\eqno(Ap10)$$ and derivatives of this
identity.

The calculation of the other bubble integrals to the same order needs
some additional manipulations but is essentially of the same type as
the one outlined above.

\head{Appendix B.} \taghead{B.}

In this appendix we shall evaluate the scaling violations in
$\beta$--functions defined along particular lines in the space of bare
couplings of the simplest model $$ S=\int_x \left [ {1\over 2} \fhi K
\fhi +{1\over 2} m_0^2 \fhi^2 +{{g_0^2} \over {4N}} (\fhi^2 )^2 \right
]\fs \eqno(Bp1)$$ We shall work in the symmetric phase for simplicity.

Introducing the auxiliary field $\sigma={{g_0^2}\over N}\fhi^2$ we get
a new action $$ S_1 = {N\over 2} {\rm Tr} \log [K+m_0^2 +\sigma ] -
{N\over {4g_0^2 }} \int_x \sigma^2 \fs \eqno(Bp2)$$

Defining the bubble integral $$ J(m^2 )=\int_p {1\over {K(p^2 )+m^2 }}
\eqno(Bp3)$$ one obtains for the renormalized coupling $g^2$ $$ g^2
={{g_0^2}\over {1-g_0^2 J^\prime (m^2 )}}\eqno(Bp4)$$ where the prime
denotes differentiation. The mass $m^2$ is given by the saddle point
equation $$ m^2 =m_0^2 +g_0^2 J(m^2 ) \fs \eqno(Bp5)$$

Suppose now that $g_0^2 (s)$, $m_0^2 (s )$ are a parametric description
of a path in parameter space and we calculate along this path the
variation of $g^2$ with respect to $m^2$. We obtain $$ \eqalign{
{{d\log g^2 }\over {dm^2}}= & ~ g^2 J^{\prime\prime} (m^2 ) + {{d\log
g_0^2} \over {dm_0^2 }} X\cr X= & ~ {{1+g^2 J^{\prime} (m^2 ) }\over
{1+g^2 J^{\prime} (m^2 ) +J(m^2 ) g^2 {{d\log g_0^2}\over{dm_0^2 }} }}
\fs \cr}\eqno(Bp6)$$ It must be true that ${{d\log g_0^2}\over{dm_0^2
}}\ne \infty$ because one cannot get to the critical surface without
varying $m_0^2$. Let us now consider the non--linear limit $g_0^2
\rightarrow \infty$ $ m_0^2 \rightarrow -\infty $ with $g_0^2 /m_0^2
\rightarrow~~\xi$.  One can still have a path in parameter space ending
at a critical point by varying $\xi$ and it makes sense to compute
${{d\log g^2 }\over {dm^2}}$ along this path. At $g_0^2 = \infty$ one
gets $g^2 J^\prime (m^2 ) =-1$ (see \(Bp4)) and therefore $X=0$ leading
to $$ {{d\log g^2 }\over {dm^2}}=g^2 J^{\prime\prime} (m^2 ) \fs
\eqno(Bp7)$$ This answer is independent of the sequence of $s$
dependent paths we took the large $g_0^2,~ -m_0^2$ limit on, as it
should be.

Let us now compute in ordinary perturbation theory to one loop order.
One gets $$ {{d\log g^2 }\over {dm^2}} =g^2 J^{\prime\prime} (m^2 ) +
{{d\log g_0^2}\over {dm_0^2 }} \left [ 1-g^2 J(m^2 ) {{d\log
g_0^2}\over {dm_0^2 }} \right ] \eqno(Bp8)$$ with an explicit
dependence on the sequence of paths, except if one happened to choose
${{d\log g_0^2}\over{dm_0^2 }}=0$, but this is not necessary as the
following example shows explicitly:  $$ \eqalign{ m_0^2 ~ &
={{1-2\lambda}\over {s}} -8 \cr g_0^2 ~ & ={{\lambda}\over {s^\alpha
}}\fs \cr }\eqno(Bp9)$$ The nonlinear limit is attained when $\lambda
\rightarrow \infty$. We see that, unless $\alpha = 0$, we shall have an
answer that is different from the exact answer in \(Bp7) and that
higher orders in $g^2$ that have been neglected depend on $\alpha$ in
such a way as to conspire to add up to give an $\alpha$ independent
answer. The problem can be easily traced now to the fact that the
quantity $X$ in \(Bp6), which vanishes in the nonlinear limit, does so
only as a result of keeping all the powers in $g^2 J^\prime (m^2)$ that
come in from the denominator. Since $J^\prime (m^2) \propto \log m^2 $
we see that the culprit for finding a spurious $\alpha$ dependence in
the leading order expansion of an $\alpha$ independent quantity is the
truncation in the number of loops. As pointed out in section (4.3)
graphs with an infinite number of loops contribute even at leading
order in the physical coupling constant to the coefficient of the
leading correction in inverse powers of the cutoff. All we are
stressing here is that the unsolved problem of summing up the
logarithms in the subleading corrections in the inverse cutoff is a
serious problem.  Maybe the $1/N$ expansion is the right place to
tackle this issue.

\head{APPENDIX C.} \taghead{C.}

In section 2 the linear model was discussed for the Pauli--Villars
regularization and it was shown that the largest Higgs mass is obtained
in the nonlinear limit of the model, where the four-point coupling
$\lambda$ is taken to infinity. The action was restricted for
simplicity to a three dimensional parameter space instead of the four
dimensional one we really need to be concerned with (see
introduction).  In this appendix we will consider an action with four
free parameters and again show that the largest Higgs mass is obtained
in the nonlinear limit.  However, our analysis is somewhat less general
than the one in section 2 because it only covers a region close to the
critical line. In section 2 we could do a better analysis because, with
the particular Pauli--Villars regularization employed there, we had a
simple closed formula for the ``bubble'' diagram, while here we work on
the lattice and closed formulae for the ``bubbles'' are not available.
We will use the $F_4$ lattice to regularize the model.

The action of the linear model with four bare parameters can be written
as\refto{BBHN1} $$ \eqalign{ S= & ~ -2\kappa \sum_{<x,x'>} \Fhi (x)
\cdot \Fhi (x') + \sum_x \Fhi^2(x) + {\lambda \over N} \sum_x
[\Fhi^2(x)-N]^2 \cr & ~ + {\eta \over N^2} \sum_x [\Fhi^2(x)]^3 +
{\eta\prime \over N} \sum_{<x,x'>} \Fhi^2(x) \cdot \Fhi^2(x') \fs \cr}
\eqno(Cp1)$$ To make the action bilinear on the $\Fhi$ fields we
introduce (as in section 2.1) two auxiliary fields $\omega$ and
$\sigma$ by inserting $$ \prod_x \int_{-\infty}^{\infty} d\sigma (x)
{N\over{2\pi}} \int_{-i\infty}^{i\infty} d\omega (x) e^{N \omega(x)
[\sigma (x) -{{\fhi^2 (x)}\over N}]} =1 \eqno(Cp2)$$ into the
functional integral. The zero mode of the $\Fhi$ field is separated as
in \(e2p1p7) and the $N=\infty$ pion propagator can be immediately read
off. From the pion propagator and using standard conventions (see
\(e3p3p5) and \(e3p3p6)) the pion wave function renormalization
constant is found to be $$ Z_\pi={1 \over 6\kappa} \fs \eqno(Cp3)$$

As in section 2.1, to get the saddle point equations we integrate the
$H$ and $\vec \pi$ fields and take the $ N \rightarrow \infty$ limit.
The saddle point equations are $$ \eqalign{ & {1 \over 6\kappa}
J_1(\omega_s\prime)+v^2=\sigma_s  \cr & (1-2\lambda-24
\kappa)+2(\lambda+12\eta\prime)\sigma_s +3\eta\sigma_s^2= 6 \kappa
\omega_s\prime \cm \cr} \eqno(Cp4)$$ with $$ \omega_s\prime={\omega_s
\over 6 \kappa} - 4 \eqno(Cp5)$$ and with $J_1$ the $F_4$ lattice
integral defined in \(e5p1p2).  The symmetric phase is characterized by
$v^2=0$, $\omega_s\prime>0$ and the broken phase by $v^2>0$,
$\omega_s\prime=0$. The phase diagram is obtained with the method
described in sections 3.1 and 5.1. It can be drawn on a two dimensional
plane spanned by the parameters $a$ and $b$, where $$ \eqalign{ &
a={\lambda + 12 \eta\prime \over (6 \kappa )^2 } \cr & b=2-{1-2\lambda
\over 12 \kappa }-r_0{\lambda+12\eta\prime \over (6 \kappa)^2} \cr}
\eqno(Cp6)$$ with $r_0=J_1(0)$ (see \(e5p1p5) ). There is a second
order line described by $$ b=hr_0^2, \ \ \ a > -2hr_0, \ \ \ {\rm
with}\  h={3 \eta \over 2 (6 \kappa)^3} \fs \eqno(Cp7)$$ This second
order line turns into a first order line at the point
$a=-2hr_0,~b=hr_0^2$. This point is a tricritical point. We see that
the generic phase structure we obtained for the non--linear actions can
also be found in the linear case.

The physically relevant part of the phase diagram is the region near
the second order line in the broken phase. In the broken phase the
first of the saddle point equations \(Cp4) gives $$ \kappa={r_0+\fp^2
\over 6 \sigma_s} \eqno(Cp8)$$ where we have used $v^2 = Z_\pi \fp^2$
(equation \(e4p1p3) ). We use this equation to trade the parameter
$\kappa$ for $\fp$.

Following sections 2.2 and 2.3 we obtain the propagators in Fourier
space in terms of a matrix $M$ as in \(e2p3p6). $M$ is given by $$
 M=\left ( \matrix{ {g(p) (r_0+\fp^2) \over \sigma_s} & {\sqrt N} v & 0
 \cr {\sqrt N} v & -{N \sigma_s^2 I(p) \over 2 (r_0+\fp^2)^2} & - {N
 \over 2} \cr 0 & -{N \over 2} & N [ \lambda + 12 \eta\prime + 3 \eta
 \sigma_s - 3 \eta\prime g(p)] \cr} \right ) \eqno(Cp9)$$ with
$I(p)=-{1\over 16 \pi^2} log p^2 + c_1 - c_2 p^2 + O(p^4log p^2)$ (see
\(e5p2p8)), the ``bubble'' integral defined in \(e5p2p7) and $g(p)$,
the kinetic energy term on the $F_4$ lattice given in \(e2p5p5).

Close to the second order line and using an analysis similar to the one
in section 4.1 we obtain equation \(e4p1p2) with the non--universal
constant $C$ depending on $\lambda$, $\eta$, and $\eta\prime$ $$ {m_R^2
} \approx C^2 (\lambda, \eta, \eta\prime )  \exp [ - 96\pi^2  /g_R ]
\eqno(Cp10)$$ where $$ C(\lambda, \eta, \eta\prime ) = \exp\left [
8\pi^2 c_1 + {4 \pi^2 r_0^2 \over \sigma_c^2 (\lambda + 12 \eta\prime +
3\eta \sigma_c)}\right ] \fs \eqno(Cp11)$$ In \(Cp11)
$\sigma_c=\sigma_c(\lambda, \eta, \eta\prime)$, the critical value of
$\sigma_s$, is determined from equations \(Cp6,Cp7) with $\kappa_c={r_0
\over 6 \sigma_c}$.

{}From \(Cp10,Cp11) it is clear that for a given $m_R$ the maximum
$g_R=3{m_R^2 \over f_\pi^2}$ is obtained when $C(\lambda, \eta,
\eta\prime)$ is the smallest. If the factor $\lambda+12\eta\prime+3
\eta \sigma_c$, appearing in the exponent in \(Cp11), could take negative
values then $C$ would have been the smallest for those values.  However,
this factor cannot take negative values because of the second of
eq.~\(Cp7) which when written out is $$ \lambda+12\eta\prime+3 \eta
\sigma_c > 0 \fs \eqno(Cp12)$$ This is a consequence of the fact that
the physical region is the one close to the second order line and does
not include the tricritical point or the region close to the ``would
be" second order line ($ b=hr_0^2, \ a < -2hr_0 $) which corresponds to
an unstable saddle point.  Therefore for given $\eta, \eta\prime$ the
minimum value of $C(\lambda,\eta,\eta\prime)$ is obtained for $\lambda
\rightarrow \infty$. We see, therefore, that the maximal Higgs mass is
again obtained in the non--linear limit, at least as long as we
restrict the observable cutoff effects sufficiently to justify the
neglect of subleading cutoff effects in \(Cp10).

\references

\refis{PLBHNV} U. M. Heller, H. Neuberger and P. Vranas \pl B283, 1992,
335.

\refis{ROME} H.Neuberger, talk given at the workshop {\sl
``Non-perturbative Aspects of Chiral Gauge Theories''}, Rome, 1992, to
appear in the proceedings.

\refis{TALLA} H. Neuberger, in {\sl ``Lattice Higgs Workshop''}, eds.
Berg et. al., World Scientific, (1988).

\refis{PLBF4} H. Neuberger, \pl B199, 1987, 536.

\refis{CAPRI} H. Neuberger, \np B (Proc. Suppl) 17, 1990, 17.

\refis{LW} M. L\"uscher and P. Weisz, \np B318, 1989, 705.

\refis{KUTIPRL} J. Kuti, L. Lin and Y. Shen, \prl 61, 1988, 678.

\refis{WHIT} C. Whitmer, {\sl ``Monte Carlo Methods in Quantum Field
Theory''}, unpublished Princeton University dissertation (1984).

\refis{TSYP} M. M. Tsypin, in {\sl ``Lattice Higgs Workshop''}, eds.
Berg et. al., World Scientific, (1988).

\refis{BBHN2} G. Bhanot, K. Bitar, U. M. Heller and H. Neuberger, \np
B353, 1991, 551.

\refis{BBHN1} G. Bhanot, K. Bitar, U. M. Heller and H. Neuberger, \np
B343, 1990, 467.

\refis{BMB} W. Bardeen, M. Moshe and M. Bander, \prl 52, ,1983, 1188.

\refis{LATT91} U. M. Heller, M. Klomfass, H. Neuberger and P. Vranas,
\np B (Proc. Suppl) 26, 1992, 522; and work in progress.

\refis{BERGERDAVID} M. C. Berger and F. David, \pl  B135, 1984, 412.

\refis{EINH} M. Einhorn, \np B246, 1984, 75.

\refis{EINHTALLA} M. Einhorn, in {\sl ``Lattice Higgs Workshop''}, eds.
Berg et. al., World Scientific, (1988).

\refis{WEINBERG} S. Weinberg, \physica, 96A, 1979, 327.

\refis{GEORGI} H. Georgi, {\sl ``Weak Interactions and Modern Particle
Theory''}, Addison-Wesley, (1984).

\refis{COLEMAN} S. Coleman, R. Jackiw, H. D. Politzer, \pr D10, 1974,
2491.

\refis{CHIVU} R. S. Chivukula, M. J. Dugan, M. Golden, preprint
HUTP-92/A025.

\refis{LINTALLA} L. Lin, J. Kuti, Y. Shen, in {\sl ``Lattice Higgs
Workshop''}, eds. Berg et.  al., World Scientific, (1988).

\refis{FLYV} H. Flyvbjerg, F. Larssen, C Kristjansen, \np B (Proc.
Suppl) 20, 1991, 44; C. Kristjansen, H. Flyvbjerg, preprint,
NBI-HE-90-55.

\refis{LW1} M. L\"uscher and P. Weisz, \np B290[FS20], 1987, 25.

\refis{LW2} M. L\"uscher and P. Weisz, \np B295[FS21], 1988, 65.

\refis{NEUBFPI} H. Neuberger, \prl 60, 1988, 889.

\refis{HJJLLN} A. Hasenfratz, K. Jansen, J. Jersak, C. B. Lang, H.
Leutwyler and T. Neuhaus, \zfp C, 46, 1990, 257.

\refis{GOCK1} M. G{\"o}ckeler, K. Jansen, T. Neuhaus, \pl B273, 1991, 450.

\refis{GOCK2} M. G{\"o}ckeler, H. A. Kastrup, T. Neuhaus and F.
Zimmermann, preprint HLRZ 92--35/PITHA 92-21.

\refis{DASHN} R. Dashen and H. Neuberger, \prl 50, 1983, 1897.

\refis{ZIMMER} M. G{\"o}ckeler, H. A. Kastrup, T. Neuhaus and F.
Zimmermann, \np B (Proc. Suppl) 26, 1992, 516.

\endreferences

\if \epsfpreprint N \endpage

\head{Figure Captions}

\halign{#\hfill\qquad &\vtop{\parindent=0pt \hsize=5in \strut#
\strut}\cr
Figure 3.1: & \captionthreeone \cr
&\cr
Figure 3.2: & \captionthreetwo \cr
&\cr
Figure 7.1: & \captionsevenone \cr
&\cr
Figure 7.2: & \captionseventwo \cr
&\cr
Figure 7.3: & \captionseventhree \cr
&\cr
Figure 7.4: & \captionsevenfour \cr
&\cr
Figure 7.5: & \captionsevenfive \cr
&\cr
Figure 7.6: & \captionsevensix \cr
}

\fi

\endit